\documentclass[a4paper, twocolumn, 11pt, accepted=2026-09-26]{quantumarticle}
\pdfoutput=1
\usepackage[utf8]{inputenc}
\usepackage[english]{babel}
\usepackage[T1]{fontenc}
\usepackage{amsmath}
\usepackage{amssymb}
\usepackage{physics}
\usepackage[dvipsnames]{xcolor}
\usepackage{graphicx}
\usepackage{wrapfig}
\usepackage{subcaption}
\usepackage{wasysym}
\usepackage{hyperref}
\usepackage{cleveref}
\usepackage{cite}

\crefname{equation}{Eq.}{Eqs.}
\Crefname{equation}{Equation}{Equations}
\crefname{figure}{Fig.}{Figs.}
\Crefname{figure}{Figure}{Figures}
\crefname{section}{Sect.}{Sects.}
\Crefname{section}{Section}{Sections}

\newcommand{\fullsubfigref}[2]{\hyperref[#2]{\ref*{#1}(\subref*{#2})}}
\newcommand{\creffullsubfigref}[2]{Fig. \hyperref[#2]{\ref{#1}(\subref*{#2})}}
\newcommand{\Creffullsubfigref}[2]{Figure \hyperref[#2]{\Cref*{#1}(\subref*{#2})}}

\renewcommand{\op}[1]{\hat{#1}}
\renewcommand{\ket}[1]{|#1\rangle}

\begin{document}

\title{Decoding Multimode Gottesman-Kitaev-Preskill Codes with Noisy Auxiliary States}

\author[1]{Marc-Antoine Roy$^{\dagger\text{,}}$}
\author[1, 2]{Thomas Pousset$^{\dagger\text{,}}$}
\author[1]{Baptiste Royer}
\affiliation[1]{Institut Quantique and Département de Physique, Université de Sherbrooke, Sherbrooke, Québec J1K 2R1, Canada}
\affiliation[2]{Telecom Paris, Institut Polytechnique de Paris, 91120 Palaiseau, France}

\begin{abstract}
    In order to achieve fault-tolerant quantum computing, we make use of quantum error correction schemes designed to protect the logical information of the system from decoherence. A promising way to preserve such information is to use the multimode Gottesman-Kitaev-Preskill (GKP) encoding, which encodes logical qubits into several harmonic oscillators. In this work, we focus on decoding the measurements obtained from Steane-type quantum error correction protocols for multimode GKP codes. We propose a decoder that considers the noise present on the auxiliary states, more specifically by tracking the correlations between errors on different modes spreading throughout the error-correction circuit. We show that leveraging the correlations between measurement results and the actual error affecting the multimode GKP state can decrease the logical error probability by up to more than an order of magnitude, yielding more robust quantum computation.
\end{abstract}

\maketitle

\section{Introduction}
\label{sec: intro}

\def\thefootnote{$\dagger$}\footnotetext{These authors contributed equally to this work}\def\thefootnote{\arabic{footnote}}

In order to achieve fault-tolerant quantum computing, one of the major challenges is the stability of the qubits. Because we have to interact with these qubits to operate and measure them, coupling with the environment is inevitable and this open system evolution induces noise, reducing the lifetime of the qubits \cite{nisqEraQuantum,nielsen2010}. To tackle this issue, the information in the qubit is redundantly encoded in noisy physical systems of higher dimension using a quantum error correction (QEC) code. The dimension of the physical system used can either be finite for qubit and qudit codes \cite{initQubitCodes1,initQubitCodes2,initQubitCodes3,initQubitCodes4,initQubitCodes5,gottesmanThesis,initSurfaceCode} or infinite for bosonic codes \cite{initBosonicCodes1,initBosonicCodes2,initBosonicCodes3,originalGKP,initBosonicCodes4,initBosonicCodes6,initBosonicCodes7,initBosonicCodes5,initBosonicCodes8}. Hybrid codes combining both qudit codes and bosonic codes have also been studied \cite{AnalogGKPSurface,RepCatQubitCode,QECwithToricGKP,nohFaultTolerantGKP,XZZXKerrCatQubitCode,MBQCSurfaceCodeGKP,colorCodeGKP,blueprintXanaduGKP,nohLowOverheadFaultTolerantGKP,GKPAndQLDPCConstruction,RepGKPQubitCode,lin_2023a,XZZXSurfaceCodeGKP,XanaduAurora,BicycleCodesXanadu,hybridOscillatorQuditVVAlbertPaper} and experimentally realized \cite{hydridCatRepCode}.

In this work, we investigate QEC with bosonic error correcting codes, more specifically GKP codes \cite{originalGKP}. The GKP code is an error-correcting code based on grid states and is tailored to correct small shifts in the quadratures of phase space. Moreover, there is strong numerical evidence that the GKP code is optimal against the bosonic pure-loss channel \cite{albertGKPBetterCode,nohGKPBetterCode,leviant_2022}. Although many approaches have been proposed to realize such a code in experimental platforms \cite{theoricPrepGKP1,theoricPrepGKP2,theoricPrepGKP3,theoricPrepGKP4,theoricPrepGKP5,prepGKPPhaseEstimation,theoricPrepGKP7,theoricPrepGKP8}, it has only been realized recently in superconducting microwave cavities \cite{campagneGKPQEC,eickbushGKPQEC,sivakRealtimeQEC,NQGKPQEC,brockGKPQudits}, planar microwave resonators \cite{MITGKPPaper}, the motional modes of trapped ions \cite{fluhmannEncodingQubitTrappedion2019,deNeeveTrappedIonGKP,matsosTrappedIonGKP,TrappedIonGKPSensing} and in photonic systems \cite{fabreGenerationTimefrequencyGrid2020,FurusawaGKPStatePrep,XanaduSquareGKPStatePrep}. 

A more general formulation of GKP codes was also included in Ref. \cite{originalGKP}, considering the encoding of a qubit into many harmonic oscillators: multimode GKP codes. These were studied in more detail in Refs. \cite{GKPRatesHarrington,conradGKPMultimode,bapGKPMultimode}. Notably, it was shown that the stabilizer group of these codes is isomorphic to a lattice, reflecting the fact that the logical code words can be understood as multidimensional grid states in phase space. Our interest towards these codes is motivated by the fact that some of these lattices have increased code distance, encoding more robust logical qubits.

QEC of multimode GKP codes requires coupling to auxiliary states, allowing one to measure these auxiliaries without perturbing the code space of the multimode GKP code. Two main approaches were studied in the literature for the type of auxiliaries used. The first strategy uses fresh GKP states and precise quadrature measurements~\cite{originalGKP, GlancyKnillGKP, QECwithToricGKP, nohFaultTolerantGKP, nohLowOverheadFaultTolerantGKP}, and the second uses single-bit phase estimation with two-level systems such as transmon qubits \cite{prepGKPPhaseEstimation, campagneGKPQEC, deNeeveTrappedIonGKP, sivakRealtimeQEC, brockGKPQudits}. Here, we focus on the former.

For such QEC schemes, it has been shown that these codes could be corrected with closest lattice point decoding under random shifts of the oscillator modes \cite{conradGKPMultimode,lin_2023a}. More precisely, for a general multimode GKP code, the correction process involves solving the closest vector problem (CVP), for which the best known classical algorithms have an exponential runtime in the dimension of the lattice studied \cite{schnorrLatticeBasisReduction1994,agrellClosestPointSearch2002}. The direct decoding of multimode GKP codes is therefore not scalable as the number of modes increases. A way to bypass this complexity problem is to consider GKP codes concatenated with qubit codes \cite{AnalogGKPSurface,QECwithToricGKP,nohFaultTolerantGKP,MBQCSurfaceCodeGKP,colorCodeGKP,blueprintXanaduGKP,nohLowOverheadFaultTolerantGKP,GKPAndQLDPCConstruction,RepGKPQubitCode,lin_2023a,XZZXSurfaceCodeGKP,BicycleCodesXanadu}.

In this paper, we focus on QEC of multimode GKP codes, where we take into account noise on the auxiliary GKP states used in Steane-type error-correction protocols. In particular, we leverage correlations between the result of the measurement realized and the actual error affecting the multimode GKP state being corrected. We show that even with noisy auxiliaries, multimode GKP codes can be decoded as efficiently as using a decoder that does not consider this noise. Notably, the resulting decoder is also a CVP-like decoder, yielding the same classical computing complexity.

We note that using these correlations was investigated in Refs. \cite{blueprintXanaduGKP,LossToRGTNoiseModel2,XanaduAurora,BicycleCodesXanadu}, in a measurement-based quantum computing architecture where the logical information and the computation are carried through a special hybrid code called a cluster-state. Notably, the decoder introduced takes into account the correlations of the errors for the cluster-state preparation step and during the actual computation. These correlations were also studied for the same single-mode code in Ref. \cite{nohLowOverheadFaultTolerantGKP}. Here, we fully generalize these approaches to any multimode GKP code.

The paper is organized as follows. We start by giving necessary notions to understand the link between multimode GKP codes and lattices in \cref{sec: gkp_theory}. Then, we discuss the considered noise model studied in \cref{sec: noise_model}. The decoder we propose for noisy auxiliaries QEC of multimode GKP codes is presented in \cref{sec: qec_multimode_gkp}. Particularly, we present the protocol previously studied in the literature with perfect auxiliary states and noisy ones in \cref{subsec: qec_perf_aux,subsec: qec_noisy_aux}, respectively, and we show our main result in \cref{subsec: what_we_propose}. We follow with numerical simulations justifying our decoder's performance compared to previous QEC schemes in \cref{sec: res}. Conclusion and final remarks are presented in \cref{sec: conclu}.

\section{Multimode GKP codes and lattices}
\label{sec: gkp_theory}

In this section, we lay out the notation and definitions we use. Afterwards, we introduce multimode GKP codes and present them from a lattice point of view \cite{originalGKP,GKPRatesHarrington,conradGKPMultimode,bapGKPMultimode}. We follow with more practical notions for these codes such as a definition of distance, their finite-energy version, how to operate multimode GKP codes, and finally, an example of a single-mode GKP code that is crucial to the QEC protocol we discuss later in \cref{sec: qec_multimode_gkp}.

We study the physical Hilbert space of $m$ harmonic oscillators. We impose $\hbar = 1$ and define the quadrature coordinate operators of the $j$th mode as $\hat{q}_j = (\hat{a}_j + \hat{a}_j^\dagger) / \sqrt{2}$ and $\hat{p}_j = -i(\hat{a}_j - \hat{a}_j^\dagger) / \sqrt{2}$, with $\hat{a}_j$ ($\hat{a}_j^\dagger$) the annihilation (creation) operator of the corresponding mode. We choose the $\texttt{qpqp}$ ordering for the quadrature operators such that they are arranged in a quadrature coordinate vector $\hat{\mathbf{x}}=\hspace{-0.25cm}\begin{array}{ccccc}(\hat{q}_1 & \hat{p}_1 & \dots & \hat{q}_{m} & \hat{p}_m)^T. \end{array}$ From these definitions, the symplectic form $\Omega$ is represented as a $2m \times 2m$ matrix with elements $[\hat{x}_j, \hat{x}_k] = i\Omega_{j,k}$, or
\begin{align}
    \Omega = \bigoplus_{j = 1}^{m} \begin{pmatrix}
        0 & 1 \\
        -1 & 0
    \end{pmatrix},
\end{align}
describing the commutation relations between all $\hat{q}_j$ and $\hat{p}_k$ operators. We make use of the translation operator
\begin{align}
    \hat{T}(\mathbf{u}) = e^{-i\ell \mathbf{u}^T\Omega \hat{\mathbf{x}}},
    \label{eq: translation_operators}
\end{align}
that applies a translation by $\mathbf{u}$ in phase space in units of $\ell = \sqrt{2\pi}$. Its action on $\hat{\mathbf{x}}$ is given by
\begin{align}
    \hat{T}^\dagger(\mathbf{u}) \hat{\mathbf{x}}\hat{T}(\mathbf{u}) = \hat{\mathbf{x}} + \ell \mathbf{u}.
    \label{eq: translation_with_x_commute_rel}
\end{align}
The commutation relation between two translation operators is given by
\begin{align}
    \hat{T}(\mathbf{u}_1)\hat{T}(\mathbf{u}_2) &= \hat{T}(\mathbf{u}_2)\hat{T}(\mathbf{u}_1)e^{-i2\pi\omega(\mathbf{u}_1, \mathbf{u}_2)},
    \label{eq: translation_commute_rel}
\end{align}
where they commute if the symplectic product of their corresponding vectors
\begin{align}
    \omega(\mathbf{u}_1, \mathbf{u}_2) = \mathbf{u}_1^T\Omega \mathbf{u}_2 = -\omega(\mathbf{u}_2, \mathbf{u}_1),
    \label{eq: symplect_prod}
\end{align}
is an integer. Two translations can be composed by adding the corresponding phase
\begin{align}
    \hat{T}(\mathbf{u}_1)\hat{T}(\mathbf{u}_2) = \hat{T}(\mathbf{u}_1 + \mathbf{u}_2)e^{-i\pi\omega(\mathbf{u}_1, \mathbf{u}_2)}.
    \label{eq: translation_compose_rel}
\end{align}
Note that in this work, we are using the term "translation" operators instead of the more widely known displacement operators $\hat{D}_j(\alpha_j)$ on the $j$th mode \cite{gaussianQuantumInfo}, and the two are related by a factor of $\sqrt \pi$ justified by our chosen units of $\ell$ for $\hat{T}(\mathbf{u})$.

With these definitions in hand, we can now define multimode GKP codes \cite{originalGKP}. Multimode GKP codes are a type of stabilizer code where the stabilizers are translation operators defined by \cref{eq: translation_operators}. For an $m$ mode GKP code there are $2m$ independent stabilizers generating the infinite stabilizer group
\begin{align}
    \mathcal{S} = \langle \hat{T}(\mathbf{s}_1), \hat{T}(\mathbf{s}_2), \dots, \hat{T}(\mathbf{s}_{2m})\rangle.
    \label{eq: GKP_stab_group}
\end{align}
Each of these translation operators is specified by a vector $\mathbf{s}_j$, forming the linearly independent ensemble $\{\mathbf{s}_j\}$. We also define the operator generating the translation by $\mathbf{s}_j$ as
\begin{align}
    \hat{g}_j = \ell \mathbf{s}_j^T\Omega \hat{\mathbf{x}},
    \label{eq: stab_GKP}
\end{align}
such that $\hat{T}(\mathbf{s}_j) = \mathrm{e}^{-i\hat{g}_j}$. The ensemble of vectors $\{\mathbf{s}_j\}$ must respect certain constraints in order for them to define a valid multimode GKP code encoding a qubit. Details on these constraints follow shortly.

From the definition of the stabilizer group \cref{eq: GKP_stab_group}, there is an isomorphism between $\mathcal{S}$ and a lattice $\Lambda \subset \mathbb{R}^{2m}$ with basis vectors given by the ensemble $\{\mathbf{s}_j\}$. This lattice has for basis matrix
\begin{align}
    S = \begin{pmatrix}
        \mathbf{s}_1^T \\
        \mathbf{s}_2^T \\
        \vdots \\
        \mathbf{s}_{2m}^T
    \end{pmatrix},
\end{align}
with the $j$th row being the vector $\mathbf{s}_j$. Thus, $S$ is a full-rank matrix describing the stabilizer lattice $\Lambda$ of the $m$ mode GKP code. The lattice $\Lambda$ comprises the set of points in $\mathbb{R}^{2m}$ that are integer linear combinations of the vectors $\mathbf{s}_j$
\begin{align}
    \Lambda = \{S^T\mathbf{a} ~|~ \mathbf{a} \in \mathbb{Z}^{2m}\}.
\end{align}
Viewing multimode GKP as lattices has many advantages. Notably, the constraints on $\{\mathbf{s}_j\}$ can be directly expressed as constraints on the lattice $\Lambda$, chosen such as to encode a valid multimode GKP code directly from $S$.

For a stabilizer code to be valid, each stabilizer must commute with one another, requiring that the stabilizer group $\mathcal{S}$ be abelian. This defines the first constraint on $\Lambda$, or correspondingly on $S$. Its symplectic Gram matrix $A$ must be integral, or equivalently
\begin{align}
    A = S\Omega S^T \in \mathbb{Z}^{2m \times 2m}.
    \label{eq: S_gram_mat}
\end{align} 
Indeed, this can be seen directly from \cref{eq: translation_commute_rel}, where we have $A_{j,k} = \omega(\mathbf{s}_j, \mathbf{s}_k)$.

For stabilizer codes, the code space $\mathcal{C}$ is the set of states invariant by $\mathcal{S}$, meaning
\begin{align}
    \ket{\psi} \in \mathcal{C} \iff \hat{T}(\mathbf{s}_j)\ket{\psi} = \ket{\psi} ~\forall j.
    \label{eq: stab_main_condition}
\end{align}
Logical operators are given by the centralizer of the stabilizer group, here the set of all translation operators that commute with translations in $\mathcal{S}$. This way, the symplectic dual lattice, which we refer to as the logical lattice, is defined as
\begin{align}
    \Lambda^\perp = \{\boldsymbol{\lambda}^\perp ~|~ \mathbf{s}_j^T\Omega\boldsymbol{\lambda}^\perp \in \mathbb{Z} ~\forall j\}.
    \label{eq: lambda_perp_lattice_points_def}
\end{align}
Each point in $\Lambda^\perp$ corresponds to a representative for a logical operator of the multimode GKP code. From \cref{eq: translation_commute_rel,eq: lambda_perp_lattice_points_def}, we can directly compute a generator matrix for $\Lambda^\perp$ such that
\begin{align}
    S^\perp \Omega S^T = \mathbb{I} \implies S^\perp = A^{-1}S,
    \label{eq: dual_basis_gen_mat}
\end{align}
where we used the definition of $A$, \cref{eq: S_gram_mat}. Here, we have chosen a particular basis where $\omega(\mathbf{s}_j^\perp, \mathbf{s}_k) = \delta_{j, k}$ with $\mathbf{s}_j^\perp$ the rows of $S^\perp$, the basis matrix of $\Lambda^\perp$. Subsequently, a basis change can be performed with a unimodular matrix $R$ to obtain a more useful representation of $\Lambda^\perp$. Since all basis vectors in $S$ respect $\omega(\mathbf{s}_j, \mathbf{s}_k) \in \mathbb{Z}$, \emph{i.e.}, elements in $\mathcal{S}$ commute with themselves, they are necessarily included in $\Lambda^\perp$ such that $\Lambda \subseteq \Lambda^\perp$. The number of distinct logical operators is given by
\begin{align}
    \bar{n}^2_\mathcal{C} = \qty|\frac{\text{det} S}{\text{det} S^\perp}| = |\text{det}A|,
\end{align}
which specifies the number of encoded logical states $\bar{n}_\mathcal{C}$. In this work, we focus on multimode GKP codes that encode a qubit. This defines the second constraint on $\Lambda$; we impose $\bar{n}_\mathcal{C} = 2$.

Finally, we arrange the vectors generating the logical Pauli operators of the multimode GKP code in a $3 \times 2m$ matrix
\begin{align}
    P = \begin{pmatrix}
        \mathbf{x}_0^T \\
        \mathbf{y}_0^T \\
        \mathbf{z}_0^T
    \end{pmatrix},
    \label{eq: log_pauli_mat}
\end{align}
where each $\mathbf{p}_0$, with $\mathbf{p} \in \{\mathbf{x}, \mathbf{y}, \mathbf{z}\}$, is a representative of minimal Euclidean length in $\Lambda^\perp$. Methods to identify the logical vectors in $P$ are described in Refs. \cite{conradGKPMultimode,lin_2023a}.

In this work, we consider GKP qubits defined from the square, hexagonal, hypercubic (tesseract) and $\mathrm{D}_{2n}$ lattices. Their basis matrices are given in \cref{sec: appendix_multimode_gkp_code_examples}.

We also note that Ref. \cite{joe_logical_compiler} provides an analytical way of computing the Wigner function of multimode GKP logical code words $\ket{\psi} \in \mathcal{C}$.

\subsection{Distance of a multimode GKP code}
\label{subsec: distance_of_GKP_codes}

Before introducing the notion of distance for multimode GKP codes, we briefly introduce the concept of the Voronoi cell of a general lattice $\mathcal{L}$. We follow closely what is discussed in Ref.~\cite{voronoiPaper}. It is defined as
\begin{align}
    \mathcal{V}(\mathcal{L}) = \{ \mathbf{v} \in \mathbb{R}^{2m} ~|~ \forall \boldsymbol{\lambda} \in \mathcal{L} : \frac{1}{2}|\boldsymbol{\lambda}| \geqslant |\mathbf{v}_{{\boldsymbol{\lambda}}}|\},
    \label{eq: gen_voronoi_cell}
\end{align}
where $\mathbf{v}_{{\boldsymbol{\lambda}}}$ is the vector resulting from the projection of $\mathbf{v}$ onto ${\boldsymbol{\lambda}}$. In words, for a point in $\mathbb{R}^{2m}$, we verify if the projection of that point on a vector in the lattice ${\boldsymbol{\lambda}}$ is smaller than or equal to half of the length of ${\boldsymbol{\lambda}}$. If it is respected for every vector in $\mathcal{L}$, the point is included in the Voronoi cell. Equivalently, the Voronoi cell contains all the points closer to the origin than any other point in $\mathcal{L}$.

\begin{figure}[t!]
    \centering
    \includegraphics[scale=1]{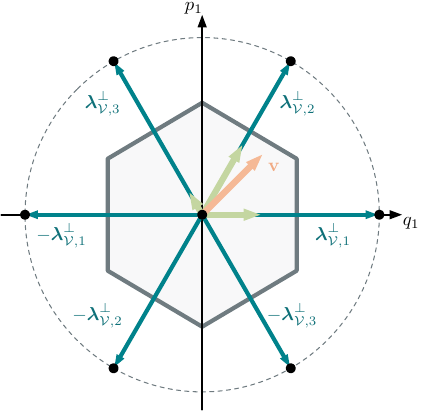}
    \caption{Logical Voronoi cell $\mathcal{V}^\perp$ (grey hexagon) for a GKP code defined on a hexagonal lattice. The black dots represent the logical lattice $\Lambda^\perp$ and the teal arrows represent all three Voronoi-relevant vectors $\boldsymbol{\lambda}_{\mathcal{V}, j}^\perp$ and their inverse. The beige arrow represents a vector $\mathbf{v}$ that has its projections $\mathbf{v}_{\boldsymbol{\lambda}_{\mathcal{V}, j}^\perp}$ on the Voronoi-relevant vectors smaller than half of their length. These projections are represented by the green vectors. As we can see, the condition \eqref{eq: gen_voronoi_cell} is respected for all Voronoi-relevant vectors, implying that $\mathbf{v}$ lies in the logical Voronoi cell. All points located in the grey hexagon also respect that condition, justifying the form of $\mathcal{V}^\perp$. Finally, we represent the distance $d$ by the radius of the dotted circle.}
    \label{fig: hexagonal_GKP_voronoi_cell}
\end{figure}

More specifically, to define the notion of distance for a multimode GKP code, we are interested in the logical Voronoi cell, the Voronoi cell of the logical lattice $\Lambda^\perp$ denoted $\mathcal{V}^\perp \equiv \mathcal{V}(\Lambda^\perp)$. The polytope representing the logical Voronoi cell has at most $2(2^{2m} - 1)$ facets, and each facet bisects a vector $\boldsymbol{\lambda}_{\mathcal{V}}^\perp$. We refer to these vectors as Voronoi-relevant vectors, and we define their ensemble as $\Lambda_{\mathcal{V}}^\perp$. In fact, it is sufficient for $\mathbf{v}$ to verify condition \eqref{eq: gen_voronoi_cell} for vectors in $\Lambda_{\mathcal{V}}^\perp$ to define completely $\mathcal{V}^\perp$. \Cref{fig: hexagonal_GKP_voronoi_cell} shows a visual example of the logical Voronoi cell for a GKP code defined on a hexagonal lattice.

Consider a translation error $\boldsymbol{\xi}$ on the multimode GKP state, $\hat{T}(\boldsymbol{\xi})$. As mentioned in Ref. \cite{GKPRatesHarrington}, if $\boldsymbol{\xi} \in \mathcal{V}^\perp$, then $\hat{T}(\boldsymbol{\xi})$ is a correctable error. From this, we can establish a link between the multimode GKP code distance and the logical Voronoi cell.

For general qubit codes, the distance is defined as the weight of the shortest non-trivial logical operator \cite{lin_2023a}. Since, by construction, $\mathcal{V}^\perp$ contains at least all halves of the minimal length logical vectors, it becomes natural to define the distance of the multimode GKP codes as the Euclidean length of the shortest non-trivial vector in $\Lambda_{\mathcal{V}}^\perp$
\begin{align}
    d = \underset{\boldsymbol{\lambda}_{\mathcal{V}}^\perp \in \Lambda_{\mathcal{V}}^\perp - \{\mathbf{0}\}}{\text{min}} |\boldsymbol{\lambda}_{\mathcal{V}}^\perp|,
    \label{eq: GKP_code_distance}
\end{align}
in units of $\ell$. An example of such a distance for a hexagonal lattice GKP code is shown in \cref{fig: hexagonal_GKP_voronoi_cell}.

\subsection{Finite-energy multimode GKP codes}
\label{sec: finite_energy_multimode_gkp}

Multimode GKP codes defined on infinite lattices must be spread throughout all phase space and be infinitely squeezed. Experimental realization of such states is therefore impossible, requiring the injection of an infinite amount of energy into the physical system used to construct the multimode GKP codes.

Finite-energy multimode GKP code words are defined by \cite{bapSBSpaper,bapGKPMultimode}
\begin{align}
    \ket{\psi_\Delta} = N_{\psi,\Delta}\hat{E}(\Delta)\ket{\psi},
\end{align}
where $\ket{\psi} \in \mathcal{C}$. Here, $N_{\psi,\Delta}$ is a normalization factor, and $\hat{E}(\Delta) = \mathrm{e}^{-\Delta^2\hat{n}}$ is a Gaussian envelope operator that exponentially reduces the probability of measuring the state $\ket{\psi}$ far off in phase space. Here, $\hat{n}$ is the total number of photons in all modes. The strength of this damping is controlled by the squeezing parameter $\Delta$, which is here the same for each mode for simplicity. In the limit $\Delta \rightarrow 0$, we retrieve the perfect infinite-energy code words $\ket{\psi}$.

In this work, the QEC protocol we study assumes the preparation of infinite-energy multimode GKP states. Nevertheless, our work also applies for finite-energy multimode GKP states. Indeed, the code words $\ket{\psi_\Delta}$ can be understood as noisier versions of the perfect code words $\ket{\psi}$ \cite{originalGKP,nohFaultTolerantGKP}. Since QEC protocols are designed to correct errors on $\ket{\psi}$, they behave in a similar way on states defined by $\ket{\psi_\Delta}$. The noise strength in the system we investigate throughout this work can be related to the squeezing parameter $\Delta$, which can be defined in units of dB \cite{nohLowOverheadFaultTolerantGKP}
\begin{align}
    \Delta_{(\text{dB})} = -10 \text{log}_{10}(2\Delta^2).
    \label{eq: finite_GKP_delta}
\end{align}
We discuss in further detail the link between finite-energy multimode GKP codes and the noise model we study in \cref{sec: noise_model}.

\subsection{Gaussian operations on multimode GKP states}
\label{subsec: operations_on_multimode_GKP}

In this work, we are investigating protocols that only require the use of Gaussian unitaries $\hat{U}_L$. These operators act on the quadrature coordinate vector following their metaplectic representation
\begin{align}
    \hat{U}_L^\dagger\hat{\mathbf{x}}\hat{U}_L = L\hat{\mathbf{x}},
    \label{eq: gaussian_with_x_commute_rel}
\end{align}
where $L$ is a symplectic matrix respecting $L^T\Omega L = \Omega$. From this, we can define their action on translation operators as
\begin{align}
    \hat{U}_{L}\hat{T}(\mathbf{u})\hat{U}_{L}^\dagger = \hat{T}(L\mathbf{u}).
    \label{eq: translation_commute_with_gaussian_rel}
\end{align}
We can also compose two Gaussian operators following
\begin{align}
    \hat{U}_{L_2}\hat{U}_{L_1} = \hat{U}_{L_2 L_1},
    \label{eq: compose_gaussian_ops}
\end{align}
which simply amounts to multiplying their corresponding symplectic matrix.

We show different Gaussian operators that are useful in this work. We first have a two-mode $qq$ quadrature coupling between mode $j$ and mode $k$
\begin{align}
    \hat{C}_{qq}^{j \rightarrow k}(\tau) = e^{-i\tau \hat{q}_j\hat{q}_k}~: \begin{array}{rl}
        \hat{q}_j \longrightarrow & \hat{q}_j \\
        \hat{p}_j \longrightarrow & \hat{p}_j - \tau\hat{q}_k \\
        \hat{q}_k \longrightarrow & \hat{q}_k \\ 
        \hat{p}_k \longrightarrow & \hat{p}_k - \tau\hat{q}_j 
    \end{array},
    \label{eq: quad_action_qq_op}
\end{align}
where we show its symplectic action on the quadrature operators. A similar two-mode $pp$ operator is defined between mode $j$ and mode $k$
\begin{align}
    \hat{C}_{pp}^{j \rightarrow k}(\tau) = e^{-i\tau \hat{p}_j\hat{p}_k}~: \begin{array}{rl}
        \hat{q}_j \longrightarrow & \hat{q}_j + \tau\hat{p}_k \\
        \hat{p}_j \longrightarrow & \hat{p}_j \\
        \hat{q}_k \longrightarrow & \hat{q}_k + \tau\hat{p}_j \\ 
        \hat{p}_k \longrightarrow & \hat{p}_k
    \end{array}.
    \label{eq: quad_action_pp_op}
\end{align}
Finally, we define a two-mode $qp$ quadrature coupling between mode $j$ and mode $k$
\begin{align}
    \hat{C}_{qp}^{j \rightarrow k}(\tau) = e^{-i\tau \hat{q}_j\hat{p}_k}~: \begin{array}{rl}
        \hat{q}_j \longrightarrow & \hat{q}_j \\
        \hat{p}_j \longrightarrow & \hat{p}_j - \tau\hat{p}_k \\
        \hat{q}_k \longrightarrow & \hat{q}_k + \tau\hat{q}_j \\ 
        \hat{p}_k \longrightarrow & \hat{p}_k
    \end{array}.
    \label{eq: quad_action_qp_op}
\end{align}
All of these operators are specified by a chosen coupling strength $\tau$. 

These gates and their physical implementation are described in more detail in Refs. \cite{twoModeSqueezingCouplingGates1,twoModeSqueezingCouplingGates2}. Notably, generating $qq$ and $pp$ quadrature coupling operators can be realized using a $qp$ coupling operation and single-mode phase shifts.

\subsection{Auxiliary GKP states}
\label{sec: aux_qunaught_description}

Here, we present a particular case of a single-mode GKP code encoding a single logical state. It was introduced in Ref. \cite{sensorStateGKP} as a sensor state and was also referenced as a qunaught state in Ref. \cite{qunaughtGKPName}. 

The qunaught state is defined on a rectangular lattice with basis matrix
\begin{align}
    S_{\varnothing, \eta} = \begin{pmatrix}
        \eta & 0 \\
        0 & \frac{1}{\eta}
    \end{pmatrix}.
    \label{eq: sensor_state_GKP}
\end{align}
The lengths of the basis vectors are chosen so that the lattice does not encode logical information ($\text{det}A_{\varnothing, \eta} = 1$). The only state associated with this lattice is given by
\begin{align}
    \ket{\varnothing_\eta} = \sum_{j = -\infty}^{\infty} \ket{\ell j\eta}_{\hat{q}} = \sum_{j = -\infty}^{\infty} \ket{\ell j/ \eta}_{\hat{p}},
    \label{eq: sensor_state_GKP_ket}
\end{align}
where we see both $\hat{q}$ and $\hat{p}$ representations.

Notably, since it is a GKP state, we can use its modular nature to realize modular measurements. As an example, suppose we translate the state by $\hat{T}(\mathbf{u})$ with $\mathbf{u} = \begin{array}{cc}(u_q & 0)^T. \end{array}$ In the $\hat{q}$ representation, the state would read $\hat{T}(\mathbf{u})\ket{\varnothing_\eta} = \sum_{j} \ket{\ell j\eta + \ell u_q}_{\hat{q}}$ and a measure of the $\hat{q}$ quadrature would yield $q_\varnothing = \ell u_q ~\text{mod}~ \ell\eta$. Thus, by tuning the squeezing of the qunaught $\eta$ and choosing the right translation $\mathbf{u}$, we can realize any desired modular measurement.

Because of this modular property and their simple lattice structure, these states are a central part of the QEC protocol we discuss in this work, serving as the auxiliaries that realize modular quadrature measurements. Their role is discussed in more detail in \cref{sec: qec_multimode_gkp}.

\section{Random Gaussian translation noise}
\label{sec: noise_model}

Because of their translational symmetry, multimode GKP codes are designed to protect against small translation errors of the quadrature operators in phase space \cite{originalGKP}. Therefore, throughout this work, the noise affecting the $m$ harmonic oscillators is considered to be the Gaussian random translation noise channel
\begin{align}
    \mathcal{N}[\boldsymbol{\mu}, \Sigma](\op{\rho}) = \int_{\mathbb{R}^{2m}} d^{2m}\boldsymbol{\xi} ~P_{\boldsymbol{\mu}, \Sigma}(\boldsymbol{\xi})\hat{T}(\boldsymbol{\xi})\hat{\rho}\hat{T}^\dag(\boldsymbol{\xi}),
    \label{eq: general_noise_model_used}
\end{align} 
with
\begin{align}
    P_{\boldsymbol{\mu}, \Sigma}(\boldsymbol{\xi}) = \frac{1}{\sqrt{(2\pi)^{2m} \text{det}\Sigma}}\mathrm{e}^{-\frac{1}{2}(\boldsymbol{\mu} - \boldsymbol{\xi})^T\Sigma^{-1}(\boldsymbol{\mu} - \boldsymbol{\xi})}.
    \label{eq: probability_of_general_noise_model_used}
\end{align}

In the Heisenberg point of view, this noise channel effects a random translation of the quadrature coordinate operators
\begin{align}
    \hat{\mathbf{x}}' = \hat{\mathbf{x}} + \ell\boldsymbol{\xi},
    \label{eq: noise_effect_on_quad_vect_x}
\end{align}
with a Gaussian probability distribution $P_{\boldsymbol{\mu}, \Sigma}(\boldsymbol{\xi})$ specified by a mean $\boldsymbol{\mu}$ and a covariance $\Sigma$ for $\boldsymbol{\xi}$.  Up to a rescaling of $\boldsymbol{\xi}$ in \cref{eq: general_noise_model_used} we retrieve the noise model usually considered for large-scale simulation of GKP codes \cite{GKPRatesHarrington,AnalogGKPSurface,nohGKPBetterCode,nohFaultTolerantGKP,blueprintXanaduGKP,MBQCSurfaceCodeGKP,conradGKPMultimode,nohLowOverheadFaultTolerantGKP, lin_2023a,BicycleCodesXanadu,GKPRatesLin}. 

For simplicity, we focus on the case of an isotropic noise where the translation errors $\hat T(\boldsymbol{\xi})$ are drawn from $\mathcal{N}[\mathbf{0}, \sigma^2\mathbb{I}](\op{\rho})$. In that situation, the covariance matrix of the noise model is diagonal, and each quadrature of each $m$ mode is randomly shifted following a one-dimensional Gaussian distribution with variance $\sigma^2$. However, most of the results presented in this work extend to the more general noise model defined by \cref{eq: general_noise_model_used}.

The effect of a Gaussian unitary $\hat U_L$ on the noise model is the following:
\begin{align}
    \hat{U}_L \Big(\mathcal{N}[\boldsymbol{\mu}, \Sigma](\op{\rho})\Big) \hat{U}_L^\dagger = \mathcal{N}[L\boldsymbol{\mu}, L\Sigma L^T](\hat{U}_L\op{\rho}\hat{U}_L^\dagger),
    \label{eq: noise_model_changing_with_symp_op}
\end{align}
which can be shown from the fact that $L$ is a linear transformation affecting a multivariate normal distribution \cite{introToMultivariateStatAnalysis}. This means that applying a Gaussian unitary on a noisy quantum state $\hat{\rho}$ is equivalent to applying a noise model with updated parameters $\boldsymbol{\mu}\rightarrow L\boldsymbol{\mu}$ and $\Sigma\rightarrow L\Sigma L^T$ on the evolved state $\hat{U}_L\op{\rho}\hat{U}_L^\dagger$.

In reality, the Gaussian error channel does not capture all possible errors in a bosonic quantum error correction setting. Practical implementations of multimode GKP codes can be done in many physical platforms, with errors originating from the physical oscillator modes (damping, dephasing, etc.) or from the finite-squeezing of the GKP states (see \cref{sec: finite_energy_multimode_gkp}) \cite{originalGKP, nohFaultTolerantGKP}. These are the errors we want our model to describe, and in some contexts it has been shown that the Gaussian noise channel represented by \cref{eq: general_noise_model_used} is a good approximation to those physical errors. 

In the case of the errors on the harmonic oscillators, any channel can be modeled by random small displacements in phase space if the interaction with the environment is weak and fast \cite{originalGKP}. Furthermore, in some particular cases, photon loss in the harmonic oscillator can be exactly modeled by a random Gaussian translation noise channel \cite{LossToRGTNoiseModel1, LossToRGTNoiseModel2}, such as its composition with the amplification channel \cite{nohGKPBetterCode}. 

The second most damaging type of errors comes from the fact that in practical implementations we work with finite-energy multimode GKP states defined by an envelope parameter $\Delta$. If $\Delta$ is sufficiently small, we can also approximate these errors by random small displacements affecting the multimode GKP state \cite{nohFaultTolerantGKP}, effectively replacing a coherent superposition of displacement errors with an incoherent mixture.

We therefore use the error model described by \cref{eq: general_noise_model_used} for its numerical simplicity and because it approximately describes real-world scenarios. We also note that, in this work, we focus on comparing two decoding methods under the same noise model rather than modeling a specific experimental setup.

\section{Quantum error correction of multimode GKP codes}
\label{sec: qec_multimode_gkp}

As for any stabilizer code, realizing QEC involves measuring the stabilizer operators of the code \cite{gottesmanThesis}. In the present case, this means measuring the generators of the infinite stabilizer group \eqref{eq: GKP_stab_group}, or, equivalently, measuring
\begin{align}
    \hat{h}_j &= \hat{g}_j~\text{mod}~2\pi,
    \label{eq: stab_quad_measurement}
\end{align}
with $\hat{g}_j$ defined by \cref{eq: stab_GKP}. Note that in this work, we use a symmetric definition of the modulo operation, meaning $\text{mod}~2\pi \in [ -\pi, \pi]$. 

Looking at \cref{eq: stab_quad_measurement}, we see that in order to measure the stabilizers of a general multimode GKP code, we must be able to measure operators of the form $\hat{x}_j$ mod $2\pi$, where $\hat{x}_j$ is a linear combination of the quadrature coordinate operators. Depending on the platform of interest, realizing such a modular measurement can be challenging experimentally \cite{originalGKP, QECwithGKPreview}. Instead, auxiliary systems are entangled with the harmonic oscillator modes so that the information of the oscillator is propagated into those auxiliary modes. They can then be destructively measured, acting only as containers for the information we wish to measure.

Here, we focus on the situation where the measurement is done via coupling with other GKP states and subsequent homodyne measurement of those auxiliary states. More specifically, we focus on Steane-type error correction. We further develop these concepts for multimode GKP codes, and we specifically take into account the effect of having noisy auxiliaries within the QEC schemes.

\begin{figure*}[t!]
    \centering
    \begin{subfigure}[t]{0.5\textwidth}
        \centering
        \caption{}
        \includegraphics[scale=1]{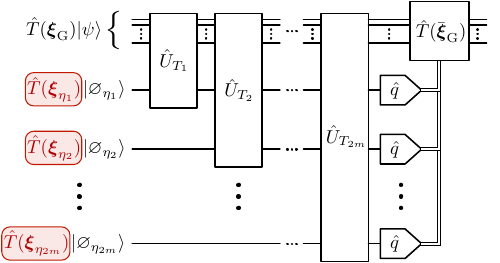}
        \label{subfig: general_mm_steane_type_QEC}
    \end{subfigure}%
    ~
    \begin{subfigure}[t]{0.5\textwidth}
        \centering
        \caption{}
        \vspace{0.75cm}
        \includegraphics[scale=1]{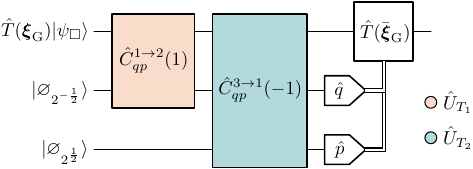}
        \vspace{0.75cm}
        \label{subfig: square_GKP_steane_type_QEC}
    \end{subfigure}
    \caption{(a) General form of a Steane-type QEC circuit for multimode GKP codes. The errors acting on the auxiliary states are shown in red to emphasize the distinction between the analysis where they are present (\cref{subsec: qec_noisy_aux}) and when they are not (\cref{subsec: qec_perf_aux}). Each $\hat{U}_{T_j}$ acts only on $\ket{\psi}$ and its corresponding auxiliary qunaught $|\varnothing_{\eta_j} \rangle$. (b) The Steane-type QEC circuit used in numerical simulations for the square GKP code, measuring operators $\hat{h}'_{\square, j}$ defined by \cref{eq: unity_stab_quad_measurement}. This circuit is the one that was originally proposed in Ref. \cite{originalGKP}, adapted with our notation. For a more detailed comparison, we refer the reader to \cref{fig: square_GKP_steane_type_QEC_original} of \cref{subsec: appendix_gen_steane_type_mm_circuit_square_gkp}. Notably, both the $\hat{q}$ and $\hat{p}$ quadratures of the auxiliaries are measured.}
    \label{fig: steane_type_QEC}
\end{figure*}

In \creffullsubfigref{fig: steane_type_QEC}{subfig: general_mm_steane_type_QEC}, we show the general Steane-type QEC circuit for any multimode GKP codes. The first thing we notice is that measuring $2m$ stabilizers of the form $\hat{h}_j$ requires $2m$ auxiliary GKP qunaught states $|\varnothing_{\eta_j} \rangle$ defined in \cref{sec: aux_qunaught_description}. Each one is a container holding the necessary information for the modulo measurement of a single $\hat{h}_j$, propagated through a Gaussian $(m+1)$-mode operator defined as $\hat{U}_{T_j}$ (see \cref{subsec: operations_on_multimode_GKP}). In \cref{subsec: appendix_gen_steane_type_mm_circuit_construct}, we show a construction to determine these operators for any multimode GKP code. We assume these gates are implemented without noise. At the end of the circuit, the qunaught states are destructively measured via homodyne detection to gather the stabilizer's information and perform QEC. To simplify the derivations, we choose to always measure the $\hat{q}$ quadrature of each auxiliary state. This is without loss of generality: measuring $\hat{p}$ instead is equivalent to applying a Gaussian unitary rotation operator by $\pi/2$ before a $\hat{q}$ measurement, which can be absorbed into the Steane-type QEC circuit and rotates the auxiliary lattice by the same angle. This is the case for the square and hexagonal Steane QEC circuits used in our numerical simulations, which follow previous conventions (for example, see Ref. \cite{originalGKP}) and measure the $\hat{p}$ quadrature of one auxiliary. Finally, we do not take into account homodyne detection errors and assume that measurements are perfect.

It is important to mention that the measurement circuits we use are not unique, in part due to the fact that the modulo form of the stabilizers measured can be scaled. More precisely, the stabilizers we choose to measure in order to obtain the main results of this paper are given by
\begin{align}
    \hat{h}'_j &= \frac{\hat{g}_j}{|\mathbf{g}_j|} ~\text{mod}~ \frac{2\pi}{|\mathbf{g}_j|}
    \label{eq: unity_stab_quad_measurement}
\end{align}
where $|\mathbf{g}_j| = |\ell \Omega \mathbf{s}_j|$. Playing with this scaling impacts the amount of squeezing that the different $\hat{U}_{T_j}$ carry and can change the way errors spread through the circuit. When the auxiliary qunaught states are perfect, all of these situations are equivalent and yield the same QEC performances. In the case where they are noisy, this does not hold true anymore. In fact, we can show that the choice of measuring stabilizers of the form of \cref{eq: stab_quad_measurement} is not the most efficient when having noisy auxiliaries. The justification for using the form of \cref{eq: unity_stab_quad_measurement} is better discussed in \cref{subsec: appendix_gen_steane_type_mm_circuit_analysis}. An example of a Steane-type QEC circuit for the square GKP code can be seen on \creffullsubfigref{fig: steane_type_QEC}{subfig: square_GKP_steane_type_QEC}, where the latter are precisely the stabilizers measured. 

Similar to what was done in Ref. \cite{nohLowOverheadFaultTolerantGKP} for the single-mode square GKP, this section aims to show that taking into account how noise correlates between the GKP and the qunaughts following the Steane-type QEC circuit greatly increases the effectiveness of the decoder.

In the rest of the paper, we refer to the $m$ mode GKP code we are realizing QEC on as the storage and all $2m$ GKP qunaught states as the auxiliaries. In total, the system is defined on $3m$ modes, such that translation operators are defined by $6m$ component vectors and Gaussian unitaries by $6m \times 6m$ matrices. The subscript G (m) accompanies every quantity that only affects the storage (measured auxiliary quadratures) subspace. The subscript S refers to quantities acting on both subspaces. Finally, no subscript simply means that we also include the quadratures of the auxiliaries that were not measured, that is the whole system.

\subsection{Noiseless auxiliaries}
\label{subsec: qec_perf_aux}

We start by analyzing the situation where the auxiliary states are ideal, infinite-energy GKP states and only the storage suffered an error $\hat{T}(\boldsymbol{\xi}_\text{G})$. In this case, the covariance matrix of the noise affecting the system is given by $\Sigma_0$, which only has non-zero entries of value $\sigma^2$ in the first $2m$ positions on the diagonal. The present subsection follows the discussion in Ref. \cite{conradGKPMultimode}.

\begin{figure*}[p!]
    \centering
    \begin{subfigure}[t]{0.5\textwidth}
        \centering
        \caption{}
        \vspace{0.3cm}
        \includegraphics[scale=1.35]{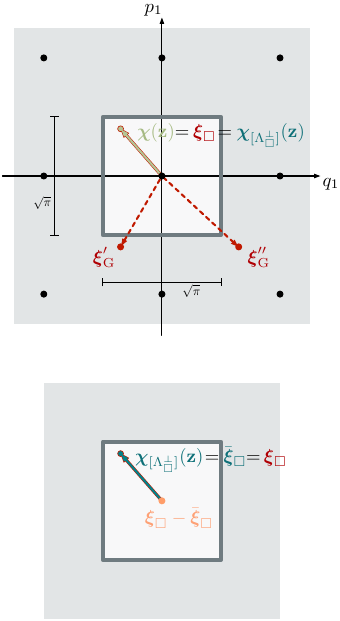}
        \label{subfig: schematic_error_vector_perfect_aux}
    \end{subfigure}%
    ~ 
    \begin{subfigure}[t]{0.5\textwidth}
        \centering
        \caption{}
        \vspace{0.3cm}
        \includegraphics[scale=1.35]{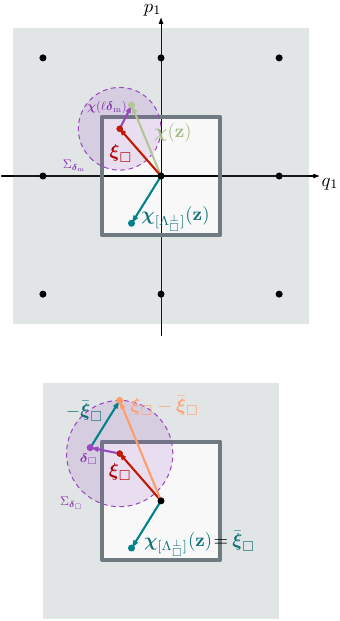}
        \label{subfig: schematic_error_vector_noisy_aux}
    \end{subfigure}
    \vspace{0.2cm}
    \caption{(a) Example scenario of the noiseless auxiliary decoding process for the square GKP that underwent a random translation $\boldsymbol{\xi}_\square$ (red arrow underneath the green arrow). The top figure shows the measurement result, and the bottom figure shows the decoding result, using the MED decoder defined in \cref{subsec: qec_perf_aux,subsec: qec_noisy_aux}. The gray square delimits the logical Voronoi cell of the square GKP storage lattice $\Lambda_\square^\perp$ (black dots). If the final displacement of the storage is located in the light gray colored areas, it leads to a logical error. For the measurement process, we identify two different quantities in the figure, $\boldsymbol{\chi}(\mathbf{z})$ and $\boldsymbol{\chi}_{[\Lambda_\square^\perp]}(\mathbf{z})$, both represented by the same green arrow. The first one represents the raw value of the measurement result and the second the measurement result modulo the logical lattice, $\boldsymbol{\chi}_{[\Lambda_\square^\perp]}(\mathbf{z}) = \boldsymbol{\chi}(\mathbf{z}) + \boldsymbol{\lambda}_\square^\perp = \boldsymbol{\chi}(\mathbf{z}) ~\text{mod}~ \Lambda^{\perp}_{\square}$. In this case, both quantities are equal since the error is inside the Voronoi cell. Thus, the measurement completely recovers the error, hence the green arrow directly on top of the red one. We also show two of the three different errors, $\boldsymbol{\xi}_\square'$ and $\boldsymbol{\xi}_\square''$ (red dotted arrow), that would yield the same measurement syndrome $\boldsymbol{\chi}(\mathbf{z})$. For the decoding process, solving the CVP, we obtain a correction given by the teal arrow. In this case, we correctly identified the error, and the total displacement after the correction (orange point) brings back the storage to the center of phase space. (b) Same example scenario with the same error, but with noisy auxiliary states. This noise shifts the measurement result and the storage by $\boldsymbol{\chi}(\ell \boldsymbol{\delta}_\text{m})$ and $\boldsymbol{\delta}_\square$, respectively (purple arrows). This is represented by sampling translations from normal distributions with covariance $\Sigma_{\boldsymbol{\delta}_{\text{m}}} = \text{Cov}\{-\boldsymbol{\chi}(\ell \boldsymbol{\delta}_\text{m})\}$ and $\Sigma_{\boldsymbol{\delta}_{\square}} = \text{Cov}\{\boldsymbol{\delta}_\square\}$, where the notation $\text{Cov}\{\mathbf{y}\}$ means the covariance matrix associated with the vector of random variables $\mathbf{y}$. These are represented by the purple circles, with the same color indicating that they are correlated. Here, the circles are arbitrarily drawn for visual purposes. Looking at the bottom figure, since $\boldsymbol{\xi}_\square-\bar{\boldsymbol{\xi}}_\square$ is outside the Voronoi cell, we see that the correction now yields a logical error. Thus, the noise present on the auxiliaries can send correctable errors to uncorrectable ones. Importantly, the shifts $\boldsymbol{\chi}(\ell \boldsymbol{\delta}_\text{m})$ and $\boldsymbol{\delta}_\square$ are correlated, which is taken into account in the new decoder presented in \cref{subsec: what_we_propose}.}
    \label{fig: schematic_error_vector}
\end{figure*}

Based on the circuit shown in \creffullsubfigref{fig: steane_type_QEC}{subfig: general_mm_steane_type_QEC}, we can define the error acting on the whole $3m$ mode system as the vector $\boldsymbol{\xi} \in \mathbb{R}^{6m}$ generating the translation $\hat{T}(\boldsymbol{\xi})$. It takes the form 
\begin{align}
    \boldsymbol{\xi} = \begin{pmatrix}
        \xi_{q, 1} \\ 
        \xi_{p, 1} \\
        \vdots \\
        \xi_{p, m} \\
        \mathbf{0}
    \end{pmatrix}
    = \begin{pmatrix}
        \boldsymbol{\xi}_\text{G} \\
        \mathbf{0}
    \end{pmatrix},
    \label{eq: error_vector_perfect_aux} 
\end{align} 
where we have unknown shifts $\xi_{q, j}$ ($\xi_{p, j}$) of the $\hat{q}_j$ ($\hat{p}_j$) quadrature coordinate of mode $j$ on the storage $\ket{\psi}$. Considering the form of $\boldsymbol{\xi}$, we simply write $\hat{T}(\boldsymbol{\xi}_\text{G})$, which acts only on the storage. The initial state is then given by
\begin{align}
    \ket{\Psi_\text{i}} = \hat{T}(\boldsymbol{\xi}_\text{G})\ket{\psi}\bigotimes_{j = 1}^{2m}\sum_{k_j \in \mathbb{Z}} \ket{\ell k_j\eta_j}_{\hat{q}_{m+j}}.
    \label{eq: steane_init_state_perfect_aux}
\end{align}
From here, we employ a lighter notation where we instead define the auxiliary states as
\begin{align}
    \bigotimes_{j = 1}^{2m}\sum_{k_j \in \mathbb{Z}} \ket{\ell k_j\eta_j}_{\hat{q}_{m+j}} = \sum_{\mathbf{k} \in \mathbb{Z}^{2m}} \ket{\ell \mathbf{k} \circ \boldsymbol{\eta}}_{\hat{\mathbf{q}}_\text{m}}.
    \label{eq: aux_new_consise_def}
\end{align}
The notation $\circ$ refers to the Hadamard product (or element-wise multiplication between two vectors) \cite{matrixAnalysis}. This way, each element of the vector $\mathbf{k} \circ \boldsymbol{\eta}$ is the $\hat{q}$ quadrature ket representation of an auxiliary state, and we omit the tensor product, which is now implied. The subscript $\hat{\mathbf{q}}_\text{m}$ indicates that the state is defined on all the $\hat{q}$ quadratures of the auxiliaries. These are the quadratures we are measuring. Feeding this state into the Steane QEC circuit, the state of the whole system right before the measurement is given by the product state (see \cref{subsec: appendix_gen_steane_type_mm_circuit_perfect_aux} for a more detailed derivation)
\begin{align}
    \ket{\Psi_{\text{f}}} &= \hat{U}_{\bar{T}}\ket{\Psi_\text{i}}\nonumber\\
    &= \hat{T}(\boldsymbol{\xi}_\text{G})\ket{\psi}\otimes \sum_{\mathbf{k} \in \mathbb{Z}^{2m}}\ket{\ell \boldsymbol{\nu} \circ S_\text{G}\Omega\boldsymbol{\xi}_\text{G} + \ell \mathbf{k} \circ \boldsymbol{\eta}}_{\hat{\mathbf{q}}_\text{m}},
    \label{eq: state_after_Steane_circuit_perfect_aux}
\end{align}
where we have defined
\begin{align}
    \hat{U}_{\bar{T}} &= \hat{U}_{T_{2m}}\hat{U}_{T_{2m-1}}\cdots\hat{U}_{T_2}\hat{U}_{T_1}, \label{eq: U_T_bar_definition}\\
    \bar{T} &= T_{2m}T_{2m-1} \cdots T_{2}T_{1},\label{eq: T_bar_definition}
\end{align}
with \cref{eq: T_bar_definition} obtained by using \cref{eq: compose_gaussian_ops}, and $\boldsymbol{\nu}$ a quantity related to the squeezing parameters in $\hat{U}_{\bar{T}}$. Now, the measurement of all the $\hat{q}$ quadratures of the auxiliaries yields the result
\begin{align}
    \mathbf{z} = \ell \boldsymbol{\nu} \circ S_\text{G}\Omega\boldsymbol{\xi}_\text{G} + \ell \mathbf{a} \circ \boldsymbol{\eta},
    \label{eq: syndrome_from_circ_perfect_aux_short_vers}
\end{align}
with $\mathbf{a}\in \mathbb{Z}^{2m}$ a random integer vector. To simplify the discussion below, we choose $\eta_j = \nu_j = \ell, ~\forall j$ so that $\mathbf{a} \circ \boldsymbol{\eta} = \ell\mathbf{a}$, and $\boldsymbol{\nu} \circ S_\text{G}\Omega\boldsymbol{\xi}_\text{G} = \ell S_\text{G}\Omega\boldsymbol{\xi}_\text{G}$. This way, the circuit enables the extraction of a syndrome given by
\begin{align}
    \mathbf{z} = \ell^2S_\text{G}\Omega\boldsymbol{\xi}_\text{G} + 2\pi \mathbf{a}.
    \label{eq: syndrome_from_circ_perfect_aux}
\end{align}
In general, any choice where $\boldsymbol{\eta} = \boldsymbol{\nu}$ works. The translation error on the storage is linked to the measurement with
\begin{align}
    \boldsymbol{\xi}_\text{G} &= -\frac{1}{\ell^2}\Omega S_\text{G}^{-1} (\mathbf{z} - 2\pi \mathbf{a}) \nonumber\\
    &= \boldsymbol{\chi}(\mathbf{z}) + (S_\text{G}^\perp)^T\mathbf{b}\nonumber\\
    &= \boldsymbol{\chi}(\mathbf{z}) + \boldsymbol{\lambda}^\perp_\text{G},
    \label{eq: error_from_circ_perfect_aux}
\end{align}
where we have defined $\boldsymbol{\chi}(\mathbf{y}) = -(\Omega S_\text{G}^\perp)^T \mathbf{y}/\ell^2$, $\boldsymbol{\lambda}^\perp_\text{G} = (S_\text{G}^\perp)^T\mathbf{b}$ and $\mathbf{b} = \Omega\mathbf{a}$ is a random vector in $\mathbb{Z}^{2m}$. From \cref{eq: error_from_circ_perfect_aux}, we see that the information contained in $\mathbf{z}$ does not directly reveal the error. It instead gives $\boldsymbol{\xi}_\text{G}$ up to an unknown logical vector in the logical lattice of the storage $\Lambda_\text{G}^\perp$. In other words, we obtain the error modulo the logical lattice of the storage, and all errors differing by a dual lattice vector yield the same syndrome $\mathbf z$. This situation is illustrated in \creffullsubfigref{fig: schematic_error_vector}{subfig: schematic_error_vector_perfect_aux}.

To bring back the state to the code space, a translation by $-\bar{\boldsymbol{\xi}}_\text{G}$ is applied to the state. In the ideal case, the decoder finds a correction translation $\bar{\boldsymbol{\xi}}_\text{G}$ that maximizes the probability that the translation $\hat{T}(\boldsymbol{\xi}_\text{G}-\bar{\boldsymbol{\xi}}_\text{G})$ is equivalent to the identity in the code space. This type of decoding is referred to as Maximum-Likelihood Decoding (MLD). More specifically, $\bar{\boldsymbol{\lambda}}^\perp_\text{G}$ is obtained by maximizing $P_{\mathbf{0}, \Sigma_0}([\boldsymbol{\xi}_\text{G}] | \tilde{\mathbf{z}})$, the probability of having all equivalent errors knowing $\tilde{\mathbf{z}}$, with $\tilde{\mathbf{z}} = \mathbf{z} / \ell$. Here, $[\boldsymbol{\xi}_\text{G}] = \{\boldsymbol{\xi}_\text{G} + \boldsymbol{\lambda}, \boldsymbol{\lambda} \in \Lambda_\text{G}\}$ refers to the equivalence class of errors that are logically equivalent, \emph{i.e.}, that differ by a stabilizer lattice vector. In other words, we must solve the following optimization problem
\begin{align}
    \bar{\boldsymbol{\lambda}}^\perp_\text{G} &= \underset{\boldsymbol{\lambda}^\perp_\text{G} \in \Lambda_\text{G}^{\perp}/\Lambda_\text{G}}{\text{arg max }} P_{\mathbf{0}, \Sigma_0}([\boldsymbol{\chi}(\mathbf{z}) + \boldsymbol{\lambda}^\perp_\text{G}] | \tilde{\mathbf{z}}),
    \label{eq: general_mld_decoder}
\end{align}
with $P_{\mathbf{0}, \Sigma_0}([\boldsymbol{\chi}(\mathbf{z}) + \boldsymbol{\lambda}^\perp_\text{G}] | \tilde{\mathbf{z}})$ given by
\begin{equation}
    P_{\mathbf{0}, \Sigma_0}([\boldsymbol{\chi}(\mathbf{z}) + \boldsymbol{\lambda}^\perp_\text{G}] | \tilde{\mathbf{z}}) \propto \sum_{\boldsymbol{\lambda} \in \Lambda_\text{G}} P_{\mathbf{0}, \Sigma_0}(\boldsymbol{\chi}(\mathbf{z}) + \boldsymbol{\lambda}^\perp_\text{G} + \boldsymbol{\lambda}),
    \label{eq: general_mld_decoder_prob}
\end{equation}
where we ignored normalization constants that do not affect the maximization and $P_{\mathbf{0}, \Sigma_0}(\mathbf{u})$ is defined following \cref{eq: probability_of_general_noise_model_used}. Generally, there are no analytical solutions to this exponentially hard problem. 

While \cref{eq: general_mld_decoder} is difficult to solve, we notice that in the limit of low noise where $\sigma^2 \rightarrow 0$, the sum in \cref{eq: general_mld_decoder_prob} is dominated by the most likely error $\boldsymbol{\xi}_\text{G}$, which is the one that has minimal Euclidean distance. In this case, one decoding technique is to find the $\boldsymbol{\lambda}^\perp_\text{G}$ that minimizes said distance, which amounts to finding the solution to the Closest-Vector Problem (CVP) \cite{lin_2023a}:
\begin{align}
    \bar{\boldsymbol{\lambda}}^\perp_\text{G}
    &= \underset{\boldsymbol{\lambda}^\perp_\text{G} \in \Lambda_\text{G}^{\perp}}{\text{arg min }} |\boldsymbol{\chi}(\mathbf{z}) + \boldsymbol{\lambda}^\perp_\text{G}|.
    \label{eq: med_cvp_decoder}
\end{align}
In this regime, we perform what is called Minimum Energy Decoding (MED), where the weight of each equivalent error is not taken into account. In \cref{sec: res}, when comparing with the decoder we developed, we refer to it as the MED decoder. Finally, the correction vector we apply is given by
\begin{align}
    \bar{\boldsymbol{\xi}}_\text{G} = \boldsymbol{\chi}(\mathbf{z}) + \bar{\boldsymbol{\lambda}}^\perp_\text{G}.
    \label{eq: error_estimated_after_mld}
\end{align}
Knowing the most likely error on our system, we can displace the storage back into the code space applying a translation of $-\bar{\boldsymbol{\xi}}_\text{G}$. In \creffullsubfigref{fig: schematic_error_vector}{subfig: schematic_error_vector_perfect_aux}, we show every element of such a correction scheme using the MED decoder based on solving the CVP.

Initially, the quadrature operator describing the storage is $\hat{\mathbf{x}}_\text{G}$. The error then shifts this operator following \cref{eq: noise_effect_on_quad_vect_x}, so that after the initial error, we have
\begin{align}
    \hat{\mathbf{x}}_{\text{G}, \xi} = \hat{\mathbf{x}}_\text{G} + \ell\boldsymbol{\xi}_\text{G}.
\end{align}
Then, using \cref{eq: translation_with_x_commute_rel}, we are applying a correction translation $\hat{T}(-\bar{\boldsymbol{\xi}}_\text{G})$ such that
\begin{align}
    \hat{\mathbf{x}}_{\text{G},c} = \hat{\mathbf{x}}_\text{G} + \ell\boldsymbol{\xi}_\text{G} - \ell \bar{\boldsymbol{\xi}}_\text{G},
\end{align}
and knowing the form of $\boldsymbol{\xi}_\text{G}$ and $\bar{\boldsymbol{\xi}}_\text{G}$ from \cref{eq: error_from_circ_perfect_aux} and \cref{eq: error_estimated_after_mld}, respectively, the storage quadrature operator becomes
\begin{align}
    \hat{\mathbf{x}}_{\text{G},c} &= \hat{\mathbf{x}}_\text{G} + \ell\boldsymbol{\lambda}^\perp_\text{G} - \ell\bar{\boldsymbol{\lambda}}^\perp_\text{G} \nonumber \\
    &= \hat{\mathbf{x}}_\text{G} + \ell\boldsymbol{\lambda}^\perp_c,
\end{align}
where we define $\boldsymbol{\lambda}^\perp_c = \boldsymbol{\lambda}^\perp_\text{G} - \bar{\boldsymbol{\lambda}}^\perp_\text{G} \in \Lambda_\text{G}^\perp$. We see that we retrieve the initial uncorrupted quadrature operator $\hat{\mathbf{x}}_\text{G}$ up to a logical Pauli operator. 

Therefore, we have a successful error correction step if $\boldsymbol{\lambda}^\perp_c \in \Lambda_\text{G}$ so that the storage does not accumulate an extra Pauli operator. In simulations, this can be verified by computing the commutation relations of $\boldsymbol{\lambda}^\perp_c$ with each logical Pauli vector defined in \cref{eq: log_pauli_mat}, making sure that it commutes with all of them. Essentially, using \cref{eq: translation_commute_rel}, we compute $\boldsymbol{\rho}(\boldsymbol{\lambda}^\perp_c)$, where
\begin{align}
    \boldsymbol{\rho}(\mathbf{v}) = P\Omega \mathbf{v}~~\text{mod}~1,
    \label{eq: pauli_check_after_correction}
\end{align}

\noindent
with $P$ defined by \cref{eq: log_pauli_mat} and the modulo applied element-wise. If $\boldsymbol{\rho}(\boldsymbol{\lambda}^\perp_c) = \mathbf{0}$, a successful error correction was applied; otherwise, an undetectable logical error was applied during the error correction step.

\subsection{Noisy auxiliaries}
\label{subsec: qec_noisy_aux}

Now, let us analyze the situation where both the storage and the auxiliary modes are noisy, meaning that all of the $3m$ modes suffer translation errors. Following what was discussed in \cref{sec: noise_model}, we take the covariance matrix of the noise affecting the system to be defined by $\Gamma_0$, having non-zero entries of value $\sigma^2$ on all its diagonal. This way, $\boldsymbol{\xi}$ is defined such that
\begin{align}
    \boldsymbol{\xi} = \begin{pmatrix}
        \boldsymbol{\xi}_\text{G} \\
        \boldsymbol{\xi}_\eta
    \end{pmatrix} = \begin{pmatrix}
        \boldsymbol{\xi}_\text{G} \\
        \boldsymbol{\xi}_{\eta_1} \\
        \boldsymbol{\xi}_{\eta_2} \\
        \vdots \\
        \boldsymbol{\xi}_{\eta_{2m}}
    \end{pmatrix},
    \label{eq: xi_noisy_case}
\end{align}
where for the first $m$ modes, we still have the error on the storage $\ket{\psi}$ defined as $\hat{T}(\boldsymbol{\xi}_\text{G})$. Now, each of the auxiliary states takes the form $\hat{T}(\boldsymbol{\xi}_{\eta_j})\ket{\varnothing_{\eta_j}}$, as we can see from \creffullsubfigref{fig: steane_type_QEC}{subfig: general_mm_steane_type_QEC}.

We start with a similar initial state as in \cref{subsec: qec_perf_aux} given by \cref{eq: steane_init_state_perfect_aux}, except that now, $\boldsymbol{\xi}$ is defined by \cref{eq: xi_noisy_case}, so that
\begin{align}
    \ket{\Psi_\text{i}} = \hat{T}(\boldsymbol{\xi}_\text{G})\ket{\psi}\otimes\sum_{\mathbf{k} \in \mathbb{Z}^{2m}} \hat{T}(\boldsymbol{\xi}_\eta)\ket{\ell \mathbf{k} \circ \boldsymbol{\eta}}_{\hat{\mathbf{q}}_\text{m}}.
\end{align}
After the circuit of \creffullsubfigref{fig: steane_type_QEC}{subfig: general_mm_steane_type_QEC}, the state right before the measurement is, up to an irrelevant global phase, given by
\begin{equation}
    \begin{aligned}
    \ket{\Psi_\text{f}} &= \hat{T}(\Pi_\text{G}\bar{T}\boldsymbol{\xi})\ket{\psi}\\
    &\otimes \sum_{\mathbf{k} \in \mathbb{Z}^{2m}} \mathrm{e}^{i\ell^2 \boldsymbol{\xi}^T\bar{T}^T\Pi_{\text{p}}^T \mathbf{k} \circ \boldsymbol{\eta}}\ket{\ell\Pi_\text{m}\bar{T}\boldsymbol{\xi} + \ell \mathbf{k} \circ \boldsymbol{\eta}}_{\hat{\mathbf{q}}_\text{m}}.
    \end{aligned}
    \label{eq: steane_init_state_noisy_aux_short_vers}
\end{equation}
Here we have introduced three projectors that act on the whole system and select particular subspaces. First, we have $\Pi_\text{G}$ which projects onto the storage modes, meaning the first $m$ modes of the system. Second, we have $\Pi_{\text{m}}$, the projector onto the $\hat{q}$ quadrature of all auxiliary states we measure. Third, we have $\Pi_{\text{p}}$, the projector onto the $\hat{p}$ quadrature of all auxiliary states. Defining $\varphi(\mathbf{k}) = \ell^2 \boldsymbol{\xi}^T\bar{T}^T\Pi_{\text{p}}^T \mathbf{k} \circ \boldsymbol{\eta}$, we can rewrite this state as
\begin{align}
    \ket{&\Psi_\text{f}} = \hat{T}(\boldsymbol{\xi}_\text{G} + \boldsymbol{\delta}_\text{G})\ket{\psi} \nonumber\\
    &\otimes \sum_{\mathbf{k} \in \mathbb{Z}^{2m}}\mathrm{e}^{i\varphi(\mathbf{k})} \ket{\ell \boldsymbol{\nu}\circ S_\text{G}\Omega\boldsymbol{\xi}_\text{G} + \ell\boldsymbol{\delta}_\text{m} + \ell \mathbf{k} \circ \boldsymbol{\eta}}_{\hat{\mathbf{q}}_\text{m}}.
    \label{eq: state_before_measurement_noisy_aux}
\end{align}
Here, the vectors $\boldsymbol{\delta}_\text{G}$ and $\boldsymbol{\delta}_\text{m} $ are linear combinations of the different components of the errors $\boldsymbol{\xi}_{\eta_j}$. This result shows that the Steane QEC circuit does not entangle the storage with the auxiliaries, as is justified by the product state. This was true for the noiseless case (see \cref{eq: state_after_Steane_circuit_perfect_aux}), and it remains true even in the presence of noise on the auxiliaries. For more details on the derivation to obtain $\ket{\Psi_\text{f}}$, we refer the reader to \cref{subsec: appendix_gen_steane_type_mm_circuit_noisy_aux}. The measurement of the auxiliaries yields a syndrome
\begin{align}
    \mathbf{z}
    &= \ell \boldsymbol{\nu} \circ S_\text{G}\Omega \boldsymbol{\xi}_\text{G} + \ell \boldsymbol{\delta}_\text{m} + \ell \mathbf{a} \circ \boldsymbol{\eta},
\end{align}
with $\mathbf{a}\in \mathbb{Z}^{2m}$. Finally, applying the same ideas as in the previous section and choosing $\eta_j = \nu_j = \ell$, $\forall j$, we get 
\begin{align}
    \mathbf{z}
    &= \ell^2 S_\text{G}\Omega \boldsymbol{\xi}_\text{G} + \ell \boldsymbol{\delta}_\text{m} + 2\pi \mathbf{a}.
    \label{eq: syndrome_from_circ_noisy_aux}
\end{align}
Comparing this result with the syndrome obtained in the case of noiseless auxiliaries, \cref{eq: syndrome_from_circ_perfect_aux}, we see that the noise on the auxiliaries corrupts the measurement, hiding the actual error inflicted on the storage with the vector $\boldsymbol{\delta}_\text{m}$.

Now, we can deal with the decoding part of the QEC protocol. Here, let us suppose we are applying the same decoding technique as in \cref{subsec: qec_perf_aux}. We isolate $\boldsymbol{\xi}_\text{G}$ to get
\begin{align}
    \boldsymbol{\xi}_\text{G} &= -\frac{1}{\ell^2}\Omega S_\text{G}^{-1} (\mathbf{z} - \ell\boldsymbol{\delta}_\text{m} - 2\pi \mathbf{a}) \nonumber\\
    &= \boldsymbol{\chi}(\mathbf{z}) + (S_\text{G}^\perp)^T \mathbf{b} - \boldsymbol{\chi}(\ell\boldsymbol{\delta}_\text{m}) \nonumber\\
    &= \boldsymbol{\chi}(\mathbf{z}) + \boldsymbol{\lambda}^\perp_\text{G} - \boldsymbol{\chi}(\ell\boldsymbol{\delta}_\text{m}).
    \label{eq: error_from_circ_noisy_aux}
\end{align}
The next step is to apply the MLD or MED decoder and extract a logical vector $\bar{\boldsymbol{\lambda}}^\perp_\text{G}$ based on our knowledge of the noise model. But, since we can't isolate the contribution of $\boldsymbol{\chi}(\ell\boldsymbol{\delta}_\text{m})$ from that of $\boldsymbol{\chi}(\mathbf{z})$, one strategy is to proceed with standard decoding assuming $\boldsymbol{\delta}_{\text{m}} = \mathbf{0}$. Taking this into consideration, we apply a correction vector on the storage $-\bar{\boldsymbol{\xi}}_\text{G}$, where $\bar{\boldsymbol{\xi}}_\text{G}$ is defined as
\begin{align}
    \bar{\boldsymbol{\xi}}_\text{G} = \boldsymbol{\chi}(\mathbf{z}) + \bar{\boldsymbol{\lambda}}^\perp_\text{G}.
\end{align}
Looking at the overall effect of the error correction procedure on the quadrature operator $\hat{\mathbf{x}}_\text{G}$ of the storage, we notice that the correction does not bring $\ket{\psi}$ back into the code space $\mathcal{C}_\text{G}$. We show an example of this situation on \creffullsubfigref{fig: schematic_error_vector}{subfig: schematic_error_vector_noisy_aux} using the same MED decoder as in \cref{subsec: qec_perf_aux}. In fact, tracking the storage quadrature operators $\hat{\mathbf{x}}_\text{G}$, after the correction we have
\begin{align}
    \hat{\mathbf{x}}_{\text{G},c} = \hat{\mathbf{x}}_\text{G} + \ell\boldsymbol{\lambda}^\perp_c + \ell \boldsymbol{\delta}_c,
    \label{eq: final_gkp_quad_after_correction}
\end{align}
where we define $\boldsymbol{\delta}_c = \boldsymbol{\delta}_\text{G} - \boldsymbol{\chi}(\ell\boldsymbol{\delta}_\text{m})$. Thus, we effectively have $\boldsymbol{\lambda}^\perp_c \in \Lambda^{\perp}_\text{G}$, but $\boldsymbol{\delta}_c \notin \Lambda^{\perp}_\text{G}$. For the correction to be effective, we must verify two conditions. First, following \cref{eq: pauli_check_after_correction}, we need $\boldsymbol{\rho}(\boldsymbol{\lambda}^\perp_c) = \mathbf{0}$, ensuring $\boldsymbol{\lambda}^\perp_c$ is a logical identity. Also, we must have $|\boldsymbol{\delta}_c| \ll d_\text{G}$, where $d_\text{G}$ is the distance of the multimode GKP code encoded by the storage and defined by \cref{eq: GKP_code_distance}. Indeed, this would mean that the error would slightly shift the storage around the center of the logical Voronoi cell, keeping the final translation a correctable error by subsequent QEC cycles. 

An important thing we notice is that there are correlations embedded into the vectors $\boldsymbol{\delta}_c$ and $\boldsymbol{\lambda}_c^\perp$. In fact, both depend directly on how the Steane operators propagate errors in the circuit. Since the decoding technique we have discussed so far does not account for these correlations, we aim to do so in order to have a better decoding process.

\subsection{Noise-correlated MED decoder}
\label{subsec: what_we_propose}

In this section, we present the decoder that takes into account the correlations that are present in $\boldsymbol{\delta}_c$ and $\boldsymbol{\lambda}_c^\perp$. This way, we can estimate a better correction $\bar{\boldsymbol{\lambda}}^\perp_\text{G}$ to compensate for errors that occur due to the extra shift $\boldsymbol{\delta}_c$. In other words, we propose a decoder that considers the noise model of the complete system, storage and auxiliaries included.

It is important to mention that the formalism we introduce here is able to treat any form of initial noise in the system. This means that $\Gamma_0$ may describe a more general initial noise setting, such as a non-diagonal covariance matrix arising from non-isotropic Gaussian noise, and our results can still be established in the same manner. However, for simplicity, we follow what is presented in \cref{subsec: qec_noisy_aux} and consider errors that are sampled from the same Gaussian distribution.

Here we aim to change the decoder and not the noise model or the circuit, such that the state before the measurement is still given by \cref{eq: steane_init_state_noisy_aux_short_vers}. Defining the quantities $\boldsymbol{\xi}_{\bar{T}, \text{G}} = \Pi_\text{G}\bar{T}\boldsymbol{\xi}$ and $\boldsymbol{\xi}_{\bar{T}, \text{m}} = \Pi_{\text{m}}\bar{T}\boldsymbol{\xi}$, the circuit shifts the storage quadrature operators following
\begin{align}
    \hat{\mathbf{x}}_{\text{G}, \xi} = \hat{\mathbf{x}}_{\text{G}} + \ell\boldsymbol{\xi}_{\bar{T}, \text{G}},
    \label{eq: GKP_quad_before_correction_our_proposal}
\end{align}
and the measurement of the auxiliaries yields the syndrome
\begin{align}
    \frac{\mathbf{z}_{\bar{T}, \text{m}}}{\ell} = \tilde{\mathbf{z}}_{\bar{T}, \text{m}} &= \boldsymbol{\xi}_{\bar{T}, \text{m}} + \mathbf{a} \circ \boldsymbol{\eta} \nonumber\\
    &= \boldsymbol{\xi}_{\bar{T}, \text{m}} + \Pi_{\text{m}}(S_\eta)^T\mathbf{a}_\eta,
    \label{eq: syndrome_our_decoder}
\end{align}
with $\mathbf{a} \in \mathbb{Z}^{2m}$ and $\mathbf{a}_\eta \in \mathbb{Z}^{4m}$. Here, we rewrote $\mathbf{a} \circ \boldsymbol{\eta}$ in terms of the dual generator matrix of the auxiliary system $(S_\eta^\perp = S_\eta)$, defined in \cref{eq: full_system_S_mat}. This gives a more general way of representing the system, since in this section we don't enforce $\eta_j = \ell$, $\forall j$.

To summarize \cref{eq: syndrome_our_decoder}, the syndrome measurement of all auxiliaries gives the error $\boldsymbol{\xi}_{\bar{T}, \text{m}}$ modulo a certain logical lattice vector in $\Lambda_\text{m}$, which is the lattice generated by $S_{\text{m}} = S_{\eta}\Pi_\text{m}^T$. Thus, the storage underwent an error $\boldsymbol{\xi}_{\bar{T}, \text{G}}$, and we try to estimate it as best as we can based on the measurement result $\tilde{\mathbf{z}}_{\bar{T}, \text{m}}$. In order to realize the decoding process, we define a similar MLD decoder as \cref{eq: general_mld_decoder}, that is
\begin{align}
    \bar{\boldsymbol{\xi}}_{\bar{T}, \text{G}} &= \underset{\boldsymbol{\xi}_{\bar{T}, \text{G}} \in \mathcal{V}_\text{G}}{\text{arg max }} P_{\mathbf{0}, \Gamma}([\boldsymbol{\xi}_{\bar{T}, \text{G}}] | \tilde{\mathbf{z}}_{\bar{T}, \text{m}}).
    \label{eq: general_noise_corr_mld_decoder}
\end{align} 
Here, $P_{\mathbf{0}, \Gamma}([\boldsymbol{\xi}_{\bar{T}, \text{G}}] | \tilde{\mathbf{z}}_{\bar{T}, \text{m}})$ is the probability that the storage error is $\boldsymbol{\xi}_{\bar{T}, \text{G}}$, or any of its logically equivalent vectors, knowing the syndrome $\tilde{\mathbf{z}}_{\bar{T}, \text{m}}$. The probability distribution for the noise on the storage and measured vectors is sampled from the covariance matrix
\begin{align}
    \Gamma = \Pi_\text{S}\bar{T}\Gamma_0\bar{T}^T\Pi_\text{S}^T,
    \label{eq: gamma_def}
\end{align}
where $\Pi_\text{S} = \Pi_\text{G} + \Pi_\text{m}$. Looking at \cref{eq: general_noise_corr_mld_decoder}, we notice a few differences from \cref{eq: general_mld_decoder}. 

First, the optimization process is on a patch in phase space defined by $\mathcal{V}_\text{G} \equiv \mathcal{V}(\Lambda_\text{G})$ from \cref{eq: gen_voronoi_cell}. This is the Voronoi cell of the storage stabilizer lattice. This is because the noisy auxiliaries QEC process does not guarantee that the storage is shifted back into $\mathcal{C}_\text{G}$, as is discussed in \cref{subsec: qec_noisy_aux}. We cannot restrict to only the logically distinct vectors in $\mathcal{V}_\text{G}$, the ensemble $\Lambda^\perp_\text{G}/\Lambda_\text{G}$. Thus, we factor in all possible points in $\mathcal{V}_\text{G}$, which enables us to estimate the most likely correction $\bar{\boldsymbol{\xi}}_{\bar{T}, \text{G}}$. In the limit of noiseless auxiliaries, $P_{\mathbf{0}, \Gamma}([\boldsymbol{\xi}_{\bar{T}, \text{G}}] | \tilde{\mathbf{z}}_{\bar{T}, \text{m}})$ is nonzero only for the four vectors that differ by a dual lattice vector in $\mathcal{V}_\text{G}$, and we retrieve \cref{eq: general_mld_decoder}, as we show in \cref{subsec: appendix_noiseless_aux_case_from_noisy}.

Second, we consider noise that is sampled through the covariance matrix $\Gamma$ to include correlations between the storage and the measurement into the probability distribution. The final noise in the $\hat p_j$ quadrature coordinates of the auxiliaries does not affect the measurement result, such that we can project it out of the covariance matrix.

Expanding $P_{\mathbf{0}, \Gamma}([\boldsymbol{\xi}_{\bar{T}, \text{G}}] | \tilde{\mathbf{z}}_{\bar{T}, \text{m}})$ (see \cref{subsec: appendix_expanding_COR_MLD} for more details), we get
\begin{align}
    P_{\mathbf{0}, \Gamma}([\boldsymbol{\xi}_{\bar{T}, \text{G}}] | \tilde{\mathbf{z}}_{\bar{T}, \text{m}}) \propto \sum_{\substack{\boldsymbol{\lambda} \in \Lambda_\text{G} \\ \boldsymbol{\lambda}_\text{m} \in \Lambda_\text{m}}} P_{\mathbf{0}, \Gamma}(\boldsymbol{\xi}_{\bar{T}, \text{S}}(\boldsymbol{\lambda}, \boldsymbol{\lambda}_\text{m})),
    \label{eq: general_cor_mld_decoder_prob}
\end{align}
where we have introduced the vector
\begin{align}
    \boldsymbol{\xi}_{\bar{T}, \text{S}}(\boldsymbol{\lambda}, \boldsymbol{\lambda}_\text{m}) = \begin{pmatrix}
        \boldsymbol{\xi}_{\bar{T}, \text{G}} + \boldsymbol{\lambda} \\
        \tilde{\mathbf{z}}_{\bar{T}, \text{m}} + \boldsymbol{\lambda}_\text{m}
    \end{pmatrix},
    \label{eq: xi_t_s}
\end{align}
defined in the subspace of the projector $\Pi_\text{S}$. Comparing with the form that takes \cref{eq: general_mld_decoder_prob}, we see that not only do we factor in the stabilizer equivalence class with the sum on all $\boldsymbol{\lambda}$, but we also take into account the probabilistic nature of the measurement through $\boldsymbol{\lambda}_\text{m}$.

Although we provide an explicit form for \cref{eq: general_noise_corr_mld_decoder} in \cref{subsec: appendix_expanding_COR_MLD}, in this work we do not take into account the equivalence class of errors since it is numerically difficult to solve for multimode GKP codes with an increasing number of modes $m$. Rather, we consider an MED decoder given by
\begin{align}
    \bar{\boldsymbol{\xi}}_{\bar{T}, \text{G}} &= \underset{\boldsymbol{\xi}_{\bar{T}, \text{G}} \in \mathcal{V}_\text{G}}{\text{arg max }} P_{\mathbf{0}, \Gamma}(\boldsymbol{\xi}_{\bar{T}, \text{G}} | \tilde{\mathbf{z}}_{\bar{T}, \text{m}}).
\end{align}
In the case where the noise in the system is rather small, this decoder still gives us an advantage during the decoding while being numerically solvable. In fact, we can show that
\begin{align}
    \bar{\boldsymbol{\xi}}_{\bar{T}, \text{G}} &= -\Gamma_\text{G}^{-1}\gamma\qty(\tilde{\mathbf{z}}_{\bar{T}, \text{m}} - F^{-1}\bar{\boldsymbol{\lambda}}_{F,\text{m}}),
    \label{eq: solved_noise_corr_mld_decoder}
\end{align}
where $\bar{\boldsymbol{\lambda}}_{F,\text{m}}$ is obtained following
\begin{align}
    \bar{\boldsymbol{\lambda}}_{F, \text{m}} = \underset{\boldsymbol{\lambda}_{F,\text{m}} \in F\Lambda_\text{m}}{\text{arg min }}|F\tilde{\mathbf{z}}_{\bar{T}, \text{m}} - \boldsymbol{\lambda}_{F,\text{m}}|.
    \label{eq: logical_vect_of_corr_mld_decoder}
\end{align}
Here, $\Gamma_\text{G}$ and $\gamma$ are determined by the decomposition of $\Gamma^{-1}$ into
\begin{align}
    \Gamma^{-1} = \begin{pmatrix}
        \Gamma_\text{G} & \gamma \\
        \gamma^T & \Gamma_\text{m}
    \end{pmatrix}.
    \label{eq: gamma_cov_mat}
\end{align}
More specifically, $\Gamma_\text{G}$ represents the correlations between the storage modes, $\Gamma_\text{m}$ the correlations between all the $\hat{q}$ auxiliary modes and $\gamma$ the correlations between both of those two subsystems. In order to use the Euclidean distance for the norm in \cref{eq: logical_vect_of_corr_mld_decoder}, we introduce $F$, the Cholesky decomposition of the matrix $\bar{\Gamma} = \Gamma_\text{m} - \gamma^T \Gamma_\text{G}^{-1} \gamma$, such that $\bar{\Gamma} = F^T F$ \cite{matrixAnalysis}. We refer the reader to \cref{subsec: appendix_expanding_COR_MED} for the detailed derivation and more intuition on the new quantities introduced. In \cref{sec: res}, the decoder represented by \cref{eq: solved_noise_corr_mld_decoder} is referred to as the COR-MED decoder.

Looking at \cref{eq: solved_noise_corr_mld_decoder}, we see that the problem boils down to solving the CVP on the lattice $F\Lambda_\text{m}$, which has the same dimension as $\Lambda_\text{G}^\perp$. Therefore, the time complexity of our decoder is similar to the MED decoder, as is verified numerically in \cref{subsec: numerical_complexity_MED_vs_CORMED}.

With a correction vector $\bar{\boldsymbol{\xi}}_{\bar{T}, \text{G}}$, we can now look at its effect on the quadrature coordinates. Using \cref{eq: GKP_quad_before_correction_our_proposal}, we have
\begin{align}
    \hat{\mathbf{x}}_{\text{G}, c}
    &= \hat{\mathbf{x}}_{\text{G}} + \ell (\boldsymbol{\xi}_{\bar{T}, \text{G}} - \bar{\boldsymbol{\xi}}_{\bar{T}, \text{G}}) \nonumber\\
    &= \hat{\mathbf{x}}_{\text{G}} + \ell \boldsymbol{\lambda}^\perp_{\bar{T},c} + \ell \boldsymbol{\delta}_{\bar{T},c},
    \label{eq: cormed_residual_error}
\end{align}
where we expressed the vector $\boldsymbol{\xi}_{\bar{T}, \text{G}} - \bar{\boldsymbol{\xi}}_{\bar{T}, \text{G}}$ as a logical vector in the storage logical lattice $\Lambda^\perp_\text{G}$ with an extra residual displacement $\boldsymbol{\delta}_{\bar{T},c}$. From here, we retrieve exactly the form of \cref{eq: final_gkp_quad_after_correction}, where ideally $\boldsymbol{\rho}(\boldsymbol{\lambda}^\perp_{\bar{T},c}) = \mathbf{0}$. In the next section, we numerically show that this decoder yields $|\boldsymbol{\delta}_{\bar{T},c}| < |\boldsymbol{\delta}_c| \ll d_\mathrm{G}$, with a correction step that keeps the storage more centered in the logical Voronoi cell $\mathcal{V}^\perp_\text{G}$, protecting against larger errors.

Finally, in order for this new COR-MED decoder to bring an advantage over the usual MED decoder, the initial noise in the system and the way of implementing the Steane-type QEC circuit must respect certain conditions, which we explore further in \cref{subsec: appendix_bigGamma_conditions}.

\section{Numerical results}
\label{sec: res}

\begin{figure*}[t!]
    \centering
    \begin{subfigure}[t]{0.5\textwidth}
        \centering
        \caption{}
        \vspace{-0.25cm}
        \includegraphics[scale=1]{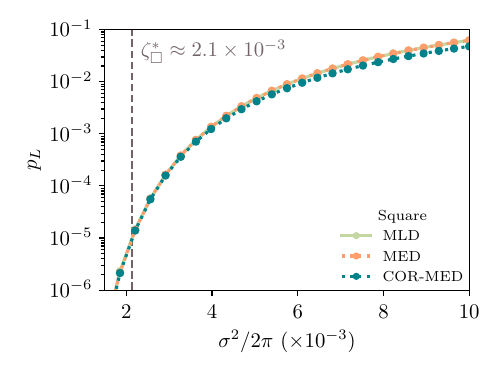}
        \label{subfig: comparing_MED_vs_CORMED_main_fig_1e9_square}
    \end{subfigure}%
    ~ 
    \begin{subfigure}[t]{0.5\textwidth}
        \centering
        \caption{}
        \vspace{-0.25cm}
        \includegraphics[scale=1]{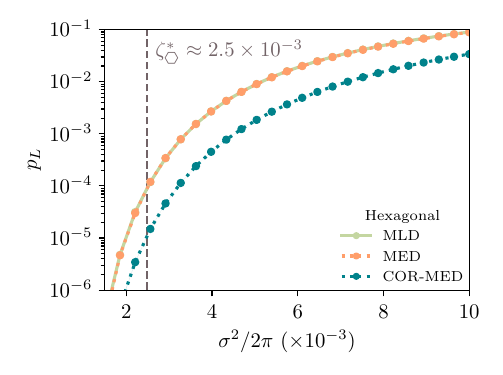}
        \label{subfig: comparing_MED_vs_CORMED_main_fig_1e9_hex}
    \end{subfigure}
    \\\vspace{-0.75cm}
    \begin{subfigure}[t]{0.5\textwidth}
        \centering
        \caption{}
        \vspace{-0.25cm}
        \includegraphics[scale=1]{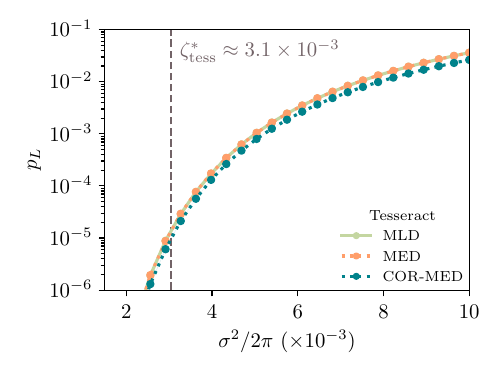}
        \label{subfig: comparing_MED_vs_CORMED_main_fig_1e9_tess}
    \end{subfigure}%
    ~
    \begin{subfigure}[t]{0.5\textwidth}
        \centering
        \caption{}
        \vspace{-0.25cm}
        \includegraphics[scale=1]{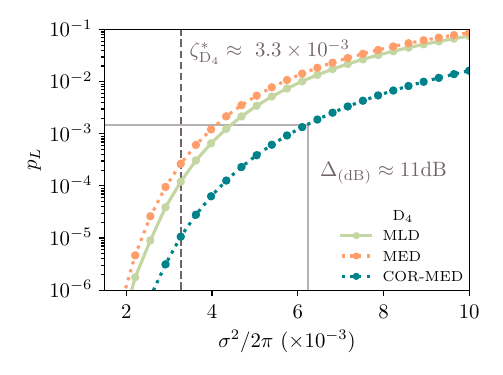}
        \label{subfig: comparing_MED_vs_CORMED_main_fig_1e9_D4}
    \end{subfigure}%
    \vspace{-0.75cm}
    \caption{Performance comparison of the MLD decoder \cref{eq: general_mld_decoder} (full green line) and the MED decoder \cref{eq: med_cvp_decoder} (dotted orange line) with our COR-MED decoder \cref{eq: solved_noise_corr_mld_decoder} (dotted teal line), the latter being the only one taking the noise of the auxiliaries into account. We show the probability of having a logical error $p_L$ as a function of the noise variance $\sigma^2/2\pi$ in the system for the (a) square GKP \cref{subsubsec: square_GKP_info}, (b) hexagonal GKP \cref{subsubsec: hex_GKP_info}, (c) tesseract GKP \cref{subsubsec: tess_GKP_info} and (d) D$_4$ GKP \cref{subsubsec: D4_GKP_info}. We show the variance $\zeta^*_\text{G}$ at which $p_L \approx 10^{-5}$ for each storage lattice by the dotted gray vertical line. For the D$_4$ GKP, we also show with light gray lines the value of $\sigma^2/2\pi$, corresponding to a $\Delta_{(\text{dB})} \approx 11$dB squeezed GKP state. Each figure is simulated over a range of $\sigma^2/2\pi = [1.5, 10] \times 10^{-3}$ and each point is obtained by a Monte-Carlo method through the sampling of $N = 10^8$ (MLD) and $N = 10^9$ (MED and COR-MED) errors $\boldsymbol{\xi}$. The standard deviation $\bar{\sigma}$ of each point is smaller than the size of the markers. Finally, in \cref{subsec: high_error_regime_decoder_behaviors}, we analyze the behavior of the three decoders at high levels of noise to show that they all converge to the expected values of $p_L$.}
    \label{fig: comparing_MED_vs_CORMED_main_fig_1e9}
\end{figure*}

In this section, we discuss the numerical simulations we performed in order to compare how the MED decoder, presented in \cref{subsec: qec_noisy_aux}, performs against our COR-MED decoder, presented in \cref{subsec: what_we_propose}. We also provide a comparison of both decoders with the MLD decoder, described by \cref{eq: general_mld_decoder}. The main results are shown on \cref{fig: comparing_MED_vs_CORMED_main_fig_1e9}. We consider the case where the noise in the system is represented by $\Gamma_0$. We first detail how the simulations are realized, and we finish by analyzing the main numerical results.

Each point appearing in the graphs of \cref{fig: comparing_MED_vs_CORMED_main_fig_1e9} is generated through a Monte-Carlo (MC) simulation, repeating the following steps $N$ times:
\begin{enumerate}
    \item Sample an error vector $\boldsymbol{\xi}$ on the system from a Gaussian noise model with covariance $\Gamma_0$
    \item Commute $\boldsymbol{\xi}$ throughout the Steane-type QEC circuit and measure the auxiliaries to obtain the syndrome $\mathbf{z}_{\bar{T}, \text{m}}$
    \item Apply the chosen decoder and extract a correction $\bar{\boldsymbol{\xi}}_{\bar{T}, \text{G}}$
    \item Verify if the resulting displaced storage, $\boldsymbol{\xi}_{\bar{T}, \text{G}} - \bar{\boldsymbol{\xi}}_{\bar{T}, \text{G}}$, is in the logical Voronoi cell $\mathcal{V}_\text{G}^\perp$
\end{enumerate}
We call this a single MC step, taking a time $t_{\boldsymbol{\xi}}$ to achieve. This MC step yields a boolean value; either the correction was successful or not. Repeating this process $N$ times for a fixed value of $\sigma^2$, defining $\Gamma_0$, enables us to build statistical data on $p_L$, the probability of having a logical error when decoding. Since each MC step is a Bernoulli trial with failure probability $p_L$, the variance of the estimated $p_L$ is given by $\bar{\sigma}^2 = \frac{p_L(1 - p_L)}{N}$ \cite{fellerProbabilityTheory}.

Then, we repeat this scheme for different strengths of initial noise $\sigma^2$, leading to the curves in \cref{fig: comparing_MED_vs_CORMED_main_fig_1e9}. Note that we take the variance in units of $\ell^2 = 2\pi$.

All the simulations performed in order to obtain our numerical results are realized with the Julia programming language \cite{Julia}. We also mention that the algorithm we use to solve the CVP in both decoders is presented in Ref. \cite{agrellClosestPointSearch2002} and implemented in Julia following the work of Ref. \cite{lin_2023a}.

Let us now analyze \cref{fig: comparing_MED_vs_CORMED_main_fig_1e9} in more detail. The four panels, Figs. \fullsubfigref{fig: comparing_MED_vs_CORMED_main_fig_1e9}{subfig: comparing_MED_vs_CORMED_main_fig_1e9_square}, \fullsubfigref{fig: comparing_MED_vs_CORMED_main_fig_1e9}{subfig: comparing_MED_vs_CORMED_main_fig_1e9_hex}, \fullsubfigref{fig: comparing_MED_vs_CORMED_main_fig_1e9}{subfig: comparing_MED_vs_CORMED_main_fig_1e9_tess} and \fullsubfigref{fig: comparing_MED_vs_CORMED_main_fig_1e9}{subfig: comparing_MED_vs_CORMED_main_fig_1e9_D4}, respectively, show $p_L(\sigma^2/2\pi)$ for the square, hexagonal, tesseract, and D$_4$ multimode GKP storage. We refer the reader to \cref{sec: appendix_multimode_gkp_code_examples} for more information about these codes. The curves corresponding to the MED and COR-MED decoders are respectively shown as dotted orange and teal lines. For all storage lattices analyzed, the COR-MED decoder gives a lower probability of having a logical error when decoding. For example, for the hexagonal GKP and in the regime where $\sigma^2 / 2\pi \in [2.5, 4.5] \times 10^{-3}$, the COR-MED decoder has a logical error probability close to an order of magnitude lower than the MED decoder. In some cases, for example the D$_4$ GKP, the performances can reach a $40$ times improvement. Notably, we remark that our COR-MED decoder is always a better choice to use, no matter the strength of the noise in the system. It gives better protection against errors for multimode GKP states, and it has the same numerical complexity as the current MED decoder studied in the literature (see \cref{subsec: numerical_complexity_MED_vs_CORMED}).

We notice that the square and the tesseract GKP codes (Figs. \fullsubfigref{fig: comparing_MED_vs_CORMED_main_fig_1e9}{subfig: comparing_MED_vs_CORMED_main_fig_1e9_square} and \fullsubfigref{fig: comparing_MED_vs_CORMED_main_fig_1e9}{subfig: comparing_MED_vs_CORMED_main_fig_1e9_tess}) have a moderate gain using the COR-MED decoder. This originates from the geometry of the measured stabilizers. Indeed, as seen in \cref{subsec: qec_noisy_aux}, the MED decoder estimates the storage error by inverting the matrix whose rows are the measured stabilizer vectors (see \cref{eq: error_from_circ_noisy_aux}). For the square and tesseract codes, these vectors are orthogonal and are all of unit norm. The inversion is then an orthogonal transformation \cite{matrixAnalysis}, which propagates the auxiliary noise into the estimate without amplifying it, so that the MED decoder is already close to optimal. In this case, the COR-MED decoder brings a modest gain by weighting each measurement according to the propagated variance in the circuit. For the hexagonal and D$_4$ codes, the stabilizers are not orthogonal, and the inversion amplifies the auxiliary noise. The COR-MED decoder avoids this amplification, which explains its larger advantage for these two codes.

In \cref{fig: comparing_MED_vs_CORMED_main_fig_1e9}, we also compare both MED decoders with an MLD decoder, the full green lines, which is realized by finding the logical vector that solves \cref{eq: general_mld_decoder} through the numerical estimation of \cref{eq: general_mld_decoder_prob}, made possible by truncating the sum up to all the first shortest vectors of $\Lambda_\mathrm{G}$. Focusing first on Figs. \fullsubfigref{fig: comparing_MED_vs_CORMED_main_fig_1e9}{subfig: comparing_MED_vs_CORMED_main_fig_1e9_square}, \fullsubfigref{fig: comparing_MED_vs_CORMED_main_fig_1e9}{subfig: comparing_MED_vs_CORMED_main_fig_1e9_hex} and \fullsubfigref{fig: comparing_MED_vs_CORMED_main_fig_1e9}{subfig: comparing_MED_vs_CORMED_main_fig_1e9_tess}, the performances of the MED and the MLD decoder are identical for the noise strengths scanned. This was expected for these codes, since most information required for the decoding process can be gathered solely by the center of phase space. In the case of the $\mathrm{D}_4$ GKP, Fig. \fullsubfigref{fig: comparing_MED_vs_CORMED_main_fig_1e9}{subfig: comparing_MED_vs_CORMED_main_fig_1e9_D4}, the performance gained with the MLD decoder is reminiscent of its highly symmetrical lattice with $12$ (up to sign) different stabilizers of minimal length \cite{bapGKPMultimode}. Taking all of them into account gives more information to the decoder, which translates into better performances. Nevertheless, this MLD decoder does not take the noise on the auxiliaries into account. Indeed, this would require solving the MLD on both the full storage mode and auxiliary modes, which is a difficult numerical task to realize. Thus, for all noise strengths, our decoder performs better, demonstrating its strength when considering a more realistically noisy system.

In Ref. \cite{nohLowOverheadFaultTolerantGKP}, concatenating single-mode square GKP qubits to construct a surface code yields a threshold of around $9.9$dB with noiseless gates. Injecting finite-energy GKP states of squeezing below threshold, for example $\Delta_{(\text{dB})} \approx 11$dB, into their proposed hybrid-QEC scheme enables the logical failure rate of the $d = 3$ surface code to be around $p_L^{(\mathrm{SC})} \approx 10^{-3}$. Our findings show that for the case of an $11$dB squeezed D$_4$ GKP code, the same probability is approximately $p_L^{(\text{D}_4)} \approx 1.5 \times 10^{-3}$, as shown in \creffullsubfigref{fig: comparing_MED_vs_CORMED_main_fig_1e9}{subfig: comparing_MED_vs_CORMED_main_fig_1e9_D4} by the light gray lines. Comparing this with their results, using the D$_4$ multimode GKP encoding yields similar performance as using a square surface-GKP code of distance $3$, since the probability of logical errors when decoding is of the same order. More importantly, the difference in hardware cost is significant, since our scheme requires a total of 6 modes, as opposed to 51 modes for the distance 3 square surface-GKP code. Therefore, using the D$_4$ GKP encoding to host the logical information is a more suitable choice performance-wise and for experimental feasibility. Furthermore, preparing such a state only requires the preparation of single-mode GKP states \cite{bapGKPMultimode}, a feat that many platforms are already able to do (see the discussion in \cref{sec: conclu}). In future work, it will be interesting to compare the performance of a D$_4$ surface-GKP code with the threshold value for the square surface-GKP code of $9.9$dB, and see if our decoder enables a lower threshold value.

On each panel of \cref{fig: comparing_MED_vs_CORMED_main_fig_1e9}, we show dotted gray vertical lines highlighting the value of $\sigma^2/2\pi$ where the COR-MED decoder achieves $p_L \approx 10^{-5}$. Enabling us to compare the performance of different storage lattices, this value of noise is denoted $\zeta^*_\text{G}$. As depicted in the figures, we have $\zeta^*_\square < \zeta^*_{\hexagon} < \zeta^*_\text{tess} < \zeta^*_{\text{D}_4}$. From this, we conclude that the D$_4$ multimode GKP code performs better against this noise model with the COR-MED decoder. This ordering can be explained by the distances of each code. Indeed, from \cref{subsubsec: square_GKP_info,subsubsec: hex_GKP_info,subsubsec: tess_GKP_info,subsubsec: D4_GKP_info}, we know that $d_\square < d_{\hexagon} < d_\text{tess} < d_{\text{D}_4}$, indicating that the D$_4$ GKP code should have the best QEC capabilities. Thus, our numerical simulations confirm that the distance is a good estimator of the QEC performance of a multimode GKP code.

To test the COR-MED decoder beyond a single QEC round, we compare it to no error correction and the MED decoder in the context of multi-round QEC for a specific noise strength $\sigma^2 / 2\pi = 5 \times 10^{-3}$ (see \cref{subsec: multi_round_QEC}). After a single round, not correcting yields the lowest logical error probability, since the final logical check perfectly corrects the errors that stay within the logical Voronoi cell, whereas the decoders occasionally misidentify the error from the noisy syndrome. From the second round onward, however, the COR-MED decoder yields the lowest logical error probability for all four GKP codes investigated. Moreover, the mean length of the residual translation modulo $\Lambda_\mathrm{G}$ left on the storage (see \cref{eq: cormed_residual_error}) is smaller for the COR-MED decoder at every round and for every code, keeping the storage mode closer to the center of its logical Voronoi cell and making it more robust against error accumulation.

Finally, we mention that the $S_\text{G}$ matrices we choose to implement $\hat{U}_{\bar{T}}$ for the hexagonal and the D$_4$ GKP are not the ones shown by \cref{eq: hexagonal_GKP,eq: d4_GKP}, respectively. Instead, to make sure the noise from the auxiliaries is spread evenly on all quadratures, we use a quadrature-symmetric form of the stabilizer matrix $S_\text{G}$. We found that both decoders perform either similarly, or better under these symmetric variants of $S_\mathrm{G}$, as was verified numerically in \cref{subsec: SG_mat_effect_on_decoders}.

\section{Discussion and outlook}
\label{sec: conclu}

In this work, we developed a decoder, which we call the COR-MED decoder, for Steane-type QEC of multimode GKP codes. We showed that by leveraging correlations between errors on the auxiliary states and the error the GKP storage underwent, the probability of having a logical error during the decoding process decreases significantly. We analyze our decoder's performance under the Gaussian random translation error channel by comparing it with another decoder based on solving the CVP \cite{conradGKPMultimode,lin_2023a}. Our results show that, for certain multimode GKP encodings, the COR-MED decoder protects the logical information against noise at least 10 times better in a relevant noise regime. This indicates that with the right decoder, some multimode GKP codes that were not of high interest because of their previous QEC performance now arise as promising candidates. This is exactly the case of the D$_4$ GKP code. With the COR-MED decoder, its decoding performances are now better than all other GKP encodings investigated in this work. Furthermore, the full single-qubit Clifford group of this code is implementable with passive Gaussian operations \cite{bapGKPMultimode}.

While the practical realization of multimode GKP codes remains a challenge, recent experimental progress in various platforms suggests that they may soon be within reach. In trapped ions, GKP states have been prepared \cite{fluhmannEncodingQubitTrappedion2019}, error-corrected \cite{deNeeveTrappedIonGKP}, and universal gate sets have been demonstrated \cite{matsosUniversalQuantumGate2025}. In circuit QED, error correction of single-mode GKP qubits \cite{campagneGKPQEC,sivakRealtimeQEC} and qudits \cite{brockGKPQudits} has been demonstrated. While homodyne detection of stationary modes in these systems is currently challenging, recent proposals suggest that high-fidelity homodyne detection may soon be possible \cite{QubitDyneSimPaper}. In photonic systems, where low-noise homodyne measurements are easier to realize \cite{goodHomodyneDetect}, GKP states have recently been generated \cite{fabreGenerationTimefrequencyGrid2020,FurusawaGKPStatePrep,XanaduSquareGKPStatePrep}.

In this work, we have focused on Steane-type error correction circuits, but we believe that the COR-MED decoder could also be adapted for different types of QEC schemes for GKP codes. For example, for the square GKP, a similar correlation tracking idea could be applied to Knill-type QEC circuits \cite{KnillQECInitpaper}, such as the quantum teleportation circuit presented in Refs. \cite{QECwithGKPreview,TelepBosonicNoisyMeasurements}. It would then be interesting to benchmark its performance against the classic circuit originally proposed in Ref. \cite{originalGKP} when considering noisy auxiliaries.

There is still a limit to which we can reduce the rate of errors. Even if we find better decoders that push this limit, we expect that the solution to achieve fault-tolerant quantum computing is to concatenate multimode GKP codes with qubit codes. This enables hybrid QEC and has already been studied for various qubit codes: mostly surface codes \cite{AnalogGKPSurface,QECwithToricGKP,nohFaultTolerantGKP,MBQCSurfaceCodeGKP,nohLowOverheadFaultTolerantGKP,RepGKPQubitCode,lin_2023a,XZZXSurfaceCodeGKP}, but also color codes \cite{colorCodeGKP}, QLDPC codes \cite{GKPAndQLDPCConstruction}, and cluster-states in the measurement-based quantum computing paradigm \cite{blueprintXanaduGKP,LossToRGTNoiseModel2,XanaduAurora,BicycleCodesXanadu}. Implementing our COR-MED decoder in a concatenated qubit code scheme is a natural next step. We expect that the gain in performance we observed in this work will translate into a lower overhead when concatenating with qubit codes.

\section*{\hspace{-0.55cm} Author Contributions}
T.P. and B.R. conceived the project. M.-A.R. and T.P. derived the mathematical framework describing the problem, verified the underlying equations, and carried out the numerical simulations required to support the analytical results. M.-A.R. produced the figures and wrote the manuscript with input from all authors. B.R. supervised the project.

The final revisions of the manuscript were carried out with the assistance of Claude Opus 5.5, developed by Anthropic. This tool was used for checking the mathematical notation, verifying the internal consistency of the manuscript, and running supplementary numerical analyses of the decoders' performance, in accordance with the authors' instructions. All content, claims, and conclusions have been reviewed and verified by the authors to ensure accuracy and originality.

\section*{\hspace{-0.55cm} Acknowledgments}
B.R. and M.-A.R. were supported by the Natural Sciences and Engineering Research Council of Canada (NSERC), the Canada First Research Excellence Fund (CFREF), the Fonds de Recherche du Québec - Nature et Technologies (FRQNT) as well as the Army Research office under grant number W911NF2310045. T.P. acknowledges funding from Institut Mines-Télécom (IMT), l'institut Carnot Télécom \& Société Numérique (TSN) and the Fondation Mines-Télécom.

\onecolumn
\appendix

\section{Multimode GKP code examples}
\label{sec: appendix_multimode_gkp_code_examples}

In this appendix, we present a few examples of multimode GKP codes that were introduced in earlier work. For each GKP qubit, we also show the corresponding logical vectors using methods described in the beginning of \cref{sec: gkp_theory} and the code distance from \cref{subsec: distance_of_GKP_codes}.

Note that in this work, we are only studying multimode codes that do not arise from their concatenation with stabilizer codes, such as repetition, surface, or QLDPC codes. More information on these constructions can be found in Refs. \cite{AnalogGKPSurface,QECwithToricGKP,nohFaultTolerantGKP,MBQCSurfaceCodeGKP,nohLowOverheadFaultTolerantGKP,lin_2023a,XZZXSurfaceCodeGKP,colorCodeGKP,GKPAndQLDPCConstruction,blueprintXanaduGKP}.

\subsection{Rectangular GKP} 

We start with a GKP code defined on a rectangular lattice GKP \cite{nohLowOverheadFaultTolerantGKP}
\begin{align}
    S_{\text{rec}, \eta} = \sqrt{2}S_{\varnothing, \eta} = \sqrt{2}\begin{pmatrix}
        \eta & 0 \\
        0 & \frac{1}{\eta}
    \end{pmatrix}.
    \label{eq: rectangular_GKP}
\end{align}
Since the squeezing in the $\hat{q}$ and $\hat{p}$ quadratures are the inverse of one another, we can still encode a qubit because $\bar{n}_{\text{rec}, \eta} = \sqrt{|\mathrm{det}A_{\mathrm{rec}, \eta}|} = 2$. The logical vectors are defined by
\begin{align}
    P_{\text{rec}, \eta} = \frac{1}{\sqrt{2}}\begin{pmatrix}
        \eta & 0 \\
        \eta & \frac{1}{\eta} \vspace{0.075cm}\\
        0 & \frac{1}{\eta}
    \end{pmatrix},
\end{align}
where the distance of this code is bounded by the smallest side of the rectangle, which depends on whether $\eta$ is smaller or larger than $1$. This means that $d_{\text{rec}, \eta} = \frac{\mathrm{min}(\eta, \eta^{-1})}{\sqrt{2}}$.

\subsection{Square GKP}
\label{subsubsec: square_GKP_info}

We retrieve the well-studied square GKP when $\eta = 1$ for the rectangular GKP 
\begin{align}
    S_\square = S_{\text{rec}, 1} = \sqrt{2}\begin{pmatrix}
        1 & 0 \\
        0 & 1
    \end{pmatrix},
    \label{eq: square_GKP}
\end{align}
originally introduced in \cite{originalGKP}. Its logical vectors are given by
\begin{align}
    P_\square = P_{\text{rec}, 1} = \frac{1}{\sqrt{2}}\begin{pmatrix}
        1 & 0 \\
        1 & 1 \\
        0 & 1
    \end{pmatrix},
\end{align}
and the distance is $d_\square = \frac{1}{\sqrt{2}} \approx 0.7071$.

\subsection{Hexagonal GKP}
\label{subsubsec: hex_GKP_info}

The code with the largest possible distance for a single mode is the hexagonal GKP code, based on a hexagonal lattice
\begin{align}
    S_{\hexagon} = \frac{2}{\sqrt[4]{3}}\begin{pmatrix}
        1 & 0 \\
        -\frac{1}{2} & \frac{\sqrt{3}}{2}
    \end{pmatrix},
    \label{eq: hexagonal_GKP}
\end{align}
also introduced in \cite{originalGKP}. It has logical vectors 
\begin{align}
    P_{\hexagon} = \frac{1}{\sqrt[4]{3}}\begin{pmatrix}
        1 & 0 \\
        \frac{1}{2} & \frac{\sqrt{3}}{2} \vspace{0.075cm}\\
        -\frac{1}{2} & \frac{\sqrt{3}}{2}
    \end{pmatrix},
\end{align}
and distance $d_{\hexagon} = \frac{1}{\sqrt[4]{3}} \approx 0.7598$.

\subsection{Tesseract GKP}
\label{subsubsec: tess_GKP_info}

Based on a lattice that has the resemblance of a 4-dimensional cube, the tesseract GKP code \cite{bapGKPMultimode}
\begin{align}
    S_{\text{tess}} = \sqrt[4]{2}\begin{pmatrix}
        1 & 0 & 0 & 0 \\
        0 & \frac{1}{\sqrt{2}} & 0 & \frac{1}{\sqrt{2}} \\
        0 & 0 & 1 & 0 \\
        0 & \frac{1}{\sqrt{2}} & 0 & -\frac{1}{\sqrt{2}}
    \end{pmatrix},
    \label{eq: tesseract_GKP}
\end{align}
with the property that $\mathbf{s}_j^T \mathbf{s}_k = 0$, $\forall j,k$ with $j \neq k$, and that $|\mathbf{s}_j| = \sqrt[4]{2}$, $\forall j$. Its logical vectors are given by
\begin{align}
    P_{\text{tess}} = \sqrt[4]{2}\begin{pmatrix}
        \frac{1}{2} & 0 & \frac{1}{2} & 0 \vspace{0.075cm}\\
        \frac{1}{2} & \frac{1}{\sqrt{2}} & \frac{1}{2} & 0 \vspace{0.075cm}\\
        0 & \frac{1}{\sqrt{2}} & 0 & 0
    \end{pmatrix},
\end{align}
and its distance is $d_\text{tess} = \frac{1}{ \sqrt[4]{2}} \approx 0.8409$.

\subsection{D$_{2n}$ GKP}
\label{subsubsec: D4_GKP_info}

Based on the D$_{2n}$ root lattices, we have a family of GKP codes that can always encode a qubit for $n \geq 2$ since again $\bar{n}_{\mathrm{D}_{2n}} = \sqrt{|\mathrm{det}A_{\mathrm{D}_{2n}}|} = 2$. The stabilizer matrix of these codes is given by
\begin{align}
    S_{\mathrm{D}_{2n}} = \begin{pmatrix}
        1 & -1 & 0 & & 0 & 0 & 0\\
        0 & 1 & -1 & \cdots & 0 & 0 & 0\\
        0 & 0 & 1 & & 0 & 0 & 0\\
        & \vdots & & \ddots & & \vdots & \\
        0 & 0 & 0 & & 1 & -1 & 0 \\
        0 & 0 & 0 & \cdots & 0 & 1 & -1 \\
        0 & 0 & 0 & & 0 & 1 & 1
    \end{pmatrix},
    \label{eq: d2n_GKP}
\end{align}
with logical operators
\begin{align}
    P_{\mathrm{D}_{2n}} = \begin{pmatrix}
        \frac{1}{2} & \frac{1}{2} & & \frac{1}{2} & \frac{1}{2} \vspace{0.075cm}\\
        -\frac{1}{2} & \frac{1}{2} & \cdots & \frac{1}{2} & \frac{1}{2} \vspace{0.075cm}\\
        1 & 0 & & 0 & 0
    \end{pmatrix}.
\end{align}
A particular case of interest is when $n = 2$, the D$_4$ code \cite{bapGKPMultimode}, which has the largest possible distance for two modes
\begin{align}
    S_{\mathrm{D}_{4}} = \begin{pmatrix}
        1 & 0 & 1 & 0 \\
        1 & 0 & 0 & -1 \\
        0 & 1 & -1 & 0 \\
        1 & 0 & 0 & 1
    \end{pmatrix},
    \label{eq: d4_GKP}
\end{align}
having the property that $|\mathbf{s}_j| = \sqrt{2}$, $\forall j$ and that $|\mathbf{x}_0| = |\mathbf{y}_0| = |\mathbf{z}_0|$, giving rise to the full single-qubit Clifford group implementable with passive Gaussian operations and a distance $d_{\mathrm{D}_{4}} = 1$.

\section{General Steane-type QEC circuit for multimode GKP codes}
\label{sec: appendix_gen_steane_type_mm_circuit}

\subsection{Circuits with noiseless auxiliary states}
\label{subsec: appendix_gen_steane_type_mm_circuit_perfect_aux}

In this section, we show that the circuit of \creffullsubfigref{fig: steane_type_QEC}{subfig: general_mm_steane_type_QEC} propagates the correct information into the auxiliary states, giving the state shown in \cref{eq: state_after_Steane_circuit_perfect_aux} right before the measurements. We restrict ourselves to the noiseless auxiliaries case.

As was argued in \cref{subsec: qec_perf_aux}, the initial state is given by
\begin{align}
    \ket{\Psi_\text{i}} = \hat{T}(\boldsymbol{\xi}_\text{G})\ket{\psi} \bigotimes_{j = 1}^{2m} \sum_{k_j \in \mathbb{Z}} \ket{\ell k_j \eta_j}_{\hat{q}_{m + j}},
\end{align}
with $\boldsymbol{\xi}_\text{G}$ justified by \cref{eq: error_vector_perfect_aux}. Here, we use this form for the initial state instead of the form that uses the definition \eqref{eq: aux_new_consise_def} since it is more convenient for the proof.

Now let's apply the Steane QEC circuit of \creffullsubfigref{fig: steane_type_QEC}{subfig: general_mm_steane_type_QEC}. During the protocol, we sequentially apply $(m+1)$-mode Gaussian operations $\hat{U}_{T_l}$. Each of these operators may consist of a combination of the operators defined in \cref{eq: quad_action_qq_op,eq: quad_action_pp_op,eq: quad_action_qp_op}. Let us analyze the result of applying $\hat{U}_{T_l}$ on $\ket{\Psi_\text{i}}$:
\begin{align}
    \hat{U}_{T_l}\ket{\Psi_\text{i}} 
    \overset{\eqref{eq: translation_commute_with_gaussian_rel}}{=} \hat{T}(T_l\boldsymbol{\xi}_\text{G})\qty[\hat{U}_{T_l}\ket{\psi}\otimes\sum_{k_l \in \mathbb{Z}} \ket{\ell k_l \eta_l}_{\hat{q}_{m + l}}]\bigotimes_{\substack{j=1 \\ j \neq l}}^{2m} \sum_{k_j \in \mathbb{Z}} \ket{\ell k_j \eta_j}_{\hat{q}_{m + j}}.
\end{align}
Since we are realizing Steane QEC, we require that the operators we use to do that correction preserve the code space of the storage and the auxiliaries \cite{SteaneQECInitpaper}. This means that we want 
\begin{align}
    \hat{U}_{T_l}\ket{\psi}\sum_{k_l \in \mathbb{Z}} \ket{\ell k_l \eta_l}_{\hat{q}_{m + l}} = \ket{\psi}\sum_{k_l \in \mathbb{Z}} \ket{\ell k_l \eta_l}_{\hat{q}_{m + l}},
    \label{eq: effect_of_Utl_on_gkp_and_aux}
\end{align}
so that all there is left to do is compute $T_l\boldsymbol{\xi}_\text{G}$. We choose $\hat{U}_{T_l}$ so that it maps the $\hat{g}_l $ operator onto the $\hat{q}_{m + l}$ quadrature. Recall that the $\hat{g}_l$ operator is given by
\begin{align}
    \hat{g}_l = \sum_{j = 1}^{2m} \kappa_{l, j}\hat{x}_{\text{G},j},
    \label{eq: GKP_g_quad_linear_map}
\end{align}
with $\kappa_{l, j} \in \mathbb{R}$ located at row $l$ and column $j$ of the $2m \times 2m$ matrix $K$. This matrix can be defined as $K = \ell \boldsymbol{\nu}' \circ S_\text{G}\Omega = \boldsymbol{\nu} \circ S_\text{G}\Omega$, where $\boldsymbol{\nu} = \ell \boldsymbol{\nu}'$. Note that here we are using the Hadamard product to denote the element-wise multiplication between each $j$th component of the vector $\boldsymbol{\nu}$ and each $j$th row of the matrix $S_\text{G}\Omega$. Here, we introduced the parameter $\boldsymbol{\nu}$ just to add a supplementary degree of freedom when measuring the stabilizers so we can more easily describe measuring $\hat{\mathbf{h}}$ from \cref{eq: stab_quad_measurement} or $\hat{\mathbf{h}}'$ from \cref{eq: unity_stab_quad_measurement}. The relation between $\boldsymbol{\nu}$ and $K$ in function of the stabilizers being measured is explained in more detail in \cref{subsec: appendix_gen_steane_type_mm_circuit_analysis}. We notice that a single $\hat{g}_l$ is parametrized by $2m$ coefficients $\kappa_{l, j}$. Therefore, a single $\hat{U}_{T_l}$ is composed at most of $2m$ two-mode squeezing operators, each with their corresponding squeezing parameter $\kappa_{l, j}$. The values of the $2m$ squeezing parameters $\kappa_{l, j}$ are given by the $l$th row of $K$, which we denote as the row vector $\boldsymbol{\kappa}_{l}^T$. In other words, we have to choose the right form and the right squeezing parameters $\boldsymbol{\kappa}_{l}$ for the operators \eqref{eq: quad_action_qq_op} to \eqref{eq: quad_action_qp_op} so that $\hat{U}_{T_l}$ has the effect
\begin{align}
    T_l\boldsymbol{\xi}_\text{G} =
    \begin{pmatrix}
        \boldsymbol{\xi}_\text{G} \\
        0 \\
        \vdots \\
        0 \\
        \boldsymbol{\kappa}_l^T\boldsymbol{\xi}_\text{G} \\
        0 \\
        \vdots \\
        0
    \end{pmatrix},
    \label{eq: effect_of_Utl_on_error}
\end{align}
where $\boldsymbol{\kappa}_l^T\boldsymbol{\xi}_\text{G}$ is located at the $2m + 2l - 1$ index of the vector. From this, we see that $\hat{T}(T_l\boldsymbol{\xi}_\text{G}) = \hat{T}(\boldsymbol{\xi}_\text{G})\hat{T}(\mathbf{g}_{\xi_{\text{G}}, l})$, with $\mathbf{g}_{\xi_{\text{G}}, l}$ a vector that has $\boldsymbol{\kappa}_l^T\boldsymbol{\xi}_\text{G}$ at index $2m + 2l - 1$ as its only non-zero entry. The initial state is now
\begin{align}
    \hat{U}_{T_l}\ket{\Psi_\text{i}} &= \hat{T}(\boldsymbol{\xi}_\text{G})\ket{\psi}\otimes \sum_{k_l \in \mathbb{Z}} \hat{T}(\mathbf{g}_{\xi_{\text{G}}, l})\ket{\ell k_l \eta_l}_{\hat{q}_{m + l}} \bigotimes_{\substack{j=1 \\ j \neq l}}^{2m} \sum_{k_j \in \mathbb{Z}} \ket{\ell k_j \eta_j}_{\hat{q}_{m + j}} \nonumber\\
    &= \hat{T}(\boldsymbol{\xi}_\text{G})\ket{\psi} \otimes\sum_{k_l \in \mathbb{Z}} \ket{\ell \boldsymbol{\kappa}_l^T\boldsymbol{\xi}_\text{G} + \ell k_l \eta_l}_{\hat{q}_{m + l}} \bigotimes_{\substack{j=1 \\ j \neq l}}^{2m} \sum_{k_j \in \mathbb{Z}} \ket{\ell k_j \eta_j}_{\hat{q}_{m + j}}.
\end{align}

Now that we know how a single $\hat{U}_{T_l}$ changes the initial state \eqref{eq: steane_init_state_perfect_aux}, we can deduce the effect of applying all $2m$ of them. Right before measurement, the final state is given by
\begin{align}
    \ket{\Psi_\text{f}} = \hat{T}(\boldsymbol{\xi}_\text{G})\ket{\psi} \bigotimes_{j = 1}^{2m} \sum_{k_j \in \mathbb{Z}} \ket{\ell\boldsymbol{\kappa}_j^T\boldsymbol{\xi}_\text{G} + \ell k_j \eta_j}_{\hat{q}_{m + j}}.
\end{align}
Finally, reintroducing the definition \cref{eq: aux_new_consise_def}, we can write
\begin{align}
    \ket{\Psi_\text{f}} = \hat{T}(\boldsymbol{\xi}_\text{G})\ket{\psi}\otimes \sum_{\mathbf{k} \in \mathbb{Z}^{2m}} \ket{\ell K\boldsymbol{\xi}_\text{G} + \ell \mathbf{k} \circ \boldsymbol{\eta}}_{\hat{\mathbf{q}}_\text{m}},
\end{align}
and developing the matrix $K$, we have
\begin{align}
    \ket{\Psi_\text{f}} = \hat{T}(\boldsymbol{\xi}_\text{G})\ket{\psi}\otimes \sum_{\mathbf{k} \in \mathbb{Z}^{2m}} \ket{\ell \boldsymbol{\nu} \circ S_\text{G}\Omega\boldsymbol{\xi}_\text{G} + \ell \mathbf{k} \circ \boldsymbol{\eta}}_{\hat{\mathbf{q}}_\text{m}},
    \label{eq: state_after_steane_circ_noiseless_aux_appendix}
\end{align}
which corresponds to the state \eqref{eq: state_after_Steane_circuit_perfect_aux}.

The core of the proof lies in the fact that we assume all of the $\hat{U}_{T_l}$ operators are well chosen to respect \eqref{eq: effect_of_Utl_on_gkp_and_aux} and \eqref{eq: effect_of_Utl_on_error} for the storage of interest. In \cref{subsec: appendix_gen_steane_type_mm_circuit_construct}, we show how to determine the form of the $T_l$ matrices so that they respect these conditions.

We emphasize that \eqref{eq: effect_of_Utl_on_error} is valid only when there is no noise on the auxiliaries. Indeed, if there were noise, the back-action would cause $\boldsymbol{\xi}_\text{G}$ to also be shifted by a certain amount based on the errors on the auxiliaries. We discuss this in more detail in the next section, appendix \ref{subsec: appendix_gen_steane_type_mm_circuit_noisy_aux}.

For the case of \cref{eq: effect_of_Utl_on_gkp_and_aux}, even if we have correct propagation of all $\hat{g}_j$ operators to the $\hat{q}$ quadrature of the qunaught $j$, we must also have that the application of $\hat{U}_{T_j}$ on the system preserves the stabilizers of the storage and the auxiliaries. In other words, to make sure that \eqref{eq: effect_of_Utl_on_gkp_and_aux} is valid, we must have 
\begin{align}
    ST_j^T = RS,~\forall j \implies S\bar{T}^T = RS,
    \label{eq: Steane_stab_preserve_condition_on_S}
\end{align}
where we use the definition \eqref{eq: T_bar_definition} and we consider $S$ as the stabilizer matrix of the storage and all the auxiliary states. Here, $R$ is a unimodular matrix representing a change of basis. More precisely, we have
\begin{align}
    S = S_\text{G}\oplus S_\eta = S_\text{G} \oplus \qty[\bigoplus_{j = 1}^{2m} S_{\varnothing, \eta_j}] = \begin{pmatrix}
        S_\text{G} & 0 & \cdots & 0 \\
        0 & S_{\varnothing,\eta_1} &  & 0 \\
        \vdots & & \ddots & \vdots \\
        0 & 0 & \cdots & S_{\varnothing,\eta_{2m}}
    \end{pmatrix}.
    \label{eq: full_system_S_mat}
\end{align}
The projector on this code space $\mathcal{C}$ is \cite{conradGKPMultimode}
\begin{align}
    \mathcal{P}_{\mathcal{C}} = \sum_{\mathbf{a} \in \mathbb{Z}^{6m}}\hat{T}(S^T \mathbf{a}),
\end{align}
where we choose the stabilizers generated by $S$ to be in the $+1$ gauge. Here, the sum is over all possible integer vectors in $\mathbb{Z}^{6m}$, therefore all elements of the stabilizer group. Using this projector, the initial state without error is 
\begin{align}
    \ket{\Phi_{\text{i}}} = \mathcal{P}_{\mathcal{C}}\ket{0}^{\otimes 3m} = \sum_{\mathbf{a} \in \mathbb{Z}^{6m}}\hat{T}(S^T \mathbf{a})\ket{0}^{\otimes 3m},
\end{align}
where $\ket{0}^{\otimes 3m}$ is the vacuum in each $3m$ mode. Here, we also note that we define the initial state as $\ket{\Phi_{\text{i}}}$, which differs from $\ket{\Psi_{\text{i}}}$ by $\ket{\Psi_{\text{i}}} = \hat{T}(\boldsymbol{\xi}_\text{G})\ket{\Phi_{\text{i}}}$. Now, when applying the Steane QEC circuit parametrized by $\bar{T}$ (see Eq. \eqref{eq: T_bar_definition}), the initial state becomes
\begin{align}
    \hat{U}_{\bar{T}}\ket{\Phi_{\text{i}}} = \hat{U}_{\bar{T}}\sum_{\mathbf{a} \in \mathbb{Z}^{6m}}\hat{T}(S^T \mathbf{a})\ket{0}^{\otimes 3m} \overset{\eqref{eq: translation_commute_with_gaussian_rel}}{=} \sum_{\mathbf{a} \in \mathbb{Z}^{6m}}\hat{T}(\bar{T}S^T \mathbf{a})\hat{U}_{\bar{T}}\ket{0}^{\otimes 3m} = \sum_{\mathbf{a} \in \mathbb{Z}^{6m}}\hat{T}((S\bar{T}^T)^T \mathbf{a})\ket{0_{\bar{T}}}.
\end{align}
If we impose Eq. \eqref{eq: Steane_stab_preserve_condition_on_S}, then
\begin{align}
    \hat{U}_{\bar{T}}\ket{\Phi_{\text{i}}} = \sum_{\mathbf{a} \in \mathbb{Z}^{6m}}\hat{T}((RS)^T \mathbf{a})\ket{0_{\bar{T}}} = \sum_{R^{-T}\mathbf{b} \in \mathbb{Z}^{6m}}\hat{T}(S^T \mathbf{b})\ket{0_{\bar{T}}},
\end{align}
and, since the sum is infinite, summing over $R^{-T}\mathbf{b}$ is the same as summing over $\mathbf{b}$. Therefore, 
\begin{align}
    \hat{U}_{\bar{T}}\ket{\Phi_{\text{i}}} = \sum_{\mathbf{b} \in \mathbb{Z}^{6m}}\hat{T}(S^T \mathbf{b})\ket{0_{\bar{T}}} = \mathcal{P}_{\mathcal{C}}\ket{0_{\bar{T}}} = \ket{\Phi_{\text{i}}},
\end{align}
where we can justify the last step by the fact that starting from vacuum or any distorted vacuum state does not matter as long as we are projecting in the end. We can also conclude that \cref{eq: effect_of_Utl_on_gkp_and_aux} must be true for a precise $\hat{U}_{T_l}$ if it is respected for $\hat{U}_{\bar{T}}$.

\subsection{Circuits with noisy auxiliary states}
\label{subsec: appendix_gen_steane_type_mm_circuit_noisy_aux}

In this section, we show that in the case where the auxiliary states are noisy, the circuit of figure \fullsubfigref{fig: steane_type_QEC}{subfig: general_mm_steane_type_QEC} yields the state \eqref{eq: state_before_measurement_noisy_aux} right before the measurement. In fact, we show that the circuit of figure \fullsubfigref{fig: steane_type_QEC}{subfig: general_mm_steane_type_QEC} still correctly propagates information from the storage to the auxiliaries, but now the errors on the auxiliary states $\boldsymbol{\xi}_{\eta_j}$ corrupt the final measurements and hide the actual error we wish to estimate.

In this scenario, we start with the initial state that was presented in section \ref{subsec: qec_noisy_aux}, that is
\begin{align}
    \ket{\Psi_\text{i}} = \hat{T}(\boldsymbol{\xi})\ket{\psi}\otimes \sum_{\mathbf{k} \in \mathbb{Z}^{2m}} \ket{\ell \mathbf{k} \circ \boldsymbol{\eta}}_{\hat{\mathbf{q}}_{\text{m}}},
\end{align}
where the error $\boldsymbol{\xi}$ is defined by \cref{eq: xi_noisy_case}. Now, we can look at how this state changes following the application of the Steane QEC circuit. Here, instead of applying a single $\hat{U}_{T_l}$ to see how it affects $\ket{\Psi_\text{i}}$, we apply them all at the same time using \cref{eq: U_T_bar_definition,eq: T_bar_definition}. Thus, the Steane circuit transforms the initial state into
\begin{align}
    \ket{\Psi_\text{f}} = \hat{U}_{\bar{T}}\ket{\Psi_\text{i}} &= \hat{U}_{\bar{T}}\hat{T}(\boldsymbol{\xi})\ket{\psi}\otimes \sum_{\mathbf{k} \in \mathbb{Z}^{2m}} \ket{\ell \mathbf{k} \circ \boldsymbol{\eta}}_{\hat{\mathbf{q}}_{\text{m}}} \nonumber \\
    &\overset{\eqref{eq: translation_commute_with_gaussian_rel}}{=} \hat{T}(\bar{T}\boldsymbol{\xi})\hat{U}_{\bar{T}}\ket{\psi}\otimes \sum_{\mathbf{k} \in \mathbb{Z}^{2m}} \ket{\ell \mathbf{k} \circ \boldsymbol{\eta}}_{\hat{\mathbf{q}}_{\text{m}}} \nonumber \\
    &\overset{\eqref{eq: effect_of_Utl_on_gkp_and_aux}}{=} \hat{T}(\bar{T}\boldsymbol{\xi})\ket{\psi}\otimes \sum_{\mathbf{k} \in \mathbb{Z}^{2m}} \ket{\ell \mathbf{k} \circ \boldsymbol{\eta}}_{\hat{\mathbf{q}}_{\text{m}}}.
    \label{eq: state_before_phase_split}
\end{align}
Now, we use the three projectors introduced in the main text. As a reminder, $\Pi_\text{G}$ is the projector onto the storage, $\Pi_{\text{m}}$ onto the $\hat{q}$ quadrature of all auxiliary states we measure, and $\Pi_{\text{p}}$ onto the $\hat{p}$ quadrature of all auxiliary states. Using these projectors, we can decompose $\hat{T}(\bar{T}\boldsymbol{\xi})$ into two different translation operators acting separately on the storage and on the auxiliaries. Since operations acting separately on these modes commute, \cref{eq: state_before_phase_split} becomes
\begin{align}
    \ket{\Psi_\text{f}} = \hat{T}(\Pi_\text{G}\bar{T}\boldsymbol{\xi})\ket{\psi} \otimes\sum_{\mathbf{k} \in \mathbb{Z}^{2m}} \hat{T}(\Pi_{\text{m}}\bar{T}\boldsymbol{\xi} + \Pi_{\text{p}}\bar{T}\boldsymbol{\xi})\ket{\ell \mathbf{k} \circ \boldsymbol{\eta}}_{\hat{\mathbf{q}}_{\text{m}}},
\end{align}
and decomposing the translation on the auxiliaries using \cref{eq: translation_compose_rel}, the final state is given by
\begin{align}
    \ket{\Psi_\text{f}} = \hat{T}(\Pi_\text{G}\bar{T}\boldsymbol{\xi})\ket{\psi} \otimes\sum_{\mathbf{k} \in \mathbb{Z}^{2m}} \hat{T}(\Pi_{\text{m}}\bar{T}\boldsymbol{\xi})\hat{T}(\Pi_{\text{p}}\bar{T}\boldsymbol{\xi})\ket{\ell \mathbf{k} \circ \boldsymbol{\eta}}_{\hat{\mathbf{q}}_{\text{m}}},
\end{align}
up to an irrelevant global phase $\phi = -\pi \boldsymbol{\xi}^T\bar{T}^T\Pi_\text{p}^T \Omega \Pi_\text{m}\bar{T}\boldsymbol{\xi}$. Since the operator $\hat{T}(\Pi_{\text{p}}\bar{T}\boldsymbol{\xi})$ is a translation of a vector with only components in the $\hat{p}_j$ quadratures of the auxiliaries; it is generated by all $\hat{q}_j$ operators of the auxiliaries. Thus,
\begin{align}
    \ket{\Psi_\text{f}} &= \hat{T}(\Pi_\text{G}\bar{T}\boldsymbol{\xi})\ket{\psi} \otimes\sum_{\mathbf{k} \in \mathbb{Z}^{2m}} \hat{T}(\Pi_{\text{m}}\bar{T}\boldsymbol{\xi})\mathrm{e}^{i\ell \boldsymbol{\xi}^T\bar{T}^T\Pi_{\text{p}}^T \hat{\mathbf{q}}_\text{m}}\ket{\ell \mathbf{k} \circ \boldsymbol{\eta}}_{\hat{\mathbf{q}}_{\text{m}}} \nonumber\\
    &= \hat{T}(\Pi_\text{G}\bar{T}\boldsymbol{\xi})\ket{\psi} \otimes\sum_{\mathbf{k} \in \mathbb{Z}^{2m}} \mathrm{e}^{i\ell^2 \boldsymbol{\xi}^T\bar{T}^T\Pi_{\text{p}}^T \mathbf{k} \circ \boldsymbol{\eta}}\ket{\ell\Pi_{\text{m}}\bar{T}\boldsymbol{\xi} + \ell \mathbf{k} \circ \boldsymbol{\eta}}_{\hat{\mathbf{q}}_{\text{m}}}
    \label{eq: noisy_aux_full_final_state}
\end{align}
and we retrieve the state of \cref{eq: steane_init_state_noisy_aux_short_vers}, as expected.

For the rest of this section, we refrain from writing the phase on the auxiliaries, as it does not change the measurement results, emphasizing that it disappears once the measurement of the auxiliaries is realized. Let us see how we can go from \cref{eq: steane_init_state_noisy_aux_short_vers} to \cref{eq: state_before_measurement_noisy_aux}. We can start by analyzing how we can simplify $\Pi_\text{G}\bar{T}\boldsymbol{\xi}$ and $\Pi_{\text{m}}\bar{T}\boldsymbol{\xi}$. When there is no noise on the auxiliaries, we have $\Pi_\text{G}\bar{T}\boldsymbol{\xi}_\text{G} = \boldsymbol{\xi}_\text{G}$ and $\Pi_{\text{m}}\bar{T}\boldsymbol{\xi}_\text{G} = K\boldsymbol{\xi}_\text{G}$, as we can deduce from \cref{eq: effect_of_Utl_on_error}. Since $\bar{T}$ is a linear operation, all the noise on the auxiliaries does is shift these two quantities such that
\begin{align}
    \Pi_\text{G}\bar{T}\boldsymbol{\xi} &= \boldsymbol{\xi}_\text{G} + \boldsymbol{\delta}_\text{G}\label{eq: noise_after_steane_on_GKP_quad} \\ 
    \Pi_{\text{m}}\bar{T}\boldsymbol{\xi} &= K\boldsymbol{\xi}_\text{G} + \boldsymbol{\delta}_{\text{m}},
    \label{eq: noise_after_steane_on_aux_quad}
\end{align}
where $\boldsymbol{\delta}_\text{G}$ and $\boldsymbol{\delta}_{\text{m}}$ are linear combinations of the random components of the vectors $\boldsymbol{\xi}_{\eta_j}$. Clearly, they are both correlated, and these are precisely the correlations we leverage in this work to achieve better decoding. Using \cref{eq: noise_after_steane_on_GKP_quad} and \cref{eq: noise_after_steane_on_aux_quad}, the final state represented by \cref{eq: noisy_aux_full_final_state} becomes
\begin{align}
    \ket{\Psi_\text{f}} &= \hat{T}(\boldsymbol{\xi}_\text{G} + \boldsymbol{\delta}_\text{G})\ket{\psi}\otimes \sum_{\mathbf{k} \in \mathbb{Z}^{2m}} \ket{\ell K\boldsymbol{\xi}_\text{G} + \ell\boldsymbol{\delta}_\text{m} + \ell \mathbf{k} \circ \boldsymbol{\eta}}_{\hat{\mathbf{q}}_\text{m}}.
\end{align}
Finally, using the definition of $K$ introduced in \cref{subsec: appendix_gen_steane_type_mm_circuit_perfect_aux}, we have
\begin{align}
    \ket{\Psi_\text{f}} &= \hat{T}(\boldsymbol{\xi}_\text{G} + \boldsymbol{\delta}_\text{G})\ket{\psi}\otimes \sum_{\mathbf{k} \in \mathbb{Z}^{2m}} \ket{\ell \boldsymbol{\nu} \circ S_\text{G} \Omega \boldsymbol{\xi}_\text{G} + \ell\boldsymbol{\delta}_\text{m} + \ell \mathbf{k} \circ \boldsymbol{\eta}}_{\hat{\mathbf{q}}_\text{m}}.
\end{align}
where we retrieve the state of \cref{eq: state_before_measurement_noisy_aux}.

In the situation where we have noise on the auxiliaries, we notice that the back-action on our system is what causes wrong QEC. Furthermore, the forms that take $\boldsymbol{\delta}_\text{G}$ and $\boldsymbol{\delta}_{\text{m}}$ depend directly on our choice of $\hat{U}_{\bar{T}}$. As we have discussed in the beginning of \cref{sec: qec_multimode_gkp}, many choices of $\hat{U}_{\bar{T}}$ lead to an equivalent measurement of the operators \eqref{eq: stab_quad_measurement}. Notably, measuring operators \eqref{eq: unity_stab_quad_measurement} yields much better results for the QEC protocol. We justify in a bit more detail why this is the case in \cref{subsec: appendix_gen_steane_type_mm_circuit_analysis}, and we show an example for the square GKP code and the tesseract GKP code in \cref{subsec: appendix_gen_steane_type_mm_circuit_square_gkp,subsec: appendix_gen_steane_type_mm_circuit_tess_gkp}, respectively.

\subsection{Construction of a general Steane-type QEC circuit}
\label{subsec: appendix_gen_steane_type_mm_circuit_construct}

In this appendix, we show how to construct a general $\hat{U}_{T_l}$ operator needed for the Steane-type QEC protocols. In order to do so, we start by determining the form of $T_l$. Note that here we assume each auxiliary is measured in the $\hat{q}$ quadrature as in the main text for simplicity.

We know that we can generally express the matrix $T_l$ as
\begin{align}
    T_l = \begin{pmatrix}
        \mathbb{I}_{2m\times 2m} & T_{l, 1} \\
        T_{l, 2} & \mathbb{I}_{4m\times 4m}
    \end{pmatrix},
    \label{eq: general_Tl_block_matrix_form}
\end{align}
where we have specified the dimensions of the identity matrices. Indeed, this operation couples the storage modes with the auxiliary modes. We can then specify the sub-matrices $T_{l, 1}$ and $T_{l, 2}$.

Let us start by describing $T_{l, 2}$, which will help understanding the form of $T_{l, 1}$. $T_{l, 2}$ is the matrix that describes how the $l$th auxiliary is affected by the storage. More specifically, $T_{l, 2}$ describes the propagation of $\hat{g}_l$ into the $\hat{q}_{m + l}$ quadrature of the $l$th auxiliary. This means that we want $\hat{q}_{m + l} \rightarrow \hat{q}_{m + l} + \hat{g}_l = \hat{q}_{m + l} + \boldsymbol{\kappa}_l^T\hat{\mathbf{x}}_\text{G}$, with $\boldsymbol{\kappa}_l^T$ defined in \cref{subsec: appendix_gen_steane_type_mm_circuit_perfect_aux}. Since the transformation only acts on the $\hat{q}_{m + l}$ quadrature of the $l$th auxiliary, $T_{l, 2}$ takes the form of a $4m \times 2m$ matrix defined such that its only non-zero row is the one containing $\boldsymbol{\kappa}_l^T$ at index $2m + 2l - 1$. This can be realized by alternating between operators \eqref{eq: quad_action_pp_op} and \eqref{eq: quad_action_qp_op} and choosing the squeezing parameters to be the components of $\boldsymbol{\kappa}_l$. More formally, we would have
\begin{align}
    \hat{U}_{T_l} = \hat{C}_{pp}^{m \rightarrow m + l}(\kappa_{l, 2m}) ~\cdots~ \hat{C}_{qp}^{2 \rightarrow m + l}(\kappa_{l, 3})\hat{C}_{pp}^{1 \rightarrow m + l}(\kappa_{l, 2})\hat{C}_{qp}^{1 \rightarrow m + l}(\kappa_{l, 1}).
    \label{eq: general_UTl}
\end{align}

Now, let us describe $T_{l, 1}$ in more detail. \Cref{eq: general_UTl} defines the desired mapping of the storage quadratures into the $l$th auxiliary. It also defines the back-action from the $l$th auxiliary onto the storage. Indeed, using \cref{eq: quad_action_pp_op,eq: quad_action_qp_op}, we can deduce that the storage quadratures are mapped following 
\begin{align}
    &\hat{q}_1 \longrightarrow \hat{q}_1 + \kappa_{l, 2}\hat{p}_{m + l} \nonumber\\
    &\hat{p}_1 \longrightarrow \hat{p}_1 - \kappa_{l, 1}\hat{p}_{m + l} \nonumber\\
    &\hat{q}_2 \longrightarrow \hat{q}_2 + \kappa_{l, 4}\hat{p}_{m + l} \\
    &\hspace{1.75cm}\vdots \nonumber\\
    &\hat{p}_m \longrightarrow \hat{p}_m - \kappa_{l, 2m - 1}\hat{p}_{m + l}\nonumber,
\end{align}
or, equivalently, $T_{l, 1}$ is defined by a $2m \times 4m$ matrix with its only non-zero column being the one containing $\Omega \boldsymbol{\kappa}_l$ at index $2l$.

More visually, we can represent $\hat{U}_{T_l}$ as
\begin{align}
    T_l = \left(\begin{array}{@{}c|c@{}}
        \mathbb{I}_{2m\times 2m}
        &
        \begin{matrix}
            0 & \cdots & 0 & \Omega\boldsymbol{\kappa}_l & \cdots & 0
        \end{matrix}  \\
        \hline
        \begin{matrix}
            0 \\
            \vdots \\
            \boldsymbol{\kappa}_l^T \\ 
            0 \\ 
            \vdots \\
            0
        \end{matrix} &
        \mathbb{I}_{4m\times 4m}
    \end{array}\right),
    \label{eq: general_Tl_mat}
\end{align}
where, again, $\Omega\boldsymbol{\kappa}_l$ is located at column $2l$ of the top right block and $\boldsymbol{\kappa}^T_l$ is located at row $2l -1$ of the bottom left block.

Knowing how we can construct a single $T_l$, we can construct $\bar{T}$ from \cref{eq: T_bar_definition} and subsequently determine the full Steane-type QEC circuit $\hat{U}_{\bar{T}}$.

Here, the operator we construct enables the correct propagation of the error defined by \cref{eq: effect_of_Utl_on_error}. In the remainder of the section, we verify that the operator also respects condition \eqref{eq: Steane_stab_preserve_condition_on_S}. To do so, we can first compute the matrix multiplication
\begin{align}
    ST_l^T = \begin{pmatrix}
        S_\text{G} & 0 \\
        0 & S_\eta
    \end{pmatrix} \begin{pmatrix}
        \mathbb{I} & T_{l, 2}^T \\
        T_{l, 1}^T & \mathbb{I}
    \end{pmatrix} = \begin{pmatrix}
        S_\text{G} & S_\text{G}T_{l, 2}^T \\
        S_\eta T_{l, 1}^T & S_\eta
    \end{pmatrix},
    \label{eq: stab_evolve_after_Tl}
\end{align}
with $S$ defined by \cref{eq: full_system_S_mat}. Let us compute $S_\text{G}T_{l, 2}^T$ and $S_\eta T_{l, 1}^T$. First, direct computation gives
\begin{align}
    S_\text{G}T_{l, 2}^T = \begin{pmatrix}
        \mathbf{s}_1^T \\
        \mathbf{s}_2^T \\
        \vdots \\
        \mathbf{s}_{2m}^T
    \end{pmatrix}
    \begin{pmatrix}
        0 & \cdots & \boldsymbol{\kappa}_l & 0 & \cdots & 0
    \end{pmatrix} = \begin{pmatrix}
        0 & \cdots & \mathbf{s}_1^T\boldsymbol{\kappa}_l & 0 & \cdots & 0 \\
        0 & \cdots & \mathbf{s}_2^T\boldsymbol{\kappa}_l & 0 & \cdots & 0 \\
        \vdots & & \vdots & \vdots & & \vdots \\
        0 & \cdots & \mathbf{s}_{2m}^T\boldsymbol{\kappa}_l & 0 & \cdots & 0
    \end{pmatrix},
\end{align}
and using the form of $\boldsymbol{\kappa}_l$ defined in \cref{subsec: appendix_gen_steane_type_mm_circuit_perfect_aux}, $\mathbf{s}_j^T \boldsymbol{\kappa}_l = -\nu_l \mathbf{s}_j^T\Omega \mathbf{s}_l = -\nu_l \omega(\mathbf{s}_j, \mathbf{s}_l) = -\nu_l A_{j, l}$ so that
\begin{align}
    S_\text{G}T_{l, 2}^T = \begin{pmatrix}
        0 & \cdots & -\boldsymbol{\nu} \circ \mathbf{A}_{\text{G}, l} & 0 & \cdots & 0
    \end{pmatrix},
    \label{eq: Tl2_res_compute}
\end{align}
where we defined $\mathbf{A}_{\text{G}, l}$ as the $l$th column of $A_\text{G}$. Second, using $-\boldsymbol{\kappa}_l^T\Omega = \nu_l \mathbf{s}_l^T$, the product $S_\eta T_{l, 1}^T$ gives
\begin{align}
    S_\eta T_{l, 1}^T = \begin{pmatrix}
        S_{\varnothing,\eta_1} & 0 & \cdots & 0 \\
        0 & S_{\varnothing,\eta_2} &  & 0 \\
        \vdots & & \ddots & \vdots \\
        0 & 0 & \cdots & S_{\varnothing,\eta_{2m}}
    \end{pmatrix}
    \begin{pmatrix}
        0 \\
        \vdots \\
        0 \\ 
        \nu_l \mathbf{s}_l^T \\ 
        \vdots \\ 
        0
    \end{pmatrix}.
    \label{eq: aux_mult_tl1}
\end{align}
Here, because $\nu_l \mathbf{s}_l^T$ is the only non-zero row located at the $2m + 2l$ index, only $S_{\varnothing,\eta_l}$ impacts the matrix multiplication. Therefore,
\begin{align}
    \begin{pmatrix}
        \eta_l & 0 \\
        0 & \frac{1}{\eta_l}
    \end{pmatrix} \begin{pmatrix}
        0 \\
        \nu_l \mathbf{s}_l^T
    \end{pmatrix} = \begin{pmatrix}
        0 \\
        \frac{\nu_l}{\eta_l}\mathbf{s}_l^T
    \end{pmatrix},
\end{align}
and, from this, \cref{eq: aux_mult_tl1} becomes
\begin{align}
    S_\eta T_{l, 1}^T = \begin{pmatrix}
        0 \\
        \vdots \\
        0 \\ 
        \frac{\nu_l}{\eta_l}\mathbf{s}_l^T \\ 
        \vdots \\ 
        0
    \end{pmatrix}.
    \label{eq: Tl1_res_compute}
\end{align}
Finally, using \cref{eq: Tl2_res_compute,eq: Tl1_res_compute} and $\eta_j = \nu_j = \ell$, $\forall j$ as in the main text, \cref{eq: stab_evolve_after_Tl} gives
\begin{align}
    ST_l^T = \left(\begin{array}{@{}c|c@{}}
        S_\text{G}
        &
        \begin{matrix}
            0 & \cdots & -\ell\mathbf{A}_{\text{G}, l} & 0 & \cdots & 0
        \end{matrix}  \\
        \hline
        \begin{matrix}
            0 \\
            \vdots \\
            0 \\ 
            \mathbf{s}_l^T \\ 
            \vdots \\
            0
        \end{matrix} &
        S_\eta
    \end{array}\right).
\end{align}
Here, because $S_\text{G}$ describes a valid multimode GKP code, $A_\text{G}$ is integral and $\mathbf{A}_{\text{G}, l} \in \mathbb{Z}^{2m}$. This means that the vector $-\ell\mathbf{A}_{\text{G}, l}$ only contains integer multiples of auxiliary stabilizers. From this, we see that the extra terms on the off-diagonal blocks of $ST_l^T$ are simply either storage stabilizers or auxiliary stabilizers. Therefore, we conclude that $ST_l^T = RS$.

\Cref{eq: general_UTl} then directly gives instructions on how to construct each $\hat{U}_{T_l}$ operator, respecting all the necessary conditions to define a valid Steane-type QEC circuit.

\subsection{Analysis of different Steane-type QEC circuits}
\label{subsec: appendix_gen_steane_type_mm_circuit_analysis}

In this appendix, we discuss in more detail the impact of the different ways we can measure the GKP code's stabilizers. More specifically, we are interested in analyzing the effect of these measurements in the presence of noisy auxiliaries.

As mentioned in the beginning of \cref{sec: qec_multimode_gkp}, the best choice to make in order to realize the modulo measurements of the stabilizers is the one described by \cref{eq: unity_stab_quad_measurement}, as opposed to \cref{eq: stab_quad_measurement}. The particularity of $\hat{h}_j'$ is that it is composed of operators that have a unit norm in terms of the quadrature operators that describe them. This becomes important when dealing with noisy auxiliaries, because it helps reduce the amount of squeezing induced in the system. 

To understand this better, let's recall how we construct the Steane-type QEC circuit from the discussion in \cref{subsec: appendix_gen_steane_type_mm_circuit_perfect_aux}. Measuring stabilizers is always done by propagating information from the storage to the auxiliary states using the operators defined by \cref{eq: quad_action_qq_op,eq: quad_action_pp_op,eq: quad_action_qp_op} and choosing the right squeezing parameter matrix $K$.

\Cref{eq: effect_of_Utl_on_error} tells us that we must have each $T_j$ propagating $\boldsymbol{\kappa}_j^T \boldsymbol{\xi}_\text{G}$ onto the $j$th $\hat{q}$ quadrature of the auxiliaries. Therefore, the squeezing parameters are given by $\boldsymbol{\kappa}_j = -\nu_j\Omega\mathbf{s}_j$. This way, the amount of squeezing necessary for a single stabilizer is $|\boldsymbol{\kappa}_j| = \nu_j|\Omega\mathbf{s}_j| = \frac{1}{\ell}\nu_j|\mathbf{g}_j|$, where we used $|\mathbf{g}_j| = |\ell \Omega \mathbf{s}_j|$. 

Now, let us look at how this amount of squeezing varies depending on the choice of stabilizers we measure. We justify this by analyzing the noiseless auxiliary case, but it is completely equivalent for the noisy auxiliaries case. Using the state after the Steane QEC protocol \eqref{eq: state_after_steane_circ_noiseless_aux_appendix}, the syndrome we extract from the measurement of the $j$th auxiliary is
\begin{align}
    z_j = \ell \nu_j \mathbf{s}^T_j\Omega\boldsymbol{\xi}_\text{G} + \ell a_j\eta_j = \ell\nu_j \mathbf{s}^T_j\Omega\boldsymbol{\xi}_\text{G} ~\text{mod}~ \ell \eta_j.
\end{align}
If we are to measure $\hat{h}_j$, stabilizers of the form \cref{eq: stab_quad_measurement}, we have $\nu_j = \eta_j = \ell$, so that $|\boldsymbol{\kappa}_j| = |\mathbf{g}_j|$. On the other hand, if we are to measure $\hat{h}_j'$, stabilizers of the form \cref{eq: unity_stab_quad_measurement}, we have $\nu_j = \eta_j = \frac{\ell}{|\mathbf{g}_j|}$, so that $|\boldsymbol{\kappa}_j| = 1$. Clearly, we inject less squeezing into the system by measuring stabilizers of the form \cref{eq: unity_stab_quad_measurement}. In fact, components of the squeezing vector $\boldsymbol{\kappa}_j$ appear in $T_j$ and thus linearly scale errors on the auxiliary into our measurements. As $|\boldsymbol{\kappa}_j|$ increases, it is more probable that a correctable error is sent to an uncorrectable one, reducing the performance of the decoder. This is also a point that was made in Ref. \cite{joe_logical_compiler}: squeezing operations on multimode GKP states increase the mean number of photons in the system, therefore increasing the energy and length of errors.

\begin{figure*}[t!]
    \centering
    \includegraphics[scale=1]{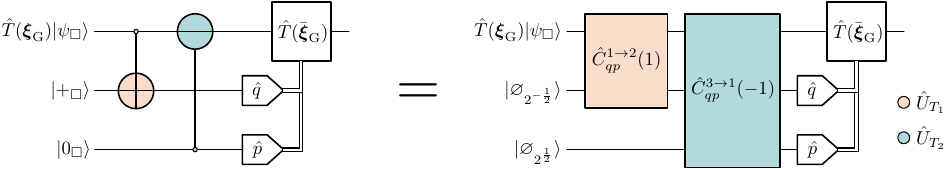}
    \caption{Steane-type QEC circuit for the square GKP code proposed in Ref. \cite{originalGKP} and measuring $\hat{\mathbf{h}}'_\square$. We have highlighted in color the different $\hat{U}_{T_j}$ operators in order to visualize them better. On the left, we see the original circuit adapted with the notation from Ref. \cite{nohLowOverheadFaultTolerantGKP} and we show the corresponding circuit using our notation for the two-mode squeezing operator \eqref{eq: quad_action_qp_op} on the right. Both are completely equivalent, as emphasized by the equal sign. Notice also that we have $\ket{+_\square} = \ket{\varnothing_{2^{-\frac{1}{2}}}}$ and $\ket{0_\square} = \ket{\varnothing_{2^{\frac{1}{2}}}}$ \cite{bapGKPMultimode}.}
    \label{fig: square_GKP_steane_type_QEC_original}
\end{figure*}

To describe the situation more quantitatively, we can compare the covariance matrix of the noise model after both circuits and determine if the circuit increased the variance of the initial noise model in each measured quadrature. Suppose the circuit measuring $\hat{\mathbf{h}}$ is parametrized by $\bar{T}$ and the circuit measuring $\hat{\mathbf{h}}'$ by $\bar{T}'$. After each circuit, the covariance matrix of the noise model changes following \cref{eq: noise_model_changing_with_symp_op}
\begin{align}
    \bar{\Sigma} = \Pi_\text{m}\bar{T}\Gamma_0\bar{T}^T\Pi_\text{m}^T ~~,~~ \bar{\Sigma}' = \Pi_\text{m}\bar{T}'\Gamma_0(\bar{T}')^T\Pi_\text{m}^T,
\end{align}
where we only care about how the squeezing affects the measurement, which dictates our correction, justifying the projector $\Pi_\text{m}$. The perfect scenario would be that there is no squeezing in our circuit so that it preserves $\Sigma$. Therefore, we compare the effect of $\bar{T}$ and $\bar{T}'$ with this scenario such that we compute the distance between these two covariance matrices as
\begin{align}
    d_{\bar{T}} = d_\text{Cov}(\bar{\Sigma}, \Pi_\text{m}\Gamma_0\Pi_\text{m}^T) ~~,~~ d_{\bar{T}'} = d_\text{Cov}(\bar{\Sigma}', \Pi_\text{m}\Gamma_0\Pi_\text{m}^T),
    \label{eq: Tbar_and_T_covariance_matrix_distance}
\end{align}
where $d_\text{Cov}(A, B)$ is defined as \cite{distBetweenCovMat}
\begin{align}
    d_\text{Cov}(A, B) = \sqrt{\sum_{k = 1}^{\mathrm{dim}(A)} \ln^2 (\mu_k(A, B))},
\end{align}
and $\mu_k(A, B)$ are the solutions to the generalized eigenvalue problem $A\mathbf{v} = \mu B \mathbf{v}$. Finally, if $d_{\bar{T}}/d_{\bar{T}'} > 1$, we deduce that the Steane-type QEC circuit represented by $\bar{T}'$ induces less squeezing in the system and is a more suitable choice. It is important to mention that this measure does not give us a quantity describing the exact change of variance from one circuit to another; it simply states that the circuit increases or decreases it. 

Now, we show a few examples of why measuring $\hat{\mathbf{h}}'$ defined by \cref{eq: unity_stab_quad_measurement} is more suitable for QEC.

\subsubsection{Application of the circuit for the single mode square GKP code}
\label{subsec: appendix_gen_steane_type_mm_circuit_square_gkp}

\begin{wrapfigure}{r!}{0.5\textwidth}
    \centering
    \vspace{-0.15cm}
    \includegraphics[scale=1]{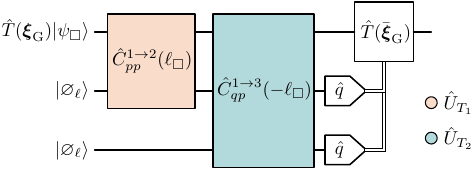}
    \caption{Steane-type QEC circuit for the square GKP code proposed measuring $\hat{\mathbf{h}}_\square$, where $\ell_\square = 2\sqrt{\pi}$. The main differences between this circuit and the one from \cref{fig: square_GKP_steane_type_QEC_original} are in the squeezing of the auxiliaries, the squeezing of the $\hat{U}_{T_j}$ operators, and the measurements, which are all made on the $\hat{q}$ quadratures.}
    \label{fig: bad_square_GKP_steane_type_QEC}
\end{wrapfigure}

In this section, we analyze the square GKP code. We compare the circuit that we obtain measuring stabilizers $\hat{\mathbf{h}}_\square$, defined by \cref{eq: stab_quad_measurement}, with the one we obtain measuring $\hat{\mathbf{h}}_\square'$, defined by \cref{eq: unity_stab_quad_measurement}. The circuit we are using to measure $\hat{\mathbf{h}}_\square$ is shown in \cref{fig: bad_square_GKP_steane_type_QEC}, and the circuit we are using for measuring $\hat{\mathbf{h}}_\square'$ is 
the same one that was originally proposed for the square GKP in Ref. \cite{originalGKP} (see \cref{fig: square_GKP_steane_type_QEC_original}). Note that we are not using the convention of always measuring the auxiliaries in $\hat{q}$ for the circuit measuring $\hat{\mathbf{h}}_\square'$ in order to follow the literature. We start by showing that both circuits are equivalent when we consider no errors on the auxiliary states.

In both cases the circuits are used to measure the stabilizers of the square GKP code. Using \cref{eq: stab_GKP} and \cref{eq: square_GKP}, the vector of operator $\hat{\mathbf{g}}$ for the square GKP code can be directly computed, giving
\begin{align}
    \hat{\mathbf{g}}_\square = 2\sqrt{\pi}\begin{pmatrix}
        \hat{p}_1 \\
        -\hat{q}_1
    \end{pmatrix}.
\end{align}
With the form of $\hat{\mathbf{g}}_\square$, we can compute the quantities of interest
\begin{align}
    \hat{\mathbf{h}}_\square = 2\sqrt{\pi}\begin{pmatrix}
        \hat{p}_1 \\
        -\hat{q}_1
    \end{pmatrix}~\text{mod}~ 2\pi,
    \label{eq: stab_quad_measurement_square_GKP}
\end{align}
and
\begin{align}
    \hat{\mathbf{h}}'_\square = \begin{pmatrix}
        \hat{p}_1\\
        -\hat{q}_1
    \end{pmatrix}~\text{mod}~ \sqrt{\pi} .
    \label{eq: unity_stab_quad_measurement_square_GKP}
\end{align}
Note that the modulo operation is applied to the vector element-wise. 

Let us compare both circuits and look at the syndrome they yield, starting with the measure of the stabilizers in \cref{eq: stab_quad_measurement_square_GKP}. For this circuit, we follow the discussion in \cref{subsec: appendix_gen_steane_type_mm_circuit_construct}. 

Because we measure stabilizers of the form \cref{eq: stab_quad_measurement}, we have $\nu_j = \eta_j = \ell$, $\forall j$. Then, the stabilizer matrix describing the system is given by  
\begin{align}
    S = \begin{pmatrix}
        S_\square & 0 & 0 \\
        0 & S_{\varnothing, \ell} & 0 \\
        0 & 0 & S_{\varnothing, \ell}
    \end{pmatrix},
\end{align}
and 
\begin{align}
    K_\square = \begin{pmatrix}
        0 & 2\sqrt{\pi} \\
        -2\sqrt{\pi} & 0
    \end{pmatrix},
\end{align}
implying $\boldsymbol{\kappa}^T_1 = \begin{array}{cc}(0 & \ell_\square) \end{array}$ and $\boldsymbol{\kappa}^T_2 = \begin{array}{cc}(-\ell_\square & 0), \end{array}$ with $\ell_\square = 2\sqrt{\pi}$. Using \cref{eq: general_UTl}, the elements in the circuit can be expressed as
\begin{align}
    &\hat{U}_{T_1} = \hat{C}_{pp}^{1 \rightarrow 2}(\kappa_{1, 2})\hat{C}_{qp}^{1 \rightarrow 2}(\kappa_{1, 1}) = \hat{C}_{pp}^{1 \rightarrow 2}(\ell_\square) \\
    &\hat{U}_{T_2} = \hat{C}_{pp}^{1 \rightarrow 3}(\kappa_{2, 2})\hat{C}_{qp}^{1 \rightarrow 3}(\kappa_{2, 1}) = \hat{C}_{qp}^{1 \rightarrow 3}(-\ell_\square),
\end{align}
where $m = 1$. This corresponds exactly to \cref{fig: bad_square_GKP_steane_type_QEC}. Then, using \cref{eq: quad_action_pp_op,eq: quad_action_qp_op}, we can write the symplectic form of the operators $\hat{U}_{T_1}$ and $\hat{U}_{T_2}$ defining the circuit such that
\begin{align}
    \bar{T} = T_2T_1 = \begin{pmatrix}
        1 & 0 & 0 & \ell_\square & 0 & 0 \\
        0 & 1 & 0 & 0 & 0 & \ell_\square \\
        0 & \ell_\square & 1 & 0 & 0 & 0 \\
        0 & 0 & 0 & 1 & 0 & 0 \\
        -\ell_\square & 0 & 0 & -\ell_\square^2 & 1 & 0 \\
        0 & 0 & 0 & 0 & 0 & 1
    \end{pmatrix},
\end{align}
respecting $S\bar{T}^T = RS$ as explained in \cref{subsec: appendix_gen_steane_type_mm_circuit_construct}. In order to extract the syndrome, all there is left to do is compute $\bar{T} \boldsymbol{\xi}_\text{G}$ with \cref{eq: effect_of_Utl_on_error}, that is 
\begin{align}
    \bar{T} \boldsymbol{\xi}_\text{G} = \begin{pmatrix}
        \boldsymbol{\xi}_\text{G} \\
        2\sqrt{\pi}\xi_{\text{G}, 2} \\
        0 \\
        -2\sqrt{\pi}\xi_{\text{G}, 1}\\
        0
    \end{pmatrix} \implies \Pi_\text{m}\bar{T} \boldsymbol{\xi}_\text{G} = \ell S_\square \Omega \boldsymbol{\xi}_\text{G},
    \label{eq: Tbar_noise_propagate_noiseless_aux}
\end{align}
where $\Pi_\text{m}$ projects on both $\hat{q}$ quadratures of the auxiliaries because of the measurement, and we have defined the $j$th component of $\boldsymbol{\xi}_\text{G}$ as $\xi_{\text{G}, j}$. Finally, the chosen value of $\boldsymbol{\eta}$ for the auxiliaries enables a syndrome of $\mathbf{z} = \ell^2 S_\square \Omega \boldsymbol{\xi}_\text{G} ~\text{mod}~ 2\pi$, as expected.

Now, let's do the same thing for the measure of the stabilizers in \cref{eq: unity_stab_quad_measurement_square_GKP}. Because the circuit does not respect our convention of always measuring $\hat{q}$, we simply construct $\bar{T}'$ by inspecting \cref{fig: square_GKP_steane_type_QEC_original}. Here, the stabilizer matrix of the system is given by
\begin{align}
    S' = \begin{pmatrix}
        S_\square & 0 & 0 \\
        0 & S_{\varnothing, 1/\sqrt{2}} & 0 \\
        0 & 0 & S_{\varnothing, \sqrt{2}}
    \end{pmatrix}.
\end{align}
We notice that the squeezing in both auxiliaries is not the same for this case. This is because we are not measuring them in the same quadratures, and both these quadratures have different scaling for the $\eta_j$ parameter. In fact, the wanted modulo value is $\sqrt{\pi}$, as we see from \cref{eq: unity_stab_quad_measurement_square_GKP}. Therefore, with $\eta_1 = \frac{1}{\sqrt{2}}$ and $\eta_2 = \sqrt{2}$, the auxiliaries are
\begin{align}
    \ket{\varnothing_{\eta_1 = 1/\sqrt{2}}} = \sum_{j = -\infty}^{\infty} \ket{j\sqrt{\pi}}_{\hat{q}_2} ~~,~~ \ket{\varnothing_{\eta_2 = \sqrt{2}}} = \sum_{j = -\infty}^{\infty} \ket{j\sqrt{\pi}}_{\hat{p}_3},
\end{align} 
giving both auxiliaries a modulo $\sqrt{\pi}$ symmetry, as needed. Then, the symplectic representation of the circuit is given by
\begin{align}
    \bar{T}' = \begin{pmatrix}
        1 & 0 & 0 & 0 & -1 & 0 \\
        0 & 1 & 0 & -1 & 0 & 0 \\
        1 & 0 & 1 & 0 & 0 & 0 \\
        0 & 0 & 0 & 1 & 0 & 0 \\
        0 & 0 & 0 & 0 & 1 & 0 \\
        0 & 1 & 0 & -1 & 0 & 1
    \end{pmatrix},
\end{align}
verifying $S'(\bar{T}')^T = R'S'$. With this circuit, $\bar{T}' \boldsymbol{\xi}_\text{G}$ now gives
\begin{align}
    \bar{T}' \boldsymbol{\xi}_\text{G} = \begin{pmatrix}
        \boldsymbol{\xi}_\text{G} \\
        \xi_{\text{G}, 1} \\
        0 \\
        0 \\
        \xi_{\text{G}, 2}
    \end{pmatrix} \implies \Pi_\text{m}'\bar{T}' \boldsymbol{\xi}_\text{G} = \boldsymbol{\xi}_\text{G}.
    \label{eq: Tbar_prime_noise_propagate_noiseless_aux}
\end{align}
As we can see, the circuit directly gives $\boldsymbol{\xi}_\text{G}$. Then, the chosen value of $\boldsymbol{\eta}$ for the auxiliaries enables a final syndrome of $\mathbf{z}' = \ell\boldsymbol{\xi}_\text{G} ~\text{mod}~ \sqrt{\pi}$. Note here that we actually have $\boldsymbol{\xi}_\text{G}$ and not $\Omega \boldsymbol{\xi}_\text{G}$ as expected from \cref{eq: unity_stab_quad_measurement_square_GKP}. This is not a problem, since we can keep track of everything in software. Thus, constructing the circuit in order for it to return directly $\Omega \boldsymbol{\xi}_\text{G}$ is not necessary, as long as we know what the circuit returns.

\begin{figure}[t!]
    \centering
    \begin{subfigure}[t]{0.5\textwidth}
        \centering
        \caption{}
        \vspace{-0.25cm}
        \includegraphics[height=2.5in, width=\textwidth]{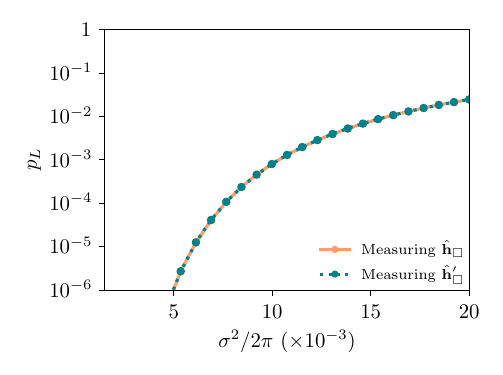}
        \label{subfig: comparison_both_square_QEC_circuit_no_noise_aux}
    \end{subfigure}%
    ~ 
    \begin{subfigure}[t]{0.5\textwidth}
        \centering
        \caption{}
        \vspace{-0.25cm}
        \includegraphics[height=2.5in, width=\textwidth]{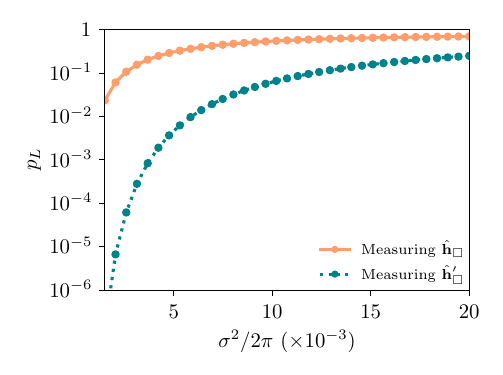}
        \label{subfig: comparison_both_square_QEC_circuit_noisy_aux}
    \end{subfigure}
    \vspace{-0.75cm}
    \caption{Comparison of decoding performances for Steane-type QEC circuits measuring $\hat{\mathbf{h}}_\square$ (orange full lines) and $\hat{\mathbf{h}}_\square'$ (teal dotted lines) for the square GKP \cref{subsubsec: square_GKP_info}. We plot the probability of having a logical error $p_L$ as a function of the noise variance $\sigma^2/2\pi$ in the system over a range of $\sigma^2/2\pi = [1.5, 20] \times 10^{-3}$ for (a) noiseless auxiliaries and (b) noisy auxiliaries. For these simulations, the MED decoder \cref{eq: med_cvp_decoder} is used, and each point is obtained by a Monte-Carlo method through the sampling of $N = 10^9$ errors $\boldsymbol{\xi}$. The standard deviation $\bar{\sigma}$ of each point is smaller than the size of the markers. For more details on the simulations, we refer the reader to \cref{sec: res}.}
    \label{fig: comparison_both_square_QEC_circuit}
    \vspace{-0.15cm}
\end{figure}

In any case, we can always isolate $\boldsymbol{\xi}_\text{G}$ and apply the decoder of our choice to perform error correction afterwards. When there is no noise on the auxiliaries, both circuits are equivalent. In \creffullsubfigref{fig: comparison_both_square_QEC_circuit}{subfig: comparison_both_square_QEC_circuit_no_noise_aux}, we show the comparison between the two circuits. We see that the curves describing the probability of having a logical error when decoding $p_L$ in function of the noise variance $\sigma^2 / 2\pi$ are exactly the same, showing that they yield the same performance.

Now, let us look at the situation where there is noise on the auxiliary states. As was argued in \cref{subsec: qec_noisy_aux} and \cref{subsec: appendix_gen_steane_type_mm_circuit_noisy_aux}, \cref{eq: Tbar_noise_propagate_noiseless_aux,eq: Tbar_prime_noise_propagate_noiseless_aux} are not valid anymore. More precisely, noise spreads to
\begin{figure}[t!]
    \centering
    \vspace{0.35cm}
    \begin{subfigure}[t]{\textwidth}
        \centering
        \caption{}
        \includegraphics[scale=1]{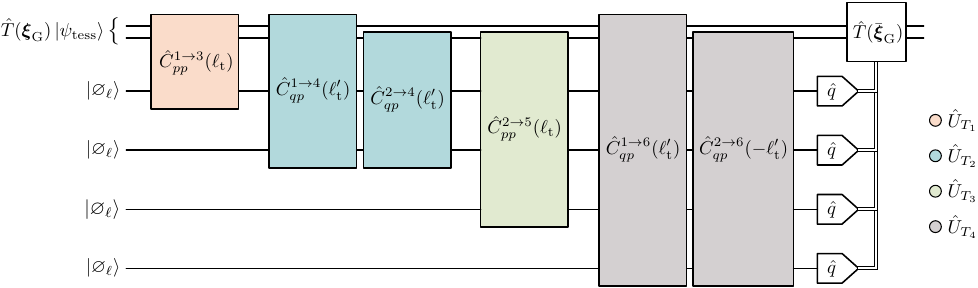}
        \label{subfig: tess_GKP_steane_type_QEC_circuits_not_optimal}
    \end{subfigure}
    \\
    \hspace{-0.75cm}
    \begin{subfigure}[t]{\textwidth}
        \centering
        \caption{}
        \includegraphics[scale=1]{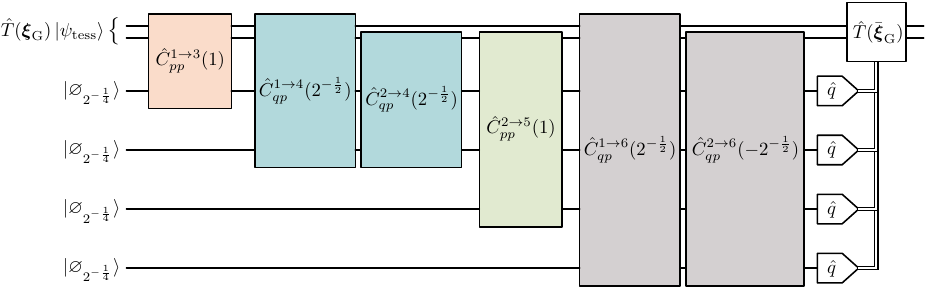}
        \label{subfig: tess_GKP_steane_type_QEC_circuits_nice}
    \end{subfigure}
    \vspace{0.5cm}
    \caption{Example of both Steane-type QEC circuits we study for the tesseract GKP code. (a) For the first circuit, we measure $\hat{\mathbf{h}}_\text{tess}$, where $\ell_\text{t} = \sqrt{2}\ell_\text{t}' = 2^{\frac{3}{4}}\sqrt{\pi}$. (b) For the second circuit, we measure $\hat{\mathbf{h}}'_\text{tess}$. This is the circuit we use to obtain the main result (see \creffullsubfigref{fig: comparing_MED_vs_CORMED_main_fig_1e9}{subfig: comparing_MED_vs_CORMED_main_fig_1e9_tess}). Looking at both circuits, we see that the differences come from the squeezing of the auxiliary states and the squeezing comprised in the $\hat{U}_{T_j}$ operators.}
    \label{fig: tess_GKP_steane_type_QEC_circuits}
\end{figure}
\begin{figure}[t!]
    \centering
    \begin{subfigure}[t]{0.5\textwidth}
        \centering
        \caption{}
        \vspace{-0.25cm}
        \includegraphics[height=2.5in, width=\textwidth]{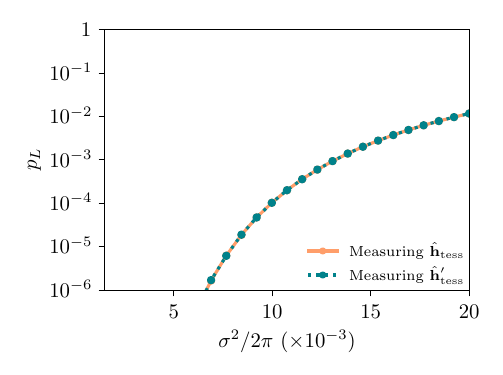}
        \label{subfig: comparison_both_tess_QEC_circuit_no_noise_aux}
    \end{subfigure}%
    ~ 
    \begin{subfigure}[t]{0.5\textwidth}
        \centering
        \caption{}
        \vspace{-0.25cm}
        \includegraphics[height=2.5in, width=\textwidth]{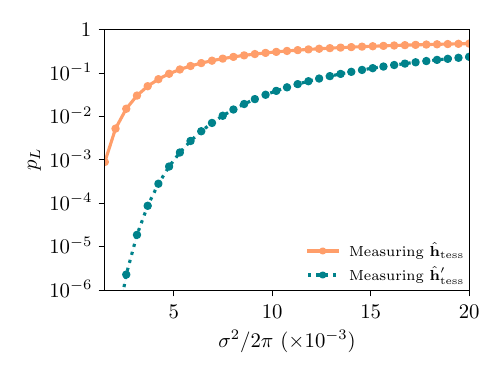}
        \label{subfig: comparison_both_tess_QEC_circuit_noisy_aux}
    \end{subfigure}
    \vspace{-0.75cm}
    \caption{Comparison of decoding performances for Steane-type QEC circuits measuring $\hat{\mathbf{h}}_\text{tess}$ (dots) and $\hat{\mathbf{h}}_\text{tess}'$ (full lines) for the tesseract GKP \cref{subsubsec: tess_GKP_info}. We plot the probability of having a logical error $p_L$ as a function of the noise variance $\sigma^2/2\pi$ in the system over a range of $\sigma^2/2\pi = [1.5, 20] \times 10^{-3}$ for (a) noiseless auxiliaries and (b) noisy auxiliaries. For these simulations, the MED decoder \cref{eq: med_cvp_decoder} is used, and each point is obtained by a Monte-Carlo method through the sampling of $N = 10^9$ errors $\boldsymbol{\xi}$. The standard deviation $\bar{\sigma}$ of each point is smaller than the size of the markers. For more details on the simulations, we refer the reader to \cref{sec: res}.}
    \label{fig: comparison_both_tess_QEC_circuit}
\end{figure}
all the quadratures and corrupts the measurements we are making in order to realize QEC. In this case, we can estimate which circuit, $\bar{T}$ or $\bar{T}'$, preserves the initial variance in a better way. Using \cref{eq: Tbar_and_T_covariance_matrix_distance}, we can show that $d_{\bar{T}} / d_{\bar{T}'} \simeq 2$, stating that the circuit measuring $\hat{\mathbf{h}}_\square'$ (see \cref{fig: square_GKP_steane_type_QEC_original}) is a better circuit choice to make in order to realize Steane-type QEC for the square GKP. To further support our claims, we compare simulations of both circuits under covariance matrix $\Gamma_0$ and we see that the circuit measuring $\hat{\mathbf{h}}_\square'$ has a much lower probability of logical errors no matter the noise strength in the system. The results are shown on \creffullsubfigref{fig: comparison_both_square_QEC_circuit}{subfig: comparison_both_square_QEC_circuit_noisy_aux}.  

\subsubsection{Application of the circuit for the two mode tesseract GKP code}
\label{subsec: appendix_gen_steane_type_mm_circuit_tess_gkp}

In this section we proceed to a similar analysis to what we did for the square GKP code in \cref{subsec: appendix_gen_steane_type_mm_circuit_square_gkp}. Now, we are interested in a two-mode GKP code, the tesseract GKP.

For this multimode GKP code, direct use of \cref{eq: stab_GKP} and \cref{eq: tesseract_GKP} gives
\begin{align}
    \hat{\mathbf{g}}_\text{tess} = 2^{\frac{3}{4}} \sqrt{\pi}\begin{pmatrix}
        \hat{p}_1 \\
        -(\hat{q}_1 + \hat{q}_2)/\sqrt{2} \\
        \hat{p}_2 \\
        -(\hat{q}_1 - \hat{q}_2)/\sqrt{2}
    \end{pmatrix}.
\end{align}
This way, we compute $\hat{\mathbf{h}}_\text{tess}$ using \cref{eq: stab_quad_measurement}
\begin{align}
    \hat{\mathbf{h}}_\text{tess} = 2^{\frac{3}{4}} \sqrt{\pi}\begin{pmatrix}
        \hat{p}_1 \\
        -(\hat{q}_1 + \hat{q}_2)/\sqrt{2} \\
        \hat{p}_2 \\
        -(\hat{q}_1 - \hat{q}_2)/\sqrt{2} 
    \end{pmatrix}~\text{mod}~ 2\pi,
\end{align}
and $\hat{\mathbf{h}}_\text{tess}'$ using \cref{eq: unity_stab_quad_measurement}
\begin{align}
    \hat{\mathbf{h}}_\text{tess}' = \begin{pmatrix}
        \hat{p}_1 \\
        -(\hat{q}_1 + \hat{q}_2)/\sqrt{2} \\
        \hat{p}_2 \\
        -(\hat{q}_1 - \hat{q}_2)/\sqrt{2} 
    \end{pmatrix} ~\text{mod}~ 2^{\frac{1}{4}}\sqrt{\pi}.
\end{align}
This gives all the quantities we need to construct the corresponding Steane-type QEC circuits. We do so by following the procedure explained in \cref{subsec: appendix_gen_steane_type_mm_circuit_construct} and demonstrated for the square GKP in \cref{subsec: appendix_gen_steane_type_mm_circuit_square_gkp}. Both circuits resulting from this approach are shown on \cref{fig: tess_GKP_steane_type_QEC_circuits}. We note that for the measure of $\hat{\mathbf{h}}_\text{tess}'$, we have $\nu_j = \eta_j = \frac{\ell}{|\mathbf{g}_j|} = 2^{-\frac{1}{4}}$, $\forall j$.

For the case of noiseless auxiliaries, since we followed the same construction as in \cref{subsec: appendix_gen_steane_type_mm_circuit_square_gkp}, both circuits have to be equivalent. This is shown by simulations on \creffullsubfigref{fig: comparison_both_tess_QEC_circuit}{subfig: comparison_both_tess_QEC_circuit_no_noise_aux}, where both curves yield the same logical error probability no matter the noise strength.

When noise is present on the auxiliaries, we estimate which circuit of the two preserves the initial variance in a better way following the same technique as in \cref{subsec: appendix_gen_steane_type_mm_circuit_square_gkp}. Using \cref{eq: Tbar_and_T_covariance_matrix_distance}, we show that $d_{\bar{T}} / d_{\bar{T}'} \simeq 1.2$. Again, this shows that the circuit of \creffullsubfigref{fig: tess_GKP_steane_type_QEC_circuits}{subfig: tess_GKP_steane_type_QEC_circuits_nice}, measuring $\hat{\mathbf{h}}_\text{tess}'$, is a better choice to make in order to realize Steane-type QEC for the tesseract GKP. We also show simulations that compare both circuits under variance $\sigma^2/2\pi$, and we confirm our predictions. Indeed, measuring $\hat{\mathbf{h}}_\text{tess}'$ yields a lower probability of logical errors no matter $\sigma^2/2\pi$. The results are shown on \creffullsubfigref{fig: comparison_both_tess_QEC_circuit}{subfig: comparison_both_tess_QEC_circuit_noisy_aux}.

\section{Noise-correlated MLD decoder}
\label{sec: appendix_solving_main_equation}

\subsection{Taking the noiseless auxiliaries limit}
\label{subsec: appendix_noiseless_aux_case_from_noisy}

In this appendix, we show how we can retrieve the MLD decoder \cref{eq: general_mld_decoder} from our noise-correlated MLD decoder \cref{eq: general_noise_corr_mld_decoder} in the limit where there is no noise on the auxiliaries.

In the case where auxiliaries are noiseless, we have $\boldsymbol{\delta}_\text{G} = \boldsymbol{\delta}_{\text{m}} = 0$, from \cref{eq: noise_after_steane_on_GKP_quad,eq: noise_after_steane_on_aux_quad}. This way, $\boldsymbol{\xi}_{\bar{T}, \text{G}} = \Pi_\text{G}\bar{T}\boldsymbol{\xi} = \boldsymbol{\xi}_{\text{G}}$ and $\boldsymbol{\xi}_{\bar{T}, \text{m}} = \Pi_{\text{m}}\bar{T}\boldsymbol{\xi} = \ell S_\text{G} \Omega \boldsymbol{\xi}_\text{G}$. Therefore, the measured syndrome $\mathbf{z}_{\bar{T}, \text{m}}$ described by \cref{eq: syndrome_our_decoder} becomes
\begin{align}
    \mathbf{z}_{\bar{T}, \text{m}} = \ell^2 S_\text{G} \Omega \boldsymbol{\xi}_\text{G} + \ell \mathbf{a} \circ \boldsymbol{\eta} = \mathbf{z} \implies \tilde{\mathbf{z}}_{\bar{T}, \text{m}} = \tilde{\mathbf{z}},
    \label{eq: z_tm_perf_aux_lim}
\end{align}
and knowing we have $\boldsymbol{\xi}_{\text{G}} = \boldsymbol{\chi}(\mathbf{z}) + \boldsymbol{\lambda}^\perp_\text{G}$ as explained in \cref{subsec: qec_perf_aux}, then 
\begin{align}
    \boldsymbol{\xi}_{\bar{T}, \text{G}} = \boldsymbol{\xi}_{\text{G}} = \boldsymbol{\chi}(\mathbf{z}) + \boldsymbol{\lambda}^\perp_\text{G}.
    \label{eq: xi_tg_perf_aux_lim}
\end{align}
We see that these two results coincide directly with the noiseless case. Thus, the combination of \cref{eq: z_tm_perf_aux_lim,eq: xi_tg_perf_aux_lim} with the decoder \cref{eq: general_noise_corr_mld_decoder} yields
\begin{align}
    \bar{\boldsymbol{\xi}}_{\text{G}} &= \underset{\boldsymbol{\xi}_{\text{G}} \in \mathcal{V}_\text{G}}{\text{arg max }} P_{\mathbf{0}, \Gamma}([\boldsymbol{\xi}_{\text{G}}] | \tilde{\mathbf{z}}).
\end{align}
Maximizing over all $\boldsymbol{\xi}_{\text{G}}$ in the Voronoi cell of $\Lambda_\text{G}$ means maximizing over all logical vectors $\boldsymbol{\lambda_\text{G}^\perp}$ in $\mathcal{V}_\text{G}$ when there is no noise on the auxiliaries. Indeed, in that situation, the noise does not corrupt the measurement and the storage so that it is guaranteed the correction sends the storage back into the code space $\mathcal{C}_\text{G}$. The only optimization left is to find the most likely logical vector in $\mathcal{V}_\text{G} = \Lambda_\text{G}^\perp / \Lambda_\text{G}$. Therefore, the decoder is given by
\begin{align}
    \bar{\boldsymbol{\lambda}}^\perp_\text{G} &= \underset{\boldsymbol{\lambda}^\perp_\text{G} \in \Lambda_\text{G}^\perp / \Lambda_\text{G}}{\text{arg max }} P_{\mathbf{0}, \Gamma}([\boldsymbol{\chi}(\mathbf{z}) + \boldsymbol{\lambda}^\perp_\text{G}] | \tilde{\mathbf{z}}).
\end{align}

All there is left to do is to show that $\Gamma = \Sigma_0$ when auxiliaries are noiseless. We start by finding a general expression for $\bar{T}$. We know from \cref{eq: general_Tl_block_matrix_form} of \cref{subsec: appendix_gen_steane_type_mm_circuit_construct} the general form that takes the different $T_j$ matrices composing $\bar{T}$. Starting by the computation of the first two, it gives
\begin{align}
    T_2T_1 &= \begin{pmatrix}
        \mathbb{I} & T_{2, 1} \\
        T_{2, 2} & \mathbb{I}
    \end{pmatrix}
    \begin{pmatrix}
        \mathbb{I} & T_{1, 1} \\
        T_{1, 2} & \mathbb{I}
    \end{pmatrix} \nonumber \\
    &= 
    \begin{pmatrix}
        \mathbb{I} + T_{2, 1}T_{1, 2} & T_{1, 1} + T_{2, 1} \\
        T_{1, 2} + T_{2, 2} & \mathbb{I} + T_{2, 2}T_{1, 1}
    \end{pmatrix},
\end{align}
where we omit the subscript describing the dimensions of the identity matrices for conciseness. We can simplify this expression further. Using the form of the different blocks in \cref{eq: general_Tl_mat}, direct calculation of $T_{2, 1}T_{1, 2}$ yields
\begin{align}
    T_{2, 1}T_{1, 2} = \begin{pmatrix}
            0 & 0 & 0 & \Omega\boldsymbol{\kappa}_2 & \cdots & 0
        \end{pmatrix}
        \begin{pmatrix}
            \boldsymbol{\kappa}_1^T \\ 
            0 \\ 
            0 \\ 
            0 \\ 
            \cdots \\ 
            0
        \end{pmatrix} = \boldsymbol{0}_{2m \times 2m},
\end{align}
with $\boldsymbol{0}_{2m \times 2m}$ being a zero matrix of size $2m \times 2m$. Then,
\begin{align}
    T_3T_2T_1 &= \begin{pmatrix}
        \mathbb{I} & T_{3, 1} \\
        T_{3, 2} & \mathbb{I}
    \end{pmatrix}
    \begin{pmatrix}
        \mathbb{I} & T_{1, 1} + T_{2, 1} \\
        T_{1, 2} + T_{2, 2} & \mathbb{I} + T_{2, 2}T_{1, 1}
    \end{pmatrix} \nonumber\\
    &= 
    \begin{pmatrix}
        \mathbb{I} + T_{3, 1}(T_{2, 2} + T_{1, 2}) & T_{1, 1} + T_{2, 1} + T_{3, 1}(\mathbb{I} + T_{2, 2}T_{1, 1}) \\
        T_{1, 2} + T_{2, 2} + T_{3, 2} & \mathbb{I} + T_{3, 2}(T_{1, 1} + T_{2, 1}) + T_{2, 2}T_{1, 1}
    \end{pmatrix},
\end{align}
and again, $T_{3, 1}T_{2, 2} = T_{3, 1}T_{1, 2} = \boldsymbol{0}_{2m \times 2m}$, giving
\begin{align}
    T_3T_2T_1 = \begin{pmatrix}
        \mathbb{I} & T_{1, 1} + T_{2, 1} + T_{3, 1} \\
        T_{1, 2} + T_{2, 2} + T_{3, 2} & \mathbb{I} + T_{3, 2}(T_{1, 1} + T_{2, 1}) + T_{2, 2}T_{1, 1}
    \end{pmatrix}.
\end{align}
By induction, $T_{j+1, 1}T_{j, 2} = \boldsymbol{0}_{2m \times 2m}$, and we deduce that $\bar{T}$ takes the form
\begin{align}
    \bar{T} = \begin{pmatrix}
        \mathbb{I} & D_1 \\
        D_2 & D_3
    \end{pmatrix},
\end{align}
with the $D_j$ matrices all depending on the different $T_{i,k}$.

Now, we can compute $\Gamma$. Using \cref{eq: gamma_def} and the fact that $\Gamma_0 = \Sigma_0$ in the situation where auxiliaries are noiseless, $\Gamma$ becomes
\begin{align}
    \Gamma &= \Pi_\text{S} \bar{T} \Sigma_0\bar{T}^T \Pi_\text{S}^T \nonumber \\
    &= \Pi_\text{S} 
    \begin{pmatrix}
        \mathbb{I} & D_1 \\
        D_2 & D_3
    \end{pmatrix} 
    \begin{pmatrix}
        \sigma^2\mathbb{I} & 0 \\
        0 & 0
    \end{pmatrix}
    \begin{pmatrix}
        \mathbb{I} & D_2^T \\
        D_1^T & D_3^T
    \end{pmatrix} \Pi_\text{S}^T \nonumber \\
    &= \Pi_\text{S} 
    \begin{pmatrix}
        \mathbb{I} & D_1 \\
        D_2 & D_3
    \end{pmatrix} 
    \begin{pmatrix}
        \sigma^2\mathbb{I} & \sigma^2D_2^T \\
        0 & 0
    \end{pmatrix}
    \Pi_\text{S}^T \nonumber \\
    &= \Pi_\text{S} 
    \begin{pmatrix}
        \sigma^2\mathbb{I} & \sigma^2D_2^T \\
        \sigma^2D_2 & \sigma^2D_2D_2^T
    \end{pmatrix}
    \Pi_\text{S}^T.
    \label{eq: gamma_mat_noiseless_aux_case}
\end{align}
As we notice, $\Gamma$ is not equal to $\Sigma_0$ since $D_2$ is a non-zero matrix. In order to retrieve $\Sigma_0$, we must only project on the storage subspace, that is 
\begin{align}
    \Pi_\text{G}\Gamma\Pi_\text{G}^T = \Pi_\text{G}\Pi_\text{S} 
    \begin{pmatrix}
        \sigma^2\mathbb{I} & \sigma^2D_2^T \\
        \sigma^2D_2 & \sigma^2D_2D_2^T
    \end{pmatrix}
    \Pi_\text{S}^T\Pi_\text{G}^T = \sigma^2\mathbb{I} = \Sigma_0.
\end{align}
In fact, the decoder that is proposed in the noiseless auxiliary case restricts only to the storage subspace. In this case, even if there is still noise spreading from the storage to the auxiliaries, no other information is gained by considering how these errors spread through the circuit. Indeed, they spread exactly the way they are supposed to; these are perfect correlations. Therefore, using \cref{eq: gamma_mat_noiseless_aux_case} or $\Sigma_0$ during decoding is completely equivalent.

\begin{figure*}[t!]
    \centering
    \begin{subfigure}[t]{0.5\textwidth}
        \centering
        \caption{}
        \vspace{-0.25cm}
        \includegraphics[scale=1]{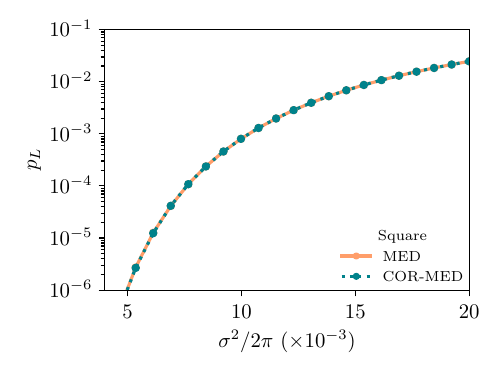}
        \label{subfig: med_decoder_comparison_noiseless_aux_square}
    \end{subfigure}%
    ~ 
    \begin{subfigure}[t]{0.5\textwidth}
        \centering
        \caption{}
        \vspace{-0.25cm}
        \includegraphics[scale=1]{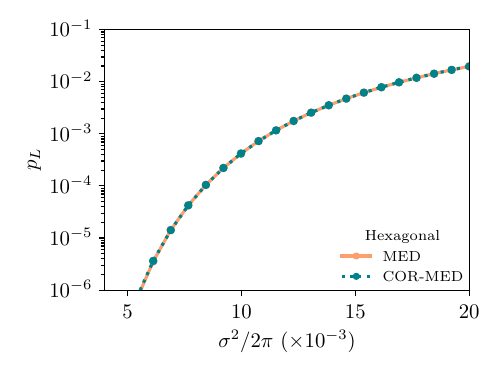}
        \label{subfig: med_decoder_comparison_noiseless_aux_hex}
    \end{subfigure}
    \\\vspace{-0.75cm}
    \begin{subfigure}[t]{0.5\textwidth}
        \centering
        \caption{}
        \vspace{-0.25cm}
        \includegraphics[scale=1]{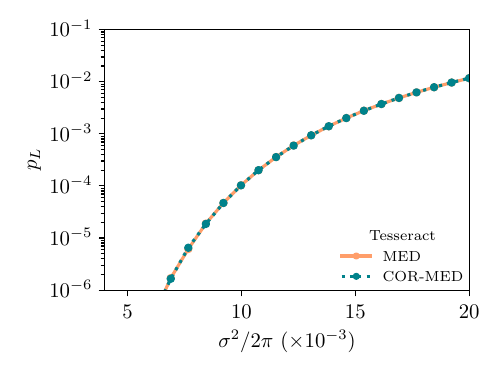}
        \label{subfig: med_decoder_comparison_noiseless_aux_tess}
    \end{subfigure}%
    ~ 
    \begin{subfigure}[t]{0.5\textwidth}
        \centering
        \caption{}
        \vspace{-0.25cm}
        \includegraphics[scale=1]{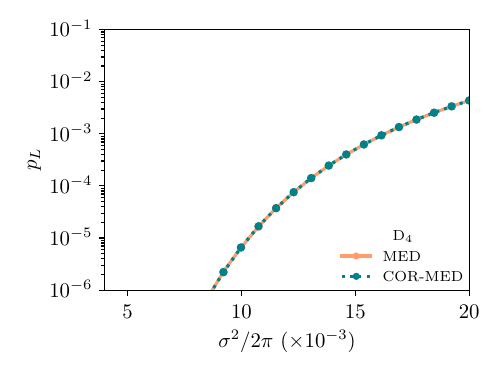}
        \label{subfig: med_decoder_comparison_noiseless_aux_d4}
    \end{subfigure}
    \vspace{-0.75cm}
    \caption{Comparison between the MED decoder presented in \cref{subsec: qec_perf_aux} with \cref{eq: med_cvp_decoder} (full orange line) and our proposed COR-MED decoder presented in \cref{subsec: what_we_propose} with \cref{eq: logical_vect_of_corr_mld_decoder} (dotted teal lines) in the limit of noiseless auxiliaries. Each point is obtained by a Monte-Carlo method through the sampling of $N = 10^9$ errors $\boldsymbol{\xi}_\mathrm{G}$, and its standard deviation $\bar{\sigma}$ is smaller than the size of the markers. For more information on the simulations, we refer the reader to \cref{sec: res}. We plot the probability of having a logical error $p_L$ as a function of the noise variance $\sigma^2 / 2\pi$ in the system over a range of $\sigma^2/2\pi = [4, 20] \times 10^{-3}$. The different panels correspond to different GKP codes: (a) the square \cref{subsubsec: square_GKP_info}, (b) the hexagonal \cref{subsubsec: hex_GKP_info}, (c) the tesseract \cref{subsubsec: tess_GKP_info}, and (d) the D$_4$ \cref{subsubsec: D4_GKP_info}.}
    \label{fig: med_decoder_comparison_noiseless_aux}
\end{figure*}

With this, our decoder reads
\begin{align}
    \bar{\boldsymbol{\lambda}}^\perp_\text{G} = \underset{\boldsymbol{\lambda}^\perp_\text{G} \in \Lambda_\text{G}^\perp / \Lambda_\text{G}}{\text{arg max }} P_{\mathbf{0}, \Sigma_0}([\boldsymbol{\chi}(\mathbf{z}) + \boldsymbol{\lambda}^\perp_\text{G}] | \tilde{\mathbf{z}}),
\end{align}
and we retrieve \cref{eq: general_mld_decoder}. 

This shows that we can retrieve the noiseless auxiliary MLD decoder \cref{eq: general_mld_decoder} from our MLD decoder \cref{eq: general_noise_corr_mld_decoder} in the limit where the noise on the auxiliaries is not present. Thus, we can assume that the same behavior is respected for the case of both decoders in the MED limit, \cref{eq: med_cvp_decoder,eq: logical_vect_of_corr_mld_decoder}. To further support this, we provide numerical simulations comparing the MED decoder and our COR-MED decoder in the case where we only have noise on the storage. These simulations can be found on \cref{fig: med_decoder_comparison_noiseless_aux}. As we can see from the figure, both decoders are in agreement for all GKP storage analyzed, irrespective of the noise strength $\sigma^2$ on the storage mode. 

Note that we are required to select a certain numerical zero value in the $\Gamma_0$ matrix when realizing these simulations. Indeed, when auxiliaries are noiseless, $\Gamma_0$ takes the form
\begin{align}
    \Gamma_0 = \Sigma_0 = \begin{pmatrix}
        \sigma^2 \mathbb{I}_{2m \times 2m} & 0 \\
        0 & 0
    \end{pmatrix}.
    \label{eq: sigma_0_mat_noiseless_aux}
\end{align}
As we will see in \cref{subsec: appendix_expanding_COR_MLD,subsec: appendix_expanding_COR_MED}, solving the COR-MED decoder requires computing $\Gamma^{-1}$, with $\Gamma$ defined by \cref{eq: gamma_def}. Since $\Gamma_0 = \Sigma_0$ is singular when auxiliaries are noiseless, so is $\Gamma$ in \cref{eq: gamma_mat_noiseless_aux_case}, and computing $\Gamma^{-1}$ amounts to inverting a singular matrix. Numerically, in order to avoid singularities in the computation of this inverse, we replace the lower right block of \cref{eq: sigma_0_mat_noiseless_aux} by numerical zeroes, here chosen to be $10^{-9}$.

\subsection{Expanding the form of our noise-correlated MLD decoder}
\label{subsec: appendix_expanding_COR_MLD}

In this appendix, we show how we can expand our proposed noise-correlated MLD decoder and find an expression we solve analytically. 

We start by looking at \cref{eq: general_noise_corr_mld_decoder} and develop the argument of the function that we must maximize $P_{\mathbf{0}, \Gamma}([\boldsymbol{\xi}_{\bar{T}, \text{G}}] | \tilde{\mathbf{z}}_{\bar{T}, \text{m}})$. Let us first try to mimic what was done in the case of noiseless auxiliaries in Ref. \cite{conradGKPMultimode}. First, we can use Bayes' law such that
\begin{align}
    P_{\mathbf{0}, \Gamma}([\boldsymbol{\xi}_{\bar{T}, \text{G}}] | \tilde{\mathbf{z}}_{\bar{T}, \text{m}}) = \frac{P_{\mathbf{0}, \Gamma}(\tilde{\mathbf{z}}_{\bar{T}, \text{m}} | [\boldsymbol{\xi}_{\bar{T}, \text{G}}]) P_{\mathbf{0}, \Gamma}([\boldsymbol{\xi}_{\bar{T}, \text{G}}])}{P_{\mathbf{0}, \Gamma}(\tilde{\mathbf{z}}_{\bar{T}, \text{m}})},
\end{align}
and expanding the equivalence class of $\boldsymbol{\xi}_{\bar{T}, \text{G}}$, it follows that
\begin{align}
    P_{\mathbf{0}, \Gamma}([\boldsymbol{\xi}_{\bar{T}, \text{G}}] | \tilde{\mathbf{z}}_{\bar{T}, \text{m}}) = \sum_{\boldsymbol{\lambda} \in \Lambda_\text{G}}\frac{P_{\mathbf{0}, \Gamma}(\tilde{\mathbf{z}}_{\bar{T}, \text{m}} | \boldsymbol{\xi}_{\bar{T}, \text{G}} + \boldsymbol{\lambda}) P_{\mathbf{0}, \Gamma}(\boldsymbol{\xi}_{\bar{T}, \text{G}} + \boldsymbol{\lambda})}{P_{\mathbf{0}, \Gamma}(\tilde{\mathbf{z}}_{\bar{T}, \text{m}})}.
    \label{eq: first_test_solving_main_decoder}
\end{align}
In the case of noiseless auxiliaries, the quantity $P_{\mathbf{0}, \Gamma}(\tilde{\mathbf{z}}_{\bar{T}, \text{m}} | \boldsymbol{\xi}_{\bar{T}, \text{G}} + \boldsymbol{\lambda})$ is always equal to one. Indeed, if we know the error that displaced our storage, then we also know with certainty the syndrome that we measured. This no longer holds true when these two quantities are corrupted by errors from the noise in the auxiliaries. Instead, we rather consider the storage and the auxiliaries as one whole correlated system, helping us to determine an expression we can solve for $P_{\mathbf{0}, \Gamma}([\boldsymbol{\xi}_{\bar{T}, \text{G}}] | \tilde{\mathbf{z}}_{\bar{T}, \text{m}})$.

Let us start with \cref{eq: general_noise_corr_mld_decoder} again. Now, using conditional probability, this equation becomes 
\begin{align}
    P_{\mathbf{0}, \Gamma}([\boldsymbol{\xi}_{\bar{T}, \text{G}}] | \tilde{\mathbf{z}}_{\bar{T}, \text{m}}) &= \frac{P_{\mathbf{0}, \Gamma}([\boldsymbol{\xi}_{\bar{T}, \text{G}}], \tilde{\mathbf{z}}_{\bar{T}, \text{m}})}{P_{\mathbf{0}, \Gamma}(\tilde{\mathbf{z}}_{\bar{T}, \text{m}})},
\end{align}
and, expanding the equivalence class of $\boldsymbol{\xi}_{\bar{T}, \text{G}}$ \cite{conradGKPMultimode}
\begin{align}
    P_{\mathbf{0}, \Gamma}([\boldsymbol{\xi}_{\bar{T}, \text{G}}] | \tilde{\mathbf{z}}_{\bar{T}, \text{m}}) &= \sum_{\boldsymbol{\lambda} \in \Lambda_\text{G}}\frac{P_{\mathbf{0}, \Gamma}(\boldsymbol{\xi}_{\bar{T}, \text{G}} + \boldsymbol{\lambda},\tilde{\mathbf{z}}_{\bar{T}, \text{m}})}{P_{\mathbf{0}, \Gamma}(\tilde{\mathbf{z}}_{\bar{T}, \text{m}})} \nonumber \\
    &\propto \sum_{\boldsymbol{\lambda} \in \Lambda_\text{G}} P_{\mathbf{0}, \Gamma}(\boldsymbol{\xi}_{\bar{T}, \text{G}} + \boldsymbol{\lambda}, \tilde{\mathbf{z}}_{\bar{T}, \text{m}}),
    \label{eq: first_cond_proba}
\end{align}
where we omitted to write $P_{\mathbf{0}, \Gamma}(\tilde{\mathbf{z}}_{\bar{T}, \text{m}})$ since it won't affect the optimization process. Let us now further develop the syndrome $\tilde{\mathbf{z}}_{\bar{T}, \text{m}}$. We know that it can be expanded such as $\tilde{\mathbf{z}}_{\bar{T}, \text{m}} = \boldsymbol{\xi}_{\bar{T}, \text{m}} + \boldsymbol{\mu}_\text{m}$, but we do not have enough information to know precisely $\boldsymbol{\mu}_\text{m} = \Pi_\text{m}(S_\eta)^T\mathbf{a}_\eta$, with $\mathbf{a}_\eta \in \mathbb{Z}^{4m}$, since it is sampled uniformly through all possible logical vectors in $\Lambda_\text{m}$ when the measurement is made. Thus, summing over all possible realizations of the measurement, \cref{eq: first_cond_proba} can be written as
\begin{align}
    P_{\mathbf{0}, \Gamma}([\boldsymbol{\xi}_{\bar{T}, \text{G}}] | \tilde{\mathbf{z}}_{\bar{T}, \text{m}}) &= \sum_{\boldsymbol{\lambda} \in \Lambda_\text{G}}\sum_{\boldsymbol{\lambda}_\text{m} \in \Lambda_\text{m}}P_{\mathbf{0}, \Gamma}(\boldsymbol{\xi}_{\bar{T}, \text{G}} + \boldsymbol{\lambda}, \boldsymbol{\xi}_{\bar{T}, \text{m}} + \boldsymbol{\mu}_\text{m} | \boldsymbol{\lambda}_\text{m}) P_{\mathbf{0}, \Gamma}(\boldsymbol{\lambda}_\text{m}),
\end{align}
and using Bayes' law
\begin{align}
    P_{\mathbf{0}, \Gamma}([\boldsymbol{\xi}_{\bar{T}, \text{G}}] | \tilde{\mathbf{z}}_{\bar{T}, \text{m}}) = \sum_{\boldsymbol{\lambda} \in \Lambda_\text{G}}\sum_{\boldsymbol{\lambda}_\text{m} \in \Lambda_\text{m}}\frac{P_{\mathbf{0}, \Gamma}(\boldsymbol{\lambda}_\text{m} | \boldsymbol{\xi}_{\bar{T}, \text{G}} + \boldsymbol{\lambda}, \boldsymbol{\xi}_{\bar{T}, \text{m}} + \boldsymbol{\mu}_\text{m}) P_{\mathbf{0}, \Gamma}(\boldsymbol{\xi}_{\bar{T}, \text{G}} + \boldsymbol{\lambda}, \boldsymbol{\xi}_{\bar{T}, \text{m}} + \boldsymbol{\mu}_\text{m})}{P_{\mathbf{0}, \Gamma}(\boldsymbol{\lambda}_\text{m})}P_{\mathbf{0}, \Gamma}(\boldsymbol{\lambda}_\text{m}).
    \label{eq: first_bayes_proba}
\end{align}
We notice that $P_{\mathbf{0}, \Gamma}(\boldsymbol{\lambda}_\text{m} | \boldsymbol{\xi}_{\bar{T}, \text{G}} + \boldsymbol{\lambda}, \boldsymbol{\xi}_{\bar{T}, \text{m}} + \boldsymbol{\mu}_\text{m})$, the probability of having $\boldsymbol{\lambda}_\text{m}$ knowing we measured $\boldsymbol{\xi}_{\bar{T}, \text{G}} + \boldsymbol{\lambda}$ and $\boldsymbol{\xi}_{\bar{T}, \text{m}} + \boldsymbol{\mu}_\text{m}$, is one only when $\boldsymbol{\mu}_\text{m} = \boldsymbol{\lambda}_\text{m}$ and has to be zero otherwise. Therefore, \cref{eq: first_bayes_proba} becomes
\begin{align}
    P_{\mathbf{0}, \Gamma}([\boldsymbol{\xi}_{\bar{T}, \text{G}}] | \tilde{\mathbf{z}}_{\bar{T}, \text{m}}) &= \sum_{\boldsymbol{\lambda} \in \Lambda_\text{G}}\sum_{\boldsymbol{\lambda}_\text{m} \in \Lambda_\text{m}} P_{\mathbf{0}, \Gamma}(\boldsymbol{\xi}_{\bar{T}, \text{G}} + \boldsymbol{\lambda}, \boldsymbol{\xi}_{\bar{T}, \text{m}} + \boldsymbol{\lambda}_\text{m}) \nonumber \\
    &= \sum_{\boldsymbol{\lambda} \in \Lambda_\text{G}}\sum_{\boldsymbol{\lambda}_\text{m} \in \Lambda_\text{m}} P_{\mathbf{0}, \Gamma}(\boldsymbol{\xi}_{\bar{T}, \text{S}}(\boldsymbol{\lambda}, \boldsymbol{\lambda}_\text{m})),
    \label{eq: final_proba}
\end{align}
where we introduce the vector $\boldsymbol{\xi}_{\bar{T}, \text{S}}(\boldsymbol{\lambda}, \boldsymbol{\lambda}_\text{m})$, defined in the main text as \cref{eq: xi_t_s}, which combines the subspace of the storage and the measured quadratures of the auxiliaries such that by definition, $P_{\mathbf{0}, \Gamma}(\boldsymbol{\xi}_{\bar{T}, \text{G}} + \boldsymbol{\lambda}, \boldsymbol{\xi}_{\bar{T}, \text{m}} + \boldsymbol{\lambda}_\text{m}) = P_{\mathbf{0}, \Gamma}(\boldsymbol{\xi}_{\bar{T}, \text{S}}(\boldsymbol{\lambda}, \boldsymbol{\lambda}_\text{m}))$. With \cref{eq: final_proba}, \cref{eq: general_noise_corr_mld_decoder} becomes
\begin{align}
    \bar{\boldsymbol{\xi}}_{\bar{T}, \text{G}} &= \underset{\boldsymbol{\xi}_{\bar{T}, \text{G}} \in \mathcal{V}_\text{G}}{\text{arg max }} \qty(\sum_{\boldsymbol{\lambda} \in \Lambda_\text{G}}\sum_{\boldsymbol{\lambda}_\text{m} \in \Lambda_\text{m}} P_{\mathbf{0}, \Gamma}(\boldsymbol{\xi}_{\bar{T}, \text{S}}(\boldsymbol{\lambda}, \boldsymbol{\lambda}_\text{m}))),
\end{align}
where we are optimizing over the Voronoi cell of the stabilizer lattice $\mathcal{V}_\text{G}$, as discussed in the main text.

Let us develop the probability $P_{\mathbf{0}, \Gamma}(\boldsymbol{\xi}_{\bar{T}, \text{S}}(\boldsymbol{\lambda}, \boldsymbol{\lambda}_\text{m}))$. This expression tells us that we need to sample the probability of having each of the vectors $\boldsymbol{\xi}_{\bar{T}, \text{S}}(\boldsymbol{\lambda}, \boldsymbol{\lambda}_\text{m})$ that have been deformed by the Steane circuit. The deformation can be directly linked to a change in the variance of our noise model. Using \eqref{eq: noise_model_changing_with_symp_op} and the fact that the projector onto the storage and measured auxiliaries is a linear transformation, we can deduce that the probability of having each $\boldsymbol{\xi}_{\bar{T}, \text{S}}(\boldsymbol{\lambda}, \boldsymbol{\lambda}_\text{m})$ is now sampled following the noise model $\mathcal{N}[\mathbf{0}, \Gamma](\hat{\rho})$, with \cite{introToMultivariateStatAnalysis}
\begin{align}
    \Gamma = \Pi_\text{S}\bar{T}\Gamma_0\bar{T}^T\Pi_\text{S}^T,
\end{align}
where $\Pi_\text{S} = \Pi_\text{G} + \Pi_\text{m}$. Thus, with \cref{eq: probability_of_general_noise_model_used}, the probability is
\begin{align}
    P_{\mathbf{0}, \Gamma}(\boldsymbol{\xi}_{\bar{T}, \text{S}}(\boldsymbol{\lambda}, \boldsymbol{\lambda}_\text{m})) = \frac{1}{\sqrt{(2\pi)^{4m}\text{det}\Gamma}} \mathrm{e}^{-\frac{1}{2} \boldsymbol{\xi}_{\bar{T}, \text{S}}(\boldsymbol{\lambda}, \boldsymbol{\lambda}_\text{m})^T \Gamma^{-1}\boldsymbol{\xi}_{\bar{T}, \text{S}}(\boldsymbol{\lambda}, \boldsymbol{\lambda}_\text{m})}.
\end{align}
Again, we omit the pre-factor in front of the exponential because it does not affect the optimization. Defining the quadratic form 
\begin{align}
    Q_\Gamma (\boldsymbol{\xi}_{\bar{T}, \text{G}}, \boldsymbol{\lambda}, \boldsymbol{\lambda}_\text{m}) = \boldsymbol{\xi}_{\bar{T}, \text{S}}(\boldsymbol{\lambda}, \boldsymbol{\lambda}_\text{m})^T \Gamma^{-1}\boldsymbol{\xi}_{\bar{T}, \text{S}}(\boldsymbol{\lambda}, \boldsymbol{\lambda}_\text{m}),
\end{align}
which depends directly on the vector we are optimizing for, $\boldsymbol{\xi}_{\bar{T}, \text{G}}$, and the two vectors $\boldsymbol{\lambda}$ and $\boldsymbol{\lambda}_\text{m}$. This way, the noise-correlated MLD decoder becomes
\begin{align}
    \bar{\boldsymbol{\xi}}_{\bar{T}, \text{G}} = \underset{\boldsymbol{\xi}_{\bar{T}, \text{G}} \in \mathcal{V}_\text{G}}{\text{arg max }} \qty(\sum_{\boldsymbol{\lambda} \in \Lambda_\text{G}}\sum_{\boldsymbol{\lambda}_\text{m} \in \Lambda_\text{m}} \mathrm{e}^{-\frac{1}{2} Q_\Gamma (\boldsymbol{\xi}_{\bar{T}, \text{G}}, \boldsymbol{\lambda}, \boldsymbol{\lambda}_\text{m})}).
    \label{eq: noise_correlated_mld_decoder}
\end{align}

Solving this equation is exponentially hard and practically not realizable for multimode GKP codes. In this work, we assume that in general noise is small and contributions from the $\boldsymbol{\lambda}$ vectors are negligible, taking only $\boldsymbol{\lambda} = \mathbf{0}$ into account. This is equivalent to an MED decoder instead of an MLD decoder. In this case, we are solving 
\begin{align}
    \bar{\boldsymbol{\xi}}_{\bar{T}, \text{G}} = \underset{\boldsymbol{\xi}_{\bar{T}, \text{G}} \in \mathcal{V}_\text{G}}{\text{arg max }} \qty(\sum_{\boldsymbol{\lambda}_\text{m} \in \Lambda_\text{m}} \mathrm{e}^{-\frac{1}{2} Q_\Gamma (\boldsymbol{\xi}_{\bar{T}, \text{G}}, \mathbf{0}, \boldsymbol{\lambda}_\text{m})}).
    \label{eq: noise_correlated_med_decoder}
\end{align}

\subsection{Expanding the form of our noise-correlated MED decoder}
\label{subsec: appendix_expanding_COR_MED}

In this appendix, we show how we can solve \cref{eq: noise_correlated_med_decoder}, giving a way to extract $\bar{\boldsymbol{\xi}}_{\bar{T}, \text{G}}$ and perform QEC on the storage modes. 

We start by fixing $\boldsymbol{\lambda}_\text{m}$ and finding the $\boldsymbol{\xi}_{\bar{T}, \text{G}}$ that maximizes $e^{-\frac{1}{2}Q_\Gamma(\boldsymbol{\xi}_{\bar{T}, \text{G}}, \mathbf{0}, \boldsymbol{\lambda}_\text{m})}$, or equivalently that minimizes $Q_\Gamma(\boldsymbol{\xi}_{\bar{T}, \text{G}}, \mathbf{0}, \boldsymbol{\lambda}_\text{m})$. We note this vector $\bar{\boldsymbol{\xi}}_{\bar{T}, \text{G}}$, as in the main text. Doing so, we see that $\bar{\boldsymbol{\xi}}_{\bar{T}, \text{G}}$ is directly set by the initial choice of $\boldsymbol{\lambda}_\text{m}$. We then find the $\bar{\boldsymbol{\lambda}}_\text{m}$ that minimizes $Q_\Gamma(\bar{\boldsymbol{\xi}}_{\bar{T}, \text{G}}, \mathbf{0}, \boldsymbol{\lambda}_\text{m})$, therefore has the most weight in the sum, enabling us to define precisely $\bar{\boldsymbol{\xi}}_{\bar{T}, \text{G}}$.

Let us start by describing $Q_\Gamma(\boldsymbol{\xi}_{\bar{T}, \text{G}}, \mathbf{0}, \boldsymbol{\lambda}_\text{m})$ in more detail. We recall that $\Gamma = \Pi_\text{S}\bar{T}\Gamma_0\bar{T}^T\Pi_\text{S}^T$. This implies that $\Gamma$ and $\Gamma^{-1}$ are symmetric because $\Gamma_0$ is. We can then explicitly define $\Gamma^{-1}$ by \cref{eq: gamma_cov_mat} of the main text, with $\Gamma_\text{G}$ and $\Gamma_\text{m}$ also symmetric. Thus, with the form of $\boldsymbol{\xi}_{\bar{T}, \text{S}}(\mathbf{0}, \boldsymbol{\lambda}_\text{m})$ given by \cref{eq: xi_t_s}, we can compute the quadratic form such that
\begin{align}
    Q_\Gamma(\boldsymbol{\xi}_{\bar{T}, \text{G}}, \mathbf{0}, \boldsymbol{\lambda}_\text{m}) &= \begin{pmatrix}
        \boldsymbol{\xi}_{\bar{T}, \text{G}} &
        \boldsymbol{\xi}_{\bar{T}, \text{m}} + \boldsymbol{\lambda}_\text{m}
    \end{pmatrix}
    \begin{pmatrix}
        \Gamma_\text{G} & \gamma \\
        \gamma^T & \Gamma_\text{m}
    \end{pmatrix}
    \begin{pmatrix}
        \boldsymbol{\xi}_{\bar{T}, \text{G}} \\
        \boldsymbol{\xi}_{\bar{T}, \text{m}} + \boldsymbol{\lambda}_\text{m}
    \end{pmatrix} \nonumber \\
    &= \boldsymbol{\xi}_{\bar{T}, \text{G}}^T \Gamma_\text{G}\boldsymbol{\xi}_{\bar{T}, \text{G}} + 2\boldsymbol{\xi}_{\bar{T}, \text{G}}^T \gamma \boldsymbol{\xi}_{\bar{T}, \text{m}}' + (\boldsymbol{\xi}_{\bar{T}, \text{m}}')^T\Gamma_\text{m}\boldsymbol{\xi}_{\bar{T}, \text{m}}',
    \label{eq: developped_Q_func}
\end{align}
where we defined $\boldsymbol{\xi}_{\bar{T}, \text{m}}' = \boldsymbol{\xi}_{\bar{T}, \text{m}} + \boldsymbol{\lambda}_\text{m}$ for conciseness. Since we want to minimize \cref{eq: developped_Q_func}, we compute
\begin{align}
    \Bigg[\frac{\partial Q_\Gamma(\boldsymbol{\xi}_{\bar{T}, \text{G}}, \mathbf{0}, \boldsymbol{\lambda}_\text{m})}{\partial \boldsymbol{\xi}_{\bar{T}, \text{G}}}\Bigg]_{\bar{\boldsymbol{\xi}}_{\bar{T}, \text{G}}} = \Bigg[\frac{\partial (\boldsymbol{\xi}_{\bar{T}, \text{G}}^T \Gamma_\text{G} \boldsymbol{\xi}_{\bar{T}, \text{G}})}{\partial \boldsymbol{\xi}_{\bar{T}, \text{G}}} + 2\frac{\partial (\boldsymbol{\xi}_{\bar{T}, \text{G}}^T \gamma \boldsymbol{\xi}_{\bar{T}, \text{m}}')}{\partial \boldsymbol{\xi}_{\bar{T}, \text{G}}}\Bigg]_{\bar{\boldsymbol{\xi}}_{\bar{T}, \text{G}}} = 0,
\end{align}
and using identities in Ref. \cite{computerVision} to differentiate quadratic forms,
\begin{align}
    (\Gamma_\text{G} + \Gamma_\text{G}^T)\bar{\boldsymbol{\xi}}_{\bar{T}, \text{G}} + 2\gamma \boldsymbol{\xi}_{\bar{T}, \text{m}}' = 2\Gamma_\text{G}\bar{\boldsymbol{\xi}}_{\bar{T}, \text{G}} + 2\gamma \boldsymbol{\xi}_{\bar{T}, \text{m}}' = 0 \implies \bar{\boldsymbol{\xi}}_{\bar{T}, \text{G}} = -\Gamma_\text{G}^{-1}\gamma (\boldsymbol{\xi}_{\bar{T}, \text{m}} + \boldsymbol{\lambda}_{\text{m}}).
    \label{eq: maximized_xi_tg}
\end{align}
We recall that $\boldsymbol{\xi}_{\bar{T}, \text{m}} + \boldsymbol{\lambda}_\text{m}$ is fixed by the value of $\tilde{\mathbf{z}}_{\bar{T}, \text{m}}$, meaning $\tilde{\mathbf{z}}_{\bar{T}, \text{m}} = \boldsymbol{\xi}_{\bar{T}, \text{m}} + \boldsymbol{\lambda}_\text{m}$, and that $\boldsymbol{\lambda}_\text{m}$ is set randomly after the measurement. Now, we need to find the $\bar{\boldsymbol{\lambda}}_{\text{m}}$ that minimizes $Q_\Gamma(\bar{\boldsymbol{\xi}}_{\bar{T}, \text{G}}, \mathbf{0}, \boldsymbol{\lambda}_\text{m})$. We can start by working on the terms in $Q_\Gamma(\bar{\boldsymbol{\xi}}_{\bar{T}, \text{G}}, \mathbf{0}, \boldsymbol{\lambda}_\text{m})$ separately. For the first term, reintroducing the notation $\boldsymbol{\xi}_{\bar{T}, \text{m}}'$, 
\begin{align}
    (\bar{\boldsymbol{\xi}}_{\bar{T}, \text{G}})^T \Gamma_\text{G}\bar{\boldsymbol{\xi}}_{\bar{T}, \text{G}} &= (-\Gamma_\text{G}^{-1}\gamma \boldsymbol{\xi}_{\bar{T}, \text{m}}')^T \Gamma_\text{G}(-\Gamma_\text{G}^{-1}\gamma \boldsymbol{\xi}_{\bar{T}, \text{m}}') \nonumber \\
    &= (\boldsymbol{\xi}_{\bar{T}, \text{m}}')^T \gamma^T \Gamma_\text{G}^{-1} \gamma \boldsymbol{\xi}_{\bar{T}, \text{m}}',
    \label{eq: first_term_of_Q_func_first_opt}
\end{align}
for the second
\begin{align}
    2(\bar{\boldsymbol{\xi}}_{\bar{T}, \text{G}})^T \gamma\boldsymbol{\xi}_{\bar{T}, \text{m}}' &= 2(-\Gamma_\text{G}^{-1}\gamma \boldsymbol{\xi}_{\bar{T}, \text{m}}')^T\gamma\boldsymbol{\xi}_{\bar{T}, \text{m}}' \nonumber \\
    &= -2(\boldsymbol{\xi}_{\bar{T}, \text{m}}')^T \gamma^T \Gamma_\text{G}^{-1} \gamma \boldsymbol{\xi}_{\bar{T}, \text{m}}',
    \label{eq: sec_term_of_Q_func_first_opt}
\end{align}
and the third stays the same. Now, the quadratic form becomes
\begin{align}
    Q_\Gamma(\bar{\boldsymbol{\xi}}_{\bar{T}, \text{G}}, \mathbf{0}, \boldsymbol{\lambda}_\text{m}) &= (\boldsymbol{\xi}_{\bar{T}, \text{m}}')^T\Gamma_\text{m}\boldsymbol{\xi}_{\bar{T}, \text{m}}' - (\boldsymbol{\xi}_{\bar{T}, \text{m}}')^T \gamma^T \Gamma_\text{G}^{-1} \gamma \boldsymbol{\xi}_{\bar{T}, \text{m}}' \nonumber \\
    &= (\boldsymbol{\xi}_{\bar{T}, \text{m}}')^T\bar{\Gamma}\boldsymbol{\xi}_{\bar{T}, \text{m}}',
\end{align}
where we have defined $\bar{\Gamma} = \Gamma_\text{m} - \gamma^T\Gamma_\text{G}^{-1} \gamma$. Then, we extract $\bar{\boldsymbol{\lambda}}_\text{m}$ by solving
\begin{align}
    \bar{\boldsymbol{\lambda}}_\text{m} &= \underset{\boldsymbol{\lambda}_\text{m} \in \Lambda_\text{m}}{\text{arg min }}Q_\Gamma(\bar{\boldsymbol{\xi}}_{\bar{T}, \text{G}}, \mathbf{0}, \boldsymbol{\lambda}_\text{m}) \nonumber \\
    &= \underset{\boldsymbol{\lambda}_\text{m} \in \Lambda_\text{m}}{\text{arg min }}\qty((\boldsymbol{\xi}_{\bar{T}, \text{m}} + \boldsymbol{\lambda}_{\text{m}})^T\bar{\Gamma}(\boldsymbol{\xi}_{\bar{T}, \text{m}} + \boldsymbol{\lambda}_{\text{m}})) \nonumber \\
    &= \underset{\boldsymbol{\lambda}_\text{m} \in \Lambda_\text{m}}{\text{arg min }}|\boldsymbol{\xi}_{\bar{T}, \text{m}} + \boldsymbol{\lambda}_{\text{m}}|_{\bar{\Gamma}}.
    \label{eq: final_cvp_solving}
\end{align}
Solving \cref{eq: final_cvp_solving} in order to find $\bar{\boldsymbol{\lambda}}_\text{m}$ requires solving the CVP with a modified distance measure $\bar \Gamma$ that takes into account the deformation of the noise induced by the measurement circuit. We can convert this $\bar{\Gamma}$ metric to a Euclidean metric provided we apply a linear transformation on the vectors. Since $\bar{\Gamma}$ is the Schur complement of $\Gamma_\text{G}$ in the positive-definite matrix $\Gamma^{-1}$, it is real, symmetric and positive definite. It therefore admits a unique Cholesky decomposition with an invertible factor $F$ \cite{matrixAnalysis}, such that
\begin{align}
    \bar{\Gamma} = F^T F \implies \mathbf{u}^T\bar{\Gamma}\mathbf{u} = \mathbf{u}^TF^TF\mathbf{u} = \mathbf{v}^T\mathbf{v},
    \label{eq: gamma_bar_cholesky_decomp}
\end{align}
where $\mathbf{v} = F\mathbf{u}$. Therefore, \cref{eq: final_cvp_solving} becomes
\begin{align}
    \bar{\boldsymbol{\lambda}}_\text{m} = F^{-1}\underset{\boldsymbol{\lambda}_{F,\text{m}} \in F\Lambda_\text{m}}{\text{arg min }}|F\boldsymbol{\xi}_{\bar{T}, \text{m}} + \boldsymbol{\lambda}_{F, \text{m}}|,
\end{align}
where we see that the CVP is now realized on a reshaped measured auxiliary lattice $F\Lambda_\text{m}$, which now encompasses the information about how the noise spreads in the measurements following the Steane QEC circuit. From this, using \cref{eq: maximized_xi_tg}, the final correction vector $\bar{\boldsymbol{\xi}}_{\bar{T}, \text{G}}$ is given by
\begin{align}
    \bar{\boldsymbol{\xi}}_{\bar{T}, \text{G}} = -\Gamma_\text{G}^{-1}\gamma\qty(\boldsymbol{\xi}_{\bar{T}, \text{m}} + F^{-1}\underset{\boldsymbol{\lambda}_{F,\text{m}} \in F\Lambda_\text{m}}{\text{arg min }}|F\boldsymbol{\xi}_{\bar{T}, \text{m}} + \boldsymbol{\lambda}_{F, \text{m}}|).
    \label{eq: simplified_noise_correlated_med_decoder}
\end{align}
Of course, since we do not have access to the actual error $\boldsymbol{\xi}_{\bar{T}, \text{m}}$, we can only use the information we have, the syndrome $\tilde{\mathbf{z}}_{\bar{T}, \text{m}}$. In fact, finding the optimal $\bar{\boldsymbol{\lambda}}_{\text{m}}$ to add to $\boldsymbol{\xi}_{\bar{T}, \text{m}}$ is the same as directly solving the equation for $\tilde{\mathbf{z}}_{\bar{T}, \text{m}}$ since we can always write 
\begin{align}
    \boldsymbol{\xi}_{\bar{T}, \text{m}} + \bar{\boldsymbol{\lambda}}_{\text{m}} &= \boldsymbol{\xi}_{\bar{T}, \text{m}} + \boldsymbol{\lambda}_{\text{m}} - \boldsymbol{\lambda}_{\text{m}}' \nonumber \\
    &= \tilde{\mathbf{z}}_{\bar{T}, \text{m}} - \boldsymbol{\lambda}_{\text{m}}', 
\end{align}
where we decomposed $\bar{\boldsymbol{\lambda}}_{\text{m}} = \boldsymbol{\lambda}_{\text{m}} - \boldsymbol{\lambda}_{\text{m}}'$. Therefore, both situations are equivalent because the result of the optimization for $\tilde{\mathbf{z}}_{\bar{T}, \text{m}}$ instead of $\boldsymbol{\xi}_{\bar{T}, \text{m}}$ would amount to $\boldsymbol{\lambda}_{\text{m}}'$ instead of $\bar{\boldsymbol{\lambda}}_{\text{m}}$. With this, we finally write
\begin{align}
    \bar{\boldsymbol{\xi}}_{\bar{T}, \text{G}} = -\Gamma_\text{G}^{-1}\gamma\qty(\tilde{\mathbf{z}}_{\bar{T}, \text{m}} - F^{-1}\underset{\boldsymbol{\lambda}_{F,\text{m}} \in F\Lambda_\text{m}}{\text{arg min }}|F\tilde{\mathbf{z}}_{\bar{T}, \text{m}} - \boldsymbol{\lambda}_{F, \text{m}}|).
\end{align}

\subsection{Numerical complexity comparison between the different decoders}
\label{subsec: numerical_complexity_MED_vs_CORMED}

In this appendix, we compare the computational time complexity of the two non-noise-correlated decoders with our new decoder. As is expected from the discussion in \cref{subsec: what_we_propose}, we show that this complexity is the same for the MED and the COR-MED decoders.

To quantify this complexity, we choose to compute the average time $t_{\boldsymbol{\xi}}$ it takes for a given decoder to decode a single error $\boldsymbol{\xi}$ following the MC simulation step defined in \cref{sec: res}, denoted as the runtime. This procedure is realized for a specific GKP code family, the $\mathrm{D}_{2n}$ GKP family, described more precisely in \cref{subsubsec: D4_GKP_info}. The choice of using the $\mathrm{D}_{2n}$ lattices is motivated by their  straightforward definition in terms of the mode number defining a GKP code $n$, giving a direct measure of the runtime in function of this parameter.

\begin{wrapfigure}{r}{0.55\textwidth}
    \centering
    \vspace{-0.5cm}
    \includegraphics[scale=1]{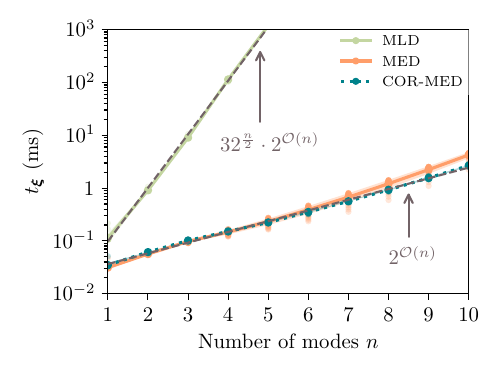}
    \vspace{-0.25cm}
    \caption{The runtime $t_{\boldsymbol{\xi}}$ a given decoder takes for an increasing number of modes in the $\mathrm{D}_{2n}$ GKP family, described in \cref{subsubsec: D4_GKP_info}. Here, we compare the MLD decoder \cref{eq: general_mld_decoder} (full green line) and the MED decoder \cref{eq: med_cvp_decoder} (full orange line) with our COR-MED decoder \cref{eq: logical_vect_of_corr_mld_decoder} (dotted teal line). For each decoder, we also show a linear fit (dotted gray line) to the runtime data, enabling us to extract a time complexity scaling. For each point $t_{\boldsymbol{\xi}}$, we average over all $N = 10^5 - 10^7$ MC simulation steps for $25 - 35$ different values of $\sigma^2/2\pi$ considered over a range of $\sigma^2/2\pi = [1.5, 10] \times 10^{-3}$, all shown with a lower opacity. The standard deviation of these averages is represented by the shaded area. These simulations were run on two AMD EPYC 7742 64-Core processors operating on Linux 5.14.0.}
    \label{fig: runtime_D2N}
\end{wrapfigure}
The results are shown on \cref{fig: runtime_D2N}. As we can see, the MED and COR-MED decoders effectively exhibit similar runtimes for all numbers of modes studied, exhibiting a computational time complexity growing exponentially in $n$. Notably, even though the MLD decoder also shows exponential growth, the COR-MED decoder remains considerably faster than the MLD decoder and enables significantly lower $p_L$ when decoding (see \cref{fig: comparing_MED_vs_CORMED_main_fig_1e9}).

Because the CVP solver used by the MED and COR-MED decoders is agnostic to the underlying structure of the lattice, we expect the runtime $t_{\boldsymbol{\xi}}$ to stay the same for any $n$-mode GKP code defined by a $2n$-dimensional lattice. This is not the case for the MLD decoder, since it is highly dependent on the structure of the GKP code's lattice $\Lambda_\mathrm{G}$. For a lattice with more smallest vectors, more terms in the sum required to solve the MLD decoder have to be taken into account.

\vspace{0.75cm}

\subsection{Condition on the Steane-type QEC circuit for the COR-MED decoder to be advantageous}
\label{subsec: appendix_bigGamma_conditions}

In this appendix, we analyze the form of $\Gamma$ in more detail to show that the unitary $\hat{U}_{\bar{T}}$ implementing the Steane-type QEC circuit must respect a particular condition for our proposed COR-MED decoder to bring an advantage over the MED decoder introduced in \cref{subsec: qec_noisy_aux}.

First, we recall that the initial noise in the relevant quadratures of the system is given by $\Gamma_\mathrm{i} = \Pi_\mathrm{S}\Gamma_0 \Pi_\mathrm{S}^T$. Here, the relevant quadratures comprise the storage's and the measured auxiliaries (see \cref{subsec: qec_noisy_aux} for more detail on this notation). Similarly, the final noise in the relevant quadratures of the system is described by $\Gamma$, defined by \cref{eq: gamma_def}. This is the noise covariance matrix that is fed to the COR-MED decoder, encompassing the correlations between the storage and the measured auxiliaries.

Now, suppose we are in the situation where 
\begin{align}
    \Gamma_\mathrm{i} = \Gamma \iff \Pi_\mathrm{S}\Gamma_0 \Pi_\mathrm{S}^T = \Pi_\mathrm{S}\bar{T}\Gamma_0 \bar{T}^T\Pi_\mathrm{S}^T.
    \label{eq: condition_for_CORMED_over_MED}
\end{align}
When condition \eqref{eq: condition_for_CORMED_over_MED} is met, using a COR-MED type decoder does not provide any more benefits over an MED type decoder since there are no correlations between the modes of interest to leverage for better performance.

If $\hat{U}_{\bar{T}}$, the corresponding Steane-type QEC circuit, is a passive Gaussian unitary, we know that $\bar{T}\bar{T}^T = \mathbb{I}_{6m \times 6m}$ \cite{gaussianQuantumInfo}. Thus, in the case where all modes are equally noisy, $\Gamma_0 = \sigma^2 \mathbb{I}_{6m \times 6m}$, and condition \cref{eq: condition_for_CORMED_over_MED} is respected, since we have $\Gamma_i = \Gamma$. However, this is not true if $\Gamma_0$ is not proportional to the identity, a situation that can happen if the auxiliaries exhibit different noise strengths than the storage modes, for example. Such a realistic scenario could still render our COR-MED decoder useful, even when dealing with passive circuits. Notably, these circuits have been proven to be of high interest for QEC with GKP states, showing significantly lower probability of errors when decoding, as was proven in Ref. \cite{nohLowOverheadFaultTolerantGKP} for the square GKP QEC teleportation scheme discovered in Ref. \cite{qunaughtGKPName}.

\subsection{Comparing decoder performances in the very high error regimes}
\label{subsec: high_error_regime_decoder_behaviors}

\begin{wrapfigure}[26]{r}{0.5\textwidth}
    \centering
    \vspace{-0.5cm}
    \includegraphics[scale=1]{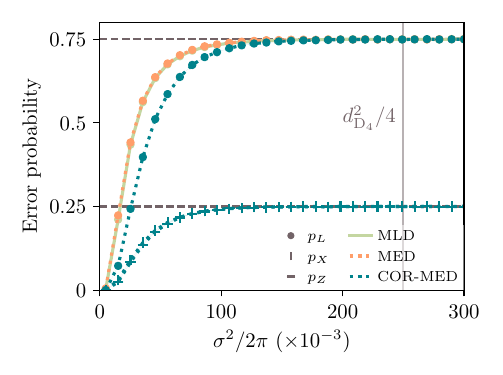}
    \vspace{-0.75cm}
    \caption{Comparison of the three decoders analyzed in this work in the high-noise regime for the D$_4$ GKP \cref{subsubsec: D4_GKP_info}. In full green line, the MLD decoder \cref{eq: general_mld_decoder}, in dotted orange line, the MED decoder \cref{eq: med_cvp_decoder} and in dotted teal line, our COR-MED decoder \cref{eq: solved_noise_corr_mld_decoder}. The latter is the only one taking the noise of the auxiliaries into account. We show the probability of having any logical errors $p_L$ (dots), $\hat{T}(\mathbf{x}_0)$ logical errors $p_{\mathbf{x}_0}$ (vertical bars) and $\hat{T}(\mathbf{z}_0)$ logical errors $p_{\mathbf{z}_0}$ (horizontal bars) as a function of the noise variance $\sigma^2/2\pi$ in the system. We have highlighted the expected theoretical values for those different probabilities with dotted gray horizontal lines and the expected value of $\sigma^2 / 2\pi$ at which $\mathcal{E}_{\sigma^2 / 2\pi \gg d_\mathrm{G}^2/4}(\hat{\rho})$ converges with a light gray vertical line. The figure is simulated over a range of $\sigma^2/2\pi = [5, 300] \times 10^{-3}$ and each point is obtained by a Monte-Carlo method through the sampling of $N = 10^9$ errors $\boldsymbol{\xi}$, and its standard deviation $\bar{\sigma}$ is smaller than the size of the markers.
}
    \label{fig: high_noise_regime_D4_pLpXpZ}
\end{wrapfigure}
In this appendix, we verify numerically that the three different decoders studied in this work behave as expected in the very high noise regime. Even though we focus on the case of noisy auxiliaries, this discussion and the conclusions that arise from it can be directly transposed to the case of noiseless auxiliaries.

We recall that the initial noise in the system is given by $\Gamma_0 = \sigma^2\mathbb{I}_{6m \times 6m}$, where $\sigma$ is the standard deviation of the Gaussian probability distribution $P_{\mathbf{0}, \Gamma_0}(\boldsymbol{\xi})$ sampling the errors $\boldsymbol{\xi}$, described more precisely in \cref{sec: noise_model}. To attain the high noise regime, we simply push $\sigma / \ell$ far outside the logical Voronoi cell $\mathcal{V}^\perp_\mathrm{G}$ of the studied GKP code, this being equivalent to taking $\sigma^2 / 2\pi \gg d_\mathrm{G}^2/4$, with $d_\mathrm{G}$ the distance of the code. This shift averages out the displacement of the storage mode so that it is as likely to be in $\mathcal{V}^\perp_\mathrm{G}$ as it is to be away from the center of $\mathcal{V}^\perp_\mathrm{G}$, closer to a random non-trivial logical representative. Then, the decoder tries to bring the storage mode back onto $\Lambda_\mathrm{G}^\perp$, yielding an effective logical depolarizing channel on the GKP code of the form
\begin{align}
    \mathcal{E}_{\sigma^2 / 2\pi \gg d_\mathrm{G}^2/4}(\hat{\rho}) &= (1 - p_L)\hat{\rho} \nonumber\\
    &+ \frac{p_L}{3}\sum_{\mathbf{p} \in \{\mathbf{x}, \mathbf{y}, \mathbf{z}\}} \hat{T}(\mathbf{p}_0) \hat{\rho} \hat{T}^\dagger(\mathbf{p}_0),
    \label{eq: maximally_mixed_log_error_chann}
\end{align}
where $p_L = \frac{3}{4}$, being the probability of having a logical error on the storage mode.

Figure \ref{fig: high_noise_regime_D4_pLpXpZ} shows the results of such a simulation for the D$_4$ GKP storage mode. As we can see, the behavior of the different decoders all end up matching directly the error channel \cref{eq: maximally_mixed_log_error_chann} above $\sigma^2/2\pi = d^2_{\mathrm{D}_4}/4 = 0.25$, shown by the light gray vertical line. Indeed, as put in evidence by the dotted gray horizontal lines, we observe a convergence of $p_L \rightarrow \frac{3}{4}$, $p_{\mathbf{x}_0} \rightarrow \frac{p_L}{3}$, $p_{\mathbf{z}_0} \rightarrow \frac{p_L}{3}$ and thus $p_{\mathbf{y}_0} = p_L - p_{\mathbf{x}_0} - p_{\mathbf{z}_0} \rightarrow \frac{p_L}{3}$. Although not shown here to prevent repetition, this was also verified for the other GKP codes of interest to this work: the square, hexagonal and tesseract GKP codes.

\subsection{Decoder comparison in the context of multi-round QEC}
\label{subsec: multi_round_QEC}

\begin{figure*}[t!]
    \centering
    \includegraphics[scale=1]{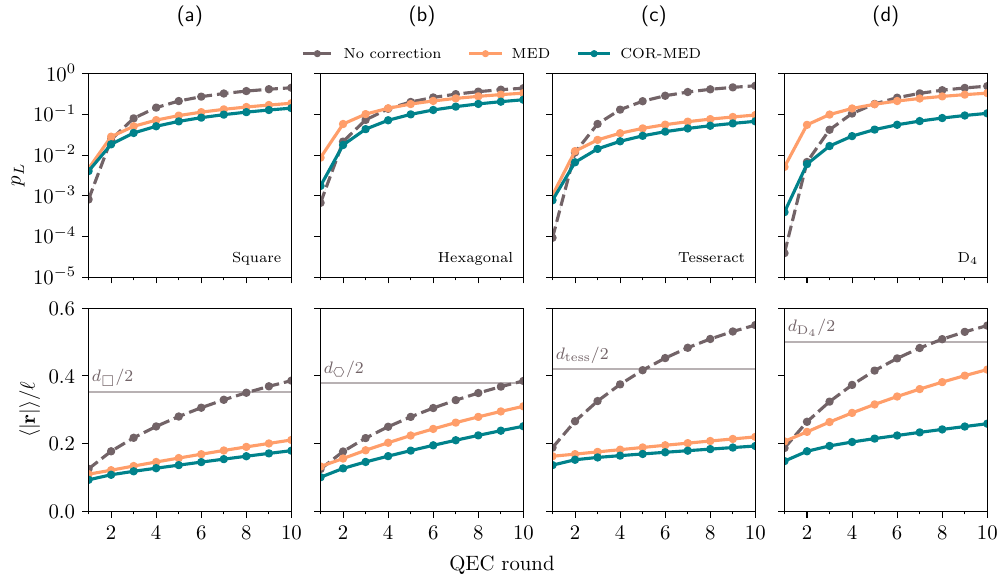}
    \caption{Performance comparison of the absence of correction and two decoders in the context of multi-round QEC: no QEC scheme (dashed gray lines), the MED decoder presented in \cref{subsec: qec_perf_aux} with \cref{eq: med_cvp_decoder} (full orange lines) and our proposed COR-MED decoder presented in \cref{subsec: what_we_propose} with \cref{eq: logical_vect_of_corr_mld_decoder} (full teal lines). Each point is obtained by a Monte-Carlo method which samples $N = 10^6$ realizations of the full sequence of rounds, with noise variance in the system of $\sigma^2 / 2\pi = 5 \times 10^{-3}$, leading to statistical uncertainties smaller than the markers. As a function of the number of QEC rounds, the top row corresponds to the logical error probability $p_L$ and the bottom row corresponds to the mean length of the residual translation modulo $\Lambda_\mathrm{G}$ after decoding. In the bottom row, we also show half the distance of the code $d_\mathrm{G} / 2$ by a light horizontal gray line. Residuals under this limit always result in no logical error. The columns correspond to simulations realized for different GKP storage lattices: (a) square \cref{subsubsec: square_GKP_info}, (b) hexagonal \cref{subsubsec: hex_GKP_info}, (c) tesseract \cref{subsubsec: tess_GKP_info}, and (d) D$_4$ \cref{subsubsec: D4_GKP_info} GKP codes.}
    \label{fig: multi_round_all_codes}
\end{figure*}

In this appendix, we analyze the performance of the decoders over repeated rounds of QEC. This complements the single-round results of \cref{sec: res}, where the logical error probability is evaluated right after one round.

We simulate up to $10$ consecutive rounds, comparing the same Steane-type QEC circuits and decoders as for \cref{fig: comparing_MED_vs_CORMED_main_fig_1e9} with no correction at all. In each round, fresh errors sampled from $\Gamma_0$ with $\sigma^2/2\pi = 5 \times 10^{-3}$ affect the storage and newly prepared auxiliaries. The storage keeps the residual translation $\boldsymbol{\xi}_{\bar{T}, \text{G}} - \bar{\boldsymbol{\xi}}_{\bar{T}, \text{G}}$ left by the previous round (see \cref{eq: cormed_residual_error}).

During simulations, the circuit is applied, the auxiliaries are measured and the decoder computes the correction. After each round, the logical check of \cref{eq: pauli_check_after_correction} is applied to the residual, without modifying it, giving $p_L$ as a function of the number of rounds. We also record the mean length of the residual translation reduced modulo $\Lambda_\mathrm{G}$, denoted as $\langle |\mathbf{r}| \rangle$. Each curve is obtained from $N = 10^6$ realizations of the full sequence of rounds, using the same noise realizations for the three strategies.

The results are shown in \cref{fig: multi_round_all_codes}. After a single round, not correcting yields the lowest $p_L$. This is because the errors of a single round mostly stay within the logical Voronoi cell, where the final logical check corrects them perfectly. The decoders, on the other hand, only have access to the noisy syndrome and occasionally misidentify the error. Without correction, however, the residual is the accumulation of all previous errors and grows with the number of rounds. Its mean length reaches $d_\mathrm{G}/2$ within $6$ to $10$ rounds, and $p_L$ quickly grows towards its saturation value. Both decoders keep the residual smaller, and from the second round onward, the COR-MED decoder yields the lowest $p_L$ for all four codes. The COR-MED decoder also leaves the smallest mean residual at every round and for every code. The difference with the MED decoder is largest for the D$_4$ GKP code, consistent with the amplification of the auxiliary noise by the MED decoder for non-orthogonal stabilizers discussed in \cref{sec: res}. We also verified that the same conclusions hold for different values of $\sigma^2 / 2\pi$.

These results show that the advantage of the COR-MED decoder is not an artifact of the single-round benchmark: by removing more of the error at each round, it keeps the storage closer to the center of its logical Voronoi cell and yields the lowest logical error probability as soon as errors accumulate.

Finally, since neither decoder accounts for the residual inherited from previous rounds, these results are a conservative estimate of what noise-correlated decoding can achieve over repeated rounds. Including the covariance of this residual in the storage block of $\Gamma_0$ would yield a round-aware COR-MED decoder, which we leave for future work.

\subsection{Effect of a symmetric $S_\mathrm{G}$ matrix to implement $\hat{U}_{\bar{T}}$ for the MED and the COR-MED decoders}
\label{subsec: SG_mat_effect_on_decoders}

\begin{figure}[t!]
    \centering
    \begin{subfigure}[t]{0.5\textwidth}
        \centering
        \caption{}
        \vspace{-0.25cm}
        \includegraphics[height=2.5in, width=\textwidth]{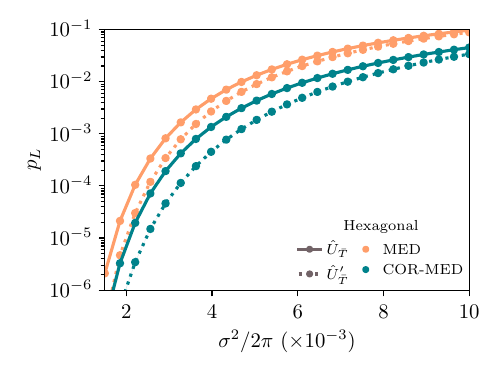}
        \label{subfig: sym_non_sym_compare_hex}
    \end{subfigure}%
    ~ 
    \begin{subfigure}[t]{0.5\textwidth}
        \centering
        \caption{}
        \vspace{-0.25cm}
        \includegraphics[height=2.5in, width=\textwidth]{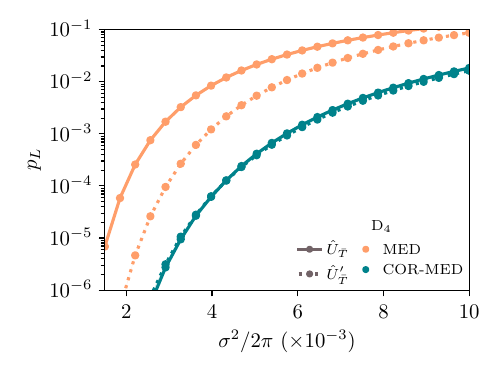}
        \label{subfig: sym_non_sym_compare_D4}
    \end{subfigure}
    \vspace{-0.75cm}
    \caption{Comparison of the MED decoder \cref{eq: med_cvp_decoder} (orange lines) and our proposed COR-MED decoder \cref{eq: logical_vect_of_corr_mld_decoder} (teal lines) under a Steane QEC circuit parametrized by the unitary $\hat{U}_{\bar{T}}$, built from a non-symmetric $S_{\mathrm{G}}$ matrix (full lines), and $\hat{U}_{\bar{T}}'$, built from a symmetric $S_{\mathrm{G}}'$ matrix (dotted lines). The two panels correspond to (a) the hexagonal GKP \cref{subsubsec: hex_GKP_info}, where $S_{\hexagon}$ and $S_{\hexagon}'$ are respectively given by \cref{eq: hexagonal_GKP,eq: stab_mat_sym_hex}, and to (b) the $\mathrm{D}_4$ GKP \cref{subsubsec: D4_GKP_info}, where $S_{\mathrm{D}_4}$ and $S_{\mathrm{D}_4}'$ are respectively given by \cref{eq: d4_GKP,eq: stab_mat_sym_D4}. We plot the probability of having a logical error $p_L$ as a function of the noise variance $\sigma^2 / 2\pi$ in the system over a range of $\sigma^2/2\pi = [1.5, 10] \times 10^{-3}$. Each point is obtained by a Monte-Carlo method through the sampling of $N = 10^9$ errors $\boldsymbol{\xi}$, and its standard deviation $\bar{\sigma}$ is smaller than the size of the markers.}
    \label{fig: sym_non_sym_comparison}
\end{figure}

In this appendix, we investigate further the impact that the structure of $\hat{U}_{\bar{T}}$ has on the decoding performances. Notably, for the two cases studied in this work, we show that building $\hat{U}_{\bar{T}}$ from a quadrature-symmetric stabilizer matrix $S_\mathrm{G}$ yields either better or the same performances for the MED and the COR-MED decoders, respectively.

From the procedure described in \cref{subsec: appendix_gen_steane_type_mm_circuit_construct}, the form of $\hat{U}_{\bar{T}}$ depends entirely on the chosen stabilizer vectors $S_\mathrm{G}$, yielding infinitely many equivalent circuits to realize QEC for a specific GKP code.

There is a subset of these basis choices where the $S_\mathrm{G}$ matrices have the same number of stabilizers with non-zero magnitude component spread across the different modes $\hat{q}_j$ and $\hat{p}_j$. We call it the quadrature-symmetric subset. Thus, a matrix in this set can be obtained by a simple change of basis on the underlying lattice $\Lambda_\mathrm{G}$. Mathematically, this is represented by applying a unimodular matrix $R_\mathrm{G}$ on $S_\text{G}$, giving a completely equivalent GKP code with stabilizer matrix $S_\text{G}' = R_\mathrm{G}S_\text{G}$ \cite{conradGKPMultimode,bapGKPMultimode,lin_2023a}. The new circuit parametrized by $\hat{U}_{\bar{T}}'$ hopefully spreads the noise more evenly on the quadratures and leads to smaller error buildups. The two examples we investigated in this work are the hexagonal and $\mathrm{D}_4$ GKP codes, as the non-orthogonality of their stabilizers made their QEC performances much more prone to variations when changing the stabilizer matrix used to construct $\hat{U}_{\bar{T}}$.

For the hexagonal GKP, we choose $R_{\hexagon}$ that applies the transformation $\mathbf{s}_1 \rightarrow \mathbf{s}_1 + \mathbf{s}_2$, such that its quadrature-symmetric stabilizer matrix used is given by
\begin{align}
    S_{\hexagon}' = R_{\hexagon}S_{\hexagon} = \frac{1}{\sqrt[4]{3}}\begin{pmatrix}
        1 & \sqrt{3} \\
        -1 & \sqrt{3}
    \end{pmatrix}.
    \label{eq: stab_mat_sym_hex}
\end{align}
In this case, we see that both $\hat{q}_1$ and $\hat{p}_1$ are distributed the same way for the two stabilizers. For the D$_4$ GKP, $R_{\mathrm{D}_4}$ applies the transformation $\mathbf{s}_4 \rightarrow \mathbf{s}_4 - \mathbf{s}_1 - \mathbf{s}_3$, yielding 
\begin{align}
    S_{\mathrm{D}_4}' = R_{\mathrm{D}_4}S_{\mathrm{D}_4} = \begin{pmatrix}
        1 & 0 & 1 & 0 \\
        1 & 0 & 0 & -1 \\
        0 & 1 & -1 & 0 \\
        0 & -1 & 0 & 1
    \end{pmatrix}.
    \label{eq: stab_mat_sym_D4}
\end{align}
Here, there are always two stabilizers that have non-zero magnitude components spread across the modes $\hat{q}_1$, $\hat{p}_1$, $\hat{q}_2$ and $\hat{p}_2$.

We compute the QEC performances of these two GKP codes for two different Steane-type QEC circuits under the MED and the COR-MED decoders, our main decoders of interest in this work. The results are shown on Figs. \fullsubfigref{fig: sym_non_sym_comparison}{subfig: sym_non_sym_compare_hex} and \fullsubfigref{fig: sym_non_sym_comparison}{subfig: sym_non_sym_compare_D4} for the hexagonal and $\mathrm{D}_4$ GKP, respectively. We see that for the range of noise strengths $\sigma^2$ studied, the quadrature-symmetric form $S_\mathrm{G}'$ generating the Steane-type circuit $\hat{U}_{\bar{T}}'$ always gives the same or lower values of $p_L$ for both the MED and the COR-MED decoder.

Thus, we choose these stabilizer matrix representatives to compare the MED decoder with the COR-MED decoder in a fair way, where we have as good performance as we can with these decoders.


\begin{thebibliography}{10}
\bibitem{nisqEraQuantum}
John Preskill.
\newblock ``{Quantum Computing in the NISQ era and beyond}''.
\newblock \href{https://dx.doi.org/10.22331/q-2018-08-06-79}{Quantum {\bf 2},
  79}~(2018).

\bibitem{nielsen2010}
Michael~A. Nielsen and Isaac~L. Chuang.
\newblock ``{Quantum Computation and Quantum Information: 10th Anniversary
  Edition}''.
\newblock \href{https://dx.doi.org/10.1017/cbo9780511976667}{Cambridge
  University Press}. ~(2010).

\bibitem{initQubitCodes1}
Peter~W. Shor.
\newblock ``{Scheme for reducing decoherence in quantum computer memory}''.
\newblock \href{https://dx.doi.org/10.1103/PhysRevA.52.R2493}{Phys. Rev. A {\bf
  52}, R2493(R)}~(1995).

\bibitem{initQubitCodes2}
Raymond Laflamme, Cesar Miquel, Juan~P. Paz, and Wojciech~H. Zurek.
\newblock ``{Perfect Quantum Error Correcting Code}''.
\newblock \href{https://dx.doi.org/10.1103/PhysRevLett.77.198}{Phys. Rev. Lett.
  {\bf 77}, 198--201}~(1996).

\bibitem{initQubitCodes3}
A.~R. Calderbank and Peter~W. Shor.
\newblock ``{Good quantum error-correcting codes exist}''.
\newblock \href{https://dx.doi.org/10.1103/PhysRevA.54.1098}{Phys. Rev. A {\bf
  54}, 1098--1105}~(1996).

\bibitem{initQubitCodes4}
Andrew~M. Steane.
\newblock ``{Multiple-particle interference and quantum error correction}''.
\newblock \href{https://dx.doi.org/10.1098/rspa.1996.0136}{Proc. R Soc. Lond.
  A. {\bf 452}, 2551--2577}~(1996).

\bibitem{initQubitCodes5}
Charles~H. Bennett, David~P. DiVincenzo, John~A. Smolin, and William~K.
  Wootters.
\newblock ``{Mixed-state entanglement and quantum error correction}''.
\newblock \href{https://dx.doi.org/10.1103/PhysRevA.54.3824}{Phys. Rev. A {\bf
  54}, 3824--3851}~(1996).

\bibitem{gottesmanThesis}
Daniel Gottesman.
\newblock ``{Stabilizer Codes and Quantum Error Correction}''.
\newblock \href{https://dx.doi.org/10.7907/rzr7-dt72}{PhD thesis}.
\newblock {California Institute of Technology}.
\newblock ~(1997).

\bibitem{initSurfaceCode}
Eric Dennis, Alexei Kitaev, Andrew Landahl, and John Preskill.
\newblock ``{Topological quantum memory}''.
\newblock \href{https://dx.doi.org/10.1063/1.1499754}{J. Math. Phys. {\bf 43},
  4452--4505}~(2002).

\bibitem{initBosonicCodes1}
Isaac~L. Chuang, Debbie~W. Leung, and Yoshihisa Yamamoto.
\newblock ``{Bosonic quantum codes for amplitude damping}''.
\newblock \href{https://dx.doi.org/10.1103/PhysRevA.56.1114}{Phys. Rev. A {\bf
  56}, 1114--1125}~(1997).

\bibitem{initBosonicCodes2}
Samuel~L. Braunstein.
\newblock ``{Error Correction for Continuous Quantum Variables}''.
\newblock \href{https://dx.doi.org/10.1103/PhysRevLett.80.4084}{Phys. Rev.
  Lett. {\bf 80}, 4084--4087}~(1998).

\bibitem{initBosonicCodes3}
P.~T. Cochrane, Gerard~J. Milburn, and William~J. Munro.
\newblock ``{Macroscopically distinct quantum-superposition states as a bosonic
  code for amplitude damping}''.
\newblock \href{https://dx.doi.org/10.1103/PhysRevA.59.2631}{Phys. Rev. A {\bf
  59}, 2631--2634}~(1999).

\bibitem{originalGKP}
Daniel Gottesman, Alexei Kitaev, and John Preskill.
\newblock ``Encoding a qubit in an oscillator''.
\newblock \href{https://dx.doi.org/10.1103/PhysRevA.64.012310}{Phys. Rev. A
  {\bf 64}, 012310}~(2001).

\bibitem{initBosonicCodes4}
Nicolas~C. Menicucci, Peter {van Loock}, Mile Gu, Christian Weedbrook,
  Timothy~C. Ralph, and Michael~A. Nielsen.
\newblock ``{Universal Quantum Computation with Continuous-Variable Cluster
  States}''.
\newblock \href{https://dx.doi.org/10.1103/PhysRevLett.97.110501}{Phys. Rev.
  Lett. {\bf 97}, 110501}~(2006).

\bibitem{initBosonicCodes6}
Zaki Leghtas, Gerhard Kirchmair, Brian Vlastakis, Robert~J. Schoelkopf,
  Michel~H. Devoret, and Mazyar Mirrahimi.
\newblock ``{Hardware-Efficient Autonomous Quantum Memory Protection}''.
\newblock \href{https://dx.doi.org/10.1103/PhysRevLett.111.120501}{Phys. Rev.
  Lett. {\bf 111}, 120501}~(2013).

\bibitem{initBosonicCodes7}
Mazyar Mirrahimi, Zaki Leghtas, Victor~V. Albert, Steven Touzard, Robert~J.
  Schoelkopf, Liang Jiang, and Michel~H. Devoret.
\newblock ``{Dynamically protected cat-qubits: a new paradigm for universal
  quantum computation}''.
\newblock \href{https://dx.doi.org/10.1088/1367-2630/16/4/045014}{New J. Phys.
  {\bf 16}, 045014}~(2014).

\bibitem{initBosonicCodes5}
Marios~H. Michael, Matti Silveri, R.~T. Brierley, Victor~V. Albert, Juha
  Salmilehto, Liang Jiang, and S.~M. Girvin.
\newblock ``{New Class of Quantum Error-Correcting Codes for a Bosonic Mode}''.
\newblock \href{https://dx.doi.org/10.1103/PhysRevX.6.031006}{Phys. Rev. X {\bf
  6}, 031006}~(2016).

\bibitem{initBosonicCodes8}
Arne~L. Grimsmo, Joshua Combes, and Ben~Q. Baragiola.
\newblock ``{Quantum Computing with Rotation-Symmetric Bosonic Codes}''.
\newblock \href{https://dx.doi.org/10.1103/PhysRevX.10.011058}{Phys. Rev. X
  {\bf 10}, 011058}~(2020).

\bibitem{AnalogGKPSurface}
Kosuke Fukui, Akihisa Tomita, Atsushi Okamoto, and Keisuke Fujii.
\newblock ``{High-Threshold Fault-Tolerant Quantum Computation with Analog
  Quantum Error Correction}''.
\newblock \href{https://dx.doi.org/10.1103/PhysRevX.8.021054}{Phys. Rev. X {\bf
  8}, 021054}~(2018).

\bibitem{RepCatQubitCode}
Jérémie Guillaud and Mazyar Mirrahimi.
\newblock ``{Repetition Cat Qubits for Fault-Tolerant Quantum Computation}''.
\newblock \href{https://dx.doi.org/10.1103/PhysRevX.9.041053}{Phys. Rev. X {\bf
  9}, 041053}~(2019).

\bibitem{QECwithToricGKP}
Christophe Vuillot, Hamed Asasi, Yang Wang, Leonid~P. Pryadko, and Barbara~M.
  Terhal.
\newblock ``{Quantum error correction with the toric Gottesman-Kitaev-Preskill
  code}''.
\newblock \href{https://dx.doi.org/10.1103/PhysRevA.99.032344}{Phys. Rev. A
  {\bf 99}, 032344}~(2019).

\bibitem{nohFaultTolerantGKP}
Kyungjoo Noh and Christopher Chamberland.
\newblock ``{Fault-tolerant bosonic quantum error correction with the
  surface-Gottesman-Kitaev-Preskill code}''.
\newblock \href{https://dx.doi.org/10.1103/PhysRevA.101.012316}{Phys. Rev. A
  {\bf 101}, 012316}~(2020).

\bibitem{XZZXKerrCatQubitCode}
Andrew~S. Darmawan, Benjamin~J. Brown, Arne~L. Grimsmo, David~K. Tuckett, and
  Shruti Puri.
\newblock ``{Practical Quantum Error Correction with the XZZX Code and Kerr-Cat
  Qubits}''.
\newblock \href{https://dx.doi.org/10.1103/PRXQuantum.2.030345}{PRX Quantum
  {\bf 2}, 030345}~(2021).

\bibitem{MBQCSurfaceCodeGKP}
Mikkel~V. Larsen, Christopher Chamberland, Kyungjoo Noh, Jonas~S.
  Neergaard-Nielsen, and Ulrik~L. Andersen.
\newblock ``{Fault-Tolerant Continuous-Variable Measurement-based Quantum
  Computation Architecture}''.
\newblock \href{https://dx.doi.org/10.1103/PRXQuantum.2.030325}{PRX Quantum
  {\bf 2}, 030325}~(2021).

\bibitem{colorCodeGKP}
Jiaxuan Zhang, Jian Zhao, Yu-Chun Wu, and Guo-Ping Guo.
\newblock ``{Quantum error correction with the color-Gottesman-Kitaev-Preskill
  code}''.
\newblock \href{https://dx.doi.org/10.1103/PhysRevA.104.062434}{Phys. Rev. A
  {\bf 104}, 062434}~(2021).

\bibitem{blueprintXanaduGKP}
J.~Eli Bourassa, Rafael~N. Alexander, Michael Vasmer, Ashlesha Patil, Ilan
  Tzitrin, Takaya Matsuura, Daiqin Su, Ben~Q. Baragiola, Saikat Guha, et~al.
\newblock ``{Blueprint for a Scalable Photonic Fault-Tolerant Quantum
  Computer}''.
\newblock \href{https://dx.doi.org/10.22331/q-2021-02-04-392}{Quantum {\bf 5},
  392}~(2021).

\bibitem{nohLowOverheadFaultTolerantGKP}
Kyungjoo Noh, Christopher Chamberland, and Fernando~G.S.L. Brandão.
\newblock ``Low-{Overhead} {Fault}-{Tolerant} {Quantum} {Error} {Correction}
  with the {Surface}-{GKP} {Code}''.
\newblock \href{https://dx.doi.org/10.1103/PRXQuantum.3.010315}{PRX Quantum
  {\bf 3}, 010315}~(2022).

\bibitem{GKPAndQLDPCConstruction}
Nithin Raveendran, Narayanan Rengaswamy, Filip Rozp\c{e}dek, Ankur Raina, Liang
  Jiang, and Bane Vasi\'{c}.
\newblock ``{Finite Rate QLDPC-GKP Coding Scheme that Surpasses the CSS Hamming
  Bound}''.
\newblock \href{https://dx.doi.org/10.22331/q-2022-07-20-767}{Quantum {\bf 6},
  767}~(2022).

\bibitem{RepGKPQubitCode}
Matthew~P. Stafford and Nicolas~C. Menicucci.
\newblock ``{Biased Gottesman-Kitaev-Preskill repetition code}''.
\newblock \href{https://dx.doi.org/10.1103/PhysRevA.108.052428}{Phys. Rev. A
  {\bf 108}, 052428}~(2023).

\bibitem{lin_2023a}
Mao Lin, Christopher Chamberland, and Kyungjoo Noh.
\newblock ``{Closest Lattice Point Decoding for Multimode
  Gottesman-Kitaev-Preskill Codes}''.
\newblock \href{https://dx.doi.org/10.1103/PRXQuantum.4.040334}{PRX Quantum
  {\bf 4}, 040334}~(2023).

\bibitem{XZZXSurfaceCodeGKP}
Jiaxuan Zhang, Yu-Chun Wu, and Guo-Ping Guo.
\newblock ``{Concatenation of the Gottesman-Kitaev-Preskill code with the XZZX
  surface code}''.
\newblock \href{https://dx.doi.org/10.1103/PhysRevA.107.062408}{Phys. Rev. A
  {\bf 107}, 062408}~(2023).

\bibitem{XanaduAurora}
Hanieh Aghaee~Rad, Thomas Ainsworth, Rafael~N. Alexander, Brandon Altieri,
  Mohsen~F. Askarani, R.~Baby, Leonardo Banchi, Ben~Q. Baragiola, J.~Eli
  Bourassa, et~al.
\newblock ``{Scaling and networking a modular photonic quantum computer}''.
\newblock \href{https://dx.doi.org/10.1038/s41586-024-08406-9}{Nature {\bf
  638}, 912--919}~(2025).

\bibitem{BicycleCodesXanadu}
Blayney~W. Walshe, Ben~Q. Baragiola, Hugo Ferretti, Jos\'e Gefaell, Michael
  Vasmer, Ryohei Weil, Takaya Matsuura, Thomas Jaeken, Giacomo Pantaleoni,
  et~al.
\newblock ``{Linear-Optical Quantum Computation with Arbitrary Error-Correcting
  Codes}''.
\newblock \href{https://dx.doi.org/10.1103/PhysRevLett.134.100602}{Phys. Rev.
  Lett. {\bf 134}, 100602}~(2025).

\bibitem{hybridOscillatorQuditVVAlbertPaper}
Sayan Chakraborty and Victor~V. Albert.
\newblock ``{Hybrid Oscillator-Qudit Quantum Processors: stabilizer states,
  stabilizer codes, symplectic operations, and non-commutative
  geometry}''~(2025).
\newblock  \href{http://arxiv.org/abs/2508.04819}{arXiv:2508.04819}.

\bibitem{hydridCatRepCode}
Harald Putterman, Kyungjoo Noh, Connor~T. Hann, Gregory~S. MacCabe, Shahriar
  Aghaeimeibodi, Rishi~N. Patel, Menyoung Lee, and William~M. Jones.
\newblock ``{Hardware-efficient quantum error correction via concatenated
  bosonic qubits}''.
\newblock \href{https://dx.doi.org/10.1038/s41586-025-08642-7}{Nature {\bf
  638}, 927--934}~(2025).

\bibitem{albertGKPBetterCode}
Victor~V. Albert, Kyungjoo Noh, Kasper Duivenvoorden, Dylan~J. Young, R.~T.
  Brierley, Philip Reinhold, Christophe Vuillot, Linshu Li, Chao Shen, et~al.
\newblock ``{Performance and structure of single-mode bosonic codes}''.
\newblock \href{https://dx.doi.org/10.1103/PhysRevA.97.032346}{Phys. Rev. A
  {\bf 97}, 032346}~(2018).

\bibitem{nohGKPBetterCode}
Kyungjoo Noh, Victor~V. Albert, and Liang Jiang.
\newblock ``{Quantum Capacity Bounds of Gaussian Thermal Loss Channels and
  Achievable Rates With Gottesman-Kitaev-Preskill Codes}''.
\newblock \href{https://dx.doi.org/10.1109/TIT.2018.2873764}{IEEE Trans. Inf.
  Theory {\bf 65}, 2563--2582}~(2019).

\bibitem{leviant_2022}
Peter Leviant, Qian Xu, Liang Jiang, and Serge Rosenblum.
\newblock ``Quantum capacity and codes for the bosonic loss-dephasing
  channel''.
\newblock \href{https://dx.doi.org/10.22331/q-2022-09-29-821}{Quantum {\bf 6},
  821}~(2022).

\bibitem{theoricPrepGKP1}
Ben~C. Travaglione and Gerard~J. Milburn.
\newblock ``{Preparing encoded states in an oscillator}''.
\newblock \href{https://dx.doi.org/10.1103/PhysRevA.66.052322}{Phys. Rev. A
  {\bf 66}, 052322}~(2002).

\bibitem{theoricPrepGKP2}
Stefano Pirandola, Stefano Mancini, David Vitali, and Paolo Tombesi.
\newblock ``{Constructing finite-dimensional codes with optical continuous
  variables}''.
\newblock \href{https://dx.doi.org/10.1209/epl/i2004-10203-9}{Europhys. Lett.
  {\bf 68}, 323--329}~(2004).

\bibitem{theoricPrepGKP3}
Stefano Pirandola, Stefano Mancini, David Vitali, and Paolo Tombesi.
\newblock ``{Continuous variable encoding by ponderomotive interaction}''.
\newblock \href{https://dx.doi.org/10.1140/epjd/e2005-00306-3}{Eur. Phys. J. D
  {\bf 37}, 283--290}~(2006).

\bibitem{theoricPrepGKP4}
Stefano Pirandola, Stefano Mancini, David Vitali, and Paolo Tombesi.
\newblock ``{Generating continuous variable quantum codewords in the near-field
  atomic lithography}''.
\newblock \href{https://dx.doi.org/10.1088/0953-4075/39/4/023}{J. Phys. B: At.
  Mol. Opt. Phys. {\bf 39}, 997}~(2006).

\bibitem{theoricPrepGKP5}
Hilma~M. Vasconcelos, Liliana Sanz, and Scott Glancy.
\newblock ``{All-optical generation of states for `Encoding a qubit in an
  oscillator`'}''.
\newblock \href{https://dx.doi.org/10.1364/OL.35.003261}{Opt. Lett. {\bf 35},
  3261--3263}~(2010).

\bibitem{prepGKPPhaseEstimation}
Barbara~M. Terhal and Daniel Weigand.
\newblock ``{Encoding a qubit into a cavity mode in circuit QED using phase
  estimation}''.
\newblock \href{https://dx.doi.org/10.1103/PhysRevA.93.012315}{Phys. Rev. A
  {\bf 93}, 012315}~(2016).

\bibitem{theoricPrepGKP7}
Keith~R. Motes, Ben~Q. Baragiola, Alexei Gilchrist, and Nicolas~C. Menicucci.
\newblock ``{Encoding qubits into oscillators with atomic ensembles and
  squeezed light}''.
\newblock \href{https://dx.doi.org/10.1103/PhysRevA.95.053819}{Phys. Rev. A
  {\bf 95}, 053819}~(2017).

\bibitem{theoricPrepGKP8}
Jacob Hastrup, Kimin Park, Jonatan~B. Brask, Radim Filip, and Ulrik~L.
  Andersen.
\newblock ``{Measurement-free preparation of grid states}''.
\newblock \href{https://dx.doi.org/10.1038/s41534-020-00353-3}{npj Quantum Inf.
  {\bf 7}, 17}~(2021).

\bibitem{campagneGKPQEC}
Philippe Campagne-Ibarcq, Alec Eickbusch, Steven Touzard, Evan Zalys-Geller,
  Nicholas~E. Frattini, Volodymyr~V. Sivak, Philip. Reinhold, Shruti Puri,
  Shyam Shankar, et~al.
\newblock ``{Quantum error correction of a qubit encoded in grid states of an
  oscillator}''.
\newblock \href{https://dx.doi.org/10.1038/s41586-020-2603-3}{Nature {\bf 584},
  368--372}~(2020).

\bibitem{eickbushGKPQEC}
Alec Eickbusch, Volodymyr Sivak, Andy~Z. Ding, Salvatore~S. Elder, Shantanu~R.
  Jha, Jayameenakshi Venkatraman, Baptiste Royer, Steven~M. Girvin, Robert~J.
  Schoelkopf, et~al.
\newblock ``{Fast universal control of an oscillator with weak dispersive
  coupling to a qubit}''.
\newblock \href{https://dx.doi.org/10.1038/s41567-022-01776-9}{Nat. Phys. {\bf
  18}, 1464--1469}~(2022).

\bibitem{sivakRealtimeQEC}
Volodymyr~V. Sivak, Alec Eickbusch, Baptiste Royer, Shraddha Singh, Ioannis
  Tsioutsios, Suhas Ganjam, Alessandro Miano, Benjamin~L. Brock, Andy~Z. Ding,
  et~al.
\newblock ``{Real-time quantum error correction beyond break-even}''.
\newblock \href{https://dx.doi.org/10.1038/s41586-023-05782-6}{Nature {\bf
  616}, 50--55}~(2023).

\bibitem{NQGKPQEC}
Dany Lachance-Quirion, Marc-Antoine Lemonde, Jean~Olivier Simoneau, Lucas
  St-Jean, Pascal Lemieux, Sara Turcotte, Wyatt Wright, Amélie Lacroix,
  Joëlle Fréchette-Viens, et~al.
\newblock ``{Autonomous Quantum Error Correction of Gottesman-Kitaev-Preskill
  States}''.
\newblock \href{https://dx.doi.org/10.1103/PhysRevLett.132.150607}{Phys. Rev.
  Lett. {\bf 132}, 150607}~(2024).

\bibitem{brockGKPQudits}
Benjamin~L. Brock, Shraddha Singh, Alec Eickbusch, Volodymyr~V. Sivak, Andy~Z.
  Ding, Luigi Frunzio, Steven~M. Girvin, and Michel~H. Devoret.
\newblock ``{Quantum error correction of qudits beyond break-even}''.
\newblock \href{https://dx.doi.org/10.1038/s41586-025-08899-y}{Nature {\bf
  641}, 612--618}~(2025).

\bibitem{MITGKPPaper}
Shantanu~R. Jha, Shoumik~D. Chowdhury, Gabriele Rolleri, Anaida Ali, Lev-Arcady
  Sellem, Réouven Assouly, David Pahl, Lukas Pahl, Junyoung An, et~al.
\newblock ``{Bosonic Error Correction with Fluxonium}''~(2026).
\newblock  \href{http://arxiv.org/abs/2609.26880}{arXiv:2609.26880}.

\bibitem{fluhmannEncodingQubitTrappedion2019}
Christa Flühmann, Thanh-Long Nguyen, Matteo Marinelli, Vlad Negnevitsky, Karan
  Mehta, and Jonathan~P. Home.
\newblock ``{Encoding a Qubit in a Trapped-Ion Mechanical Oscillator}''.
\newblock \href{https://dx.doi.org/10.1038/s41586-019-0960-6}{Nature {\bf 566},
  513--517}~(2019).

\bibitem{deNeeveTrappedIonGKP}
Brennan de~Neeve, Thanh-Long Nguyen, Tanja Behrle, and Jonathan~P. Home.
\newblock ``{Error correction of a logical grid state qubit by dissipative
  pumping}''.
\newblock \href{https://dx.doi.org/10.1038/s41567-021-01487-7}{Nat. Phys. {\bf
  18}, 296--300}~(2022).

\bibitem{matsosTrappedIonGKP}
Vassili~G. Matsos, Christophe~H. Valahu, Thomas Navickas, Arjun~D. Rao,
  Maverick~J. Millican, Xanda~C. Kolesnikow, Michael~J. Biercuk, and Ting~R.
  Tan.
\newblock ``{Robust and Deterministic Preparation of Bosonic Logical States in
  a Trapped Ion}''.
\newblock \href{https://dx.doi.org/10.1103/PhysRevLett.133.050602}{Phys. Rev.
  Lett. {\bf 133}, 050602}~(2024).

\bibitem{TrappedIonGKPSensing}
Christophe~H. Valahu, Matthew~P. Stafford, Zixin Huang, Vassili~G. Matsos,
  Maverick~J. Millican, Teerawat Chalermpusitarak, Nicolas~C. Menicucci, Joshua
  Combes, Ben~Q. Baragiola, et~al.
\newblock ``{Quantum-enhanced multiparameter sensing in a single mode}''.
\newblock \href{https://dx.doi.org/10.1126/sciadv.adw9757}{Sci. Adv. {\bf 11},
  eadw9757}~(2025).

\bibitem{fabreGenerationTimefrequencyGrid2020}
Nicolas Fabre, Giorgio Maltese, Félicien Appas, Simone Felicetti, Andreas
  Ketterer, Arne Keller, Thomas Coudreau, Florent Baboux, Maria~I. Amanti,
  et~al.
\newblock ``{Generation of time-frequency grid state with integrated biphoton
  frequency combs}''.
\newblock \href{https://dx.doi.org/10.1103/PhysRevA.102.012607}{Phys. Rev. A
  {\bf 102}, 012607}~(2020).

\bibitem{FurusawaGKPStatePrep}
Shunya Konno, Warit Asavanant, Fumiya Hanamura, Hironari Nagayoshi, Kosuke
  Fukui, Atsushi Sakaguchi, Ryuhoh Ide, Fumihiro China, Masahiro Yabuno, et~al.
\newblock ``{Logical states for fault-tolerant quantum computation with
  propagating light}''.
\newblock \href{https://dx.doi.org/10.1126/science.adk7560}{Science {\bf 383},
  289--293}~(2024).

\bibitem{XanaduSquareGKPStatePrep}
Mikkel~V. Larsen, J.~Eli Bourassa, Sacha Kocsis, Joel~F. Tasker, Robert~S.
  Chadwick, Carlos González-Arciniegas, Jacob Hastrup, Carlos~E.
  Lopetegui-González, Filippo~M. Miatto, et~al.
\newblock ``{Integrated photonic source of Gottesman–Kitaev–Preskill
  qubits}''.
\newblock \href{https://dx.doi.org/10.1038/s41586-025-09044-5}{Nature {\bf
  642}, 587--591}~(2025).

\bibitem{GKPRatesHarrington}
Jim Harrington and John Preskill.
\newblock ``{Achievable rates for the Gaussian quantum channel}''.
\newblock \href{https://dx.doi.org/10.1103/PhysRevA.64.062301}{Phys. Rev. A
  {\bf 64}, 062301}~(2001).

\bibitem{conradGKPMultimode}
Jonathan Conrad, Jens Eisert, and Francesco Arzani.
\newblock ``{Gottesman-Kitaev-Preskill codes: A lattice perspective}''.
\newblock \href{https://dx.doi.org/10.22331/q-2022-02-10-648}{Quantum {\bf 6},
  648}~(2022).

\bibitem{bapGKPMultimode}
Baptiste Royer, Shraddha Singh, and Steven~M. Girvin.
\newblock ``{Encoding Qubits in Multimode Grid States}''.
\newblock \href{https://dx.doi.org/10.1103/PRXQuantum.3.010335}{PRX Quantum
  {\bf 3}, 010335}~(2022).

\bibitem{GlancyKnillGKP}
Scott Glancy and Emanuel Knill.
\newblock ``{Error analysis for encoding a qubit in an oscillator}''.
\newblock \href{https://dx.doi.org/10.1103/PhysRevA.73.012325}{Phys. Rev. A
  {\bf 73}, 012325}~(2006).

\bibitem{schnorrLatticeBasisReduction1994}
Claus~P. Schnorr and Martin Euchner.
\newblock ``{Lattice basis reduction: Improved practical algorithms and solving
  subset sum problems}''.
\newblock \href{https://dx.doi.org/10.1007/BF01581144}{Math. Program. {\bf 66},
  181--199}~(1994).

\bibitem{agrellClosestPointSearch2002}
Erik Agrell, Thomas Eriksson, Alexander Vardy, and Kenneth Zeger.
\newblock ``Closest point search in lattices''.
\newblock \href{https://dx.doi.org/10.1109/TIT.2002.800499}{IEEE Trans. Inf.
  Theory {\bf 48}, 2201--2214}~(2002).

\bibitem{LossToRGTNoiseModel2}
Ilan Tzitrin, Takaya Matsuura, Rafael~N. Alexander, Guillaume Dauphinais,
  J.~Eli Bourassa, Krishna~K. Sabapathy, Nicolas~C. Menicucci, and Ish Dhand.
\newblock ``{Fault-Tolerant Quantum Computation with Static Linear Optics}''.
\newblock \href{https://dx.doi.org/10.1103/PRXQuantum.2.040353}{PRX Quantum
  {\bf 2}, 040353}~(2021).

\bibitem{gaussianQuantumInfo}
Christian Weedbrook, Stefano Pirandola, Raúl García-Patrón, Nicolas~J. Cerf,
  Timothy~C. Ralph, Jeffrey~H. Shapiro, and Seth Lloyd.
\newblock ``{Gaussian quantum information}''.
\newblock \href{https://dx.doi.org/10.1103/RevModPhys.84.621}{Rev. Mod. Phys.
  {\bf 84}, 621}~(2012).

\bibitem{joe_logical_compiler}
Jonathan Pelletier and Baptiste Royer.
\newblock ``{Enlarging the GKP stabilizer group for enhanced noise
  protection}''.
\newblock \href{https://dx.doi.org/10.22331/q-2026-07-08-2156}{Quantum {\bf
  10}, 2156}~(2026).

\bibitem{voronoiPaper}
Johannes Blömer and Kathlén Kohn.
\newblock ``{Voronoi Cells of Lattices with Respect to Arbitrary Norms}''.
\newblock \href{https://dx.doi.org/10.1137/17M1132045}{SIAGA {\bf 2},
  314--338}~(2018).

\bibitem{bapSBSpaper}
Baptiste Royer, Shraddha Singh, and S.~M. Girvin.
\newblock ``{Stabilization of Finite-Energy Gottesman-Kitaev-Preskill
  States}''.
\newblock \href{https://dx.doi.org/10.1103/PhysRevLett.125.260509}{Phys. Rev.
  Lett. {\bf 125}, 260509}~(2020).

\bibitem{twoModeSqueezingCouplingGates1}
Barbara Kraus, Klemens Hammerer, G{\'e}za Giedke, and Juan~I. Cirac.
\newblock ``{Entanglement generation and Hamiltonian simulation in
  continuous-variable systems}''.
\newblock \href{https://dx.doi.org/10.1103/PhysRevA.67.042314}{Phys. Rev. A
  {\bf 67}, 042314}~(2003).

\bibitem{twoModeSqueezingCouplingGates2}
Jarom{\'i}r Fiur{\'i}\v{s}ek.
\newblock ``Unitary-gate synthesis for continuous-variable systems''.
\newblock \href{https://dx.doi.org/10.1103/PhysRevA.68.022304}{Phys. Rev. A
  {\bf 68}, 022304}~(2003).

\bibitem{sensorStateGKP}
Kasper Duivenvoorden, Barbara~M. Terhal, and Daniel Weigand.
\newblock ``{Single-mode displacement sensor}''.
\newblock \href{https://dx.doi.org/10.1103/PhysRevA.95.012305}{Phys. Rev. A
  {\bf 95}, 012305}~(2017).

\bibitem{qunaughtGKPName}
Blayney~W. Walshe, Ben~Q. Baragiola, Rafael~N. Alexander, and Nicolas~C.
  Menicucci.
\newblock ``{Continuous-variable gate teleportation and bosonic-code error
  correction}''.
\newblock \href{https://dx.doi.org/10.1103/PhysRevA.102.062411}{Phys. Rev. A
  {\bf 102}, 062411}~(2020).

\bibitem{GKPRatesLin}
Mao Lin and Kyungjoo Noh.
\newblock ``{Exploring the quantum capacity of a Gaussian random-displacement
  channel using Gottesman-Kitaev-Preskill codes and maximum-likelihood
  decoding}''.
\newblock \href{https://dx.doi.org/10.1103/PhysRevA.111.052445}{Phys. Rev. A
  {\bf 111}, 052445}~(2025).

\bibitem{introToMultivariateStatAnalysis}
Theodore~W. Anderson.
\newblock ``{An Introduction to Multivariate Statistical Analysis, 3rd
  Edition}''.
\newblock
  \href{https://www.wiley.com/en-us/An+Introduction+to+Multivariate+Statistical+Analysis%2C+3rd+Edition-p-9780471360919}{John
  Wiley \& Sons} (2003).

\bibitem{LossToRGTNoiseModel1}
Kosuke Fukui, Rafael~N. Alexander, and Peter van Loock.
\newblock ``{All-optical long-distance quantum communication with
  Gottesman-Kitaev-Preskill qubits}''.
\newblock \href{https://dx.doi.org/10.1103/PhysRevResearch.3.033118}{Phys. Rev.
  Res. {\bf 3}, 033118}~(2021).

\bibitem{QECwithGKPreview}
Arne~L. Grimsmo and Shruti Puri.
\newblock ``{Quantum Error Correction with the Gottesman-Kitaev-Preskill
  Code}''.
\newblock \href{https://dx.doi.org/10.1103/PRXQuantum.2.020101}{PRX Quantum
  {\bf 2}, 020101}~(2021).

\bibitem{matrixAnalysis}
Roger~A. Horn and Charles~R. Johnson.
\newblock ``{Matrix Analysis}''.
\newblock \href{https://dx.doi.org/10.1017/CBO9781139020411}{Cambridge
  University Press}. ~(2012).

\bibitem{fellerProbabilityTheory}
William Feller.
\newblock ``{An Introduction to Probability Theory and Its Applications, 3rd
  Edition}''.
\newblock
  \href{https://www.wiley.com/en-us/shop/general-introductory-statistics/an-introduction-to-probability-theory-and-its-applications-volume-1-3rd-edition-p-9780471257080#description-section}{John
  Wiley \& Sons} (1968).

\bibitem{Julia}
Jeff Bezanson, Alan Edelman, Stefan Karpinski, and Viral~B. Shah.
\newblock ``{Julia: A fresh approach to numerical computing}''.
\newblock \href{https://dx.doi.org/10.1137/141000671}{SIREV {\bf 59},
  65--98}~(2017).

\bibitem{matsosUniversalQuantumGate2025}
Vassili~G. Matsos, Christophe~H. Valahu, Maverik~J. Millican, Thomas Navickas,
  Xanda~C. Kolesnikow, Michael~J. Biercuk, and Ting~R. Tan.
\newblock ``{Universal Quantum Gate Set for Gottesman-Kitaev-Preskill Logical
  Qubits}''.
\newblock \href{https://dx.doi.org/10.1038/s41567-025-03002-8}{Nat. Phys. {\bf
  21}, 1664--1669}~(2025).

\bibitem{QubitDyneSimPaper}
Ingrid Strandberg, Axel~M. Eriksson, Baptiste Royer, Mikael Kervinen, and
  Simone Gasparinetti.
\newblock ``{Digital Homodyne and Heterodyne Detection for Stationary Bosonic
  Modes}''.
\newblock \href{https://dx.doi.org/10.1103/PhysRevLett.133.063601}{Phys. Rev.
  Lett. {\bf 133}, 063601}~(2024).

\bibitem{goodHomodyneDetect}
Henning Vahlbruch, Moritz Mehmet, Karsten Danzmann, and Roman Schnabel.
\newblock ``{Detection of 15 dB Squeezed States of Light and their Application
  for the Absolute Calibration of Photoelectric Quantum Efficiency}''.
\newblock \href{https://dx.doi.org/10.1103/PhysRevLett.117.110801}{Phys. Rev.
  Lett. {\bf 117}, 110801}~(2016).

\bibitem{KnillQECInitpaper}
Emanuel Knill.
\newblock ``{Quantum computing with realistically noisy devices}''.
\newblock \href{https://dx.doi.org/10.1038/nature03350}{Nature {\bf 434},
  39--44}~(2005).

\bibitem{TelepBosonicNoisyMeasurements}
Timo Hillmann, Fernando Quijandría, Arne~L. Grimsmo, and Giulia Ferrini.
\newblock ``{Performance of Teleportation-Based Error-Correction Circuits for
  Bosonic Codes with Noisy Measurements}''.
\newblock \href{https://dx.doi.org/10.1103/PRXQuantum.3.020334}{PRX Quantum
  {\bf 3}, 020334}~(2022).

\bibitem{SteaneQECInitpaper}
Andrew~M. Steane.
\newblock ``{Active Stabilization, Quantum Computation, and Quantum State
  Synthesis}''.
\newblock \href{https://dx.doi.org/10.1103/PhysRevLett.78.2252}{Phys. Rev.
  Lett. {\bf 78}, 2252}~(1997).

\bibitem{distBetweenCovMat}
Wolfgang F{\"o}rstner and Boudewijn Moonen.
\newblock ``{A Metric for Covariance Matrices}''.
\newblock In Erik~W. Grafarend, Friedrich~W. Krumm, and Volker~S. Schwarze,
  editors, Geodesy -- The Challenge of the 3rd Millennium.
\newblock \href{https://dx.doi.org/10.1007/978-3-662-05296-9_31}{Pages
  299--309}.
\newblock Springer, Berlin, Heidelberg~(2003).

\bibitem{computerVision}
Simon J.~D. Prince.
\newblock ``{Computer vision: models, learning and inference}''.
\newblock \href{https://dx.doi.org/10.1017/CBO9780511996504}{Cambridge
  University Press}. ~(2012).

\end{thebibliography}
\end{document}